\documentclass[fleqn,usenatbib]{mnras}

\usepackage[T1]{fontenc}

\DeclareRobustCommand{\VAN}[3]{#2}
\let\VANthebibliography\thebibliography
\def\thebibliography{\DeclareRobustCommand{\VAN}[3]{##3}\VANthebibliography}

\usepackage{graphicx}	% Including figure files
\usepackage{amsmath}	% Advanced maths commands
\usepackage{subfigure}
\usepackage{multirow}
\usepackage{multicol}
\usepackage[normalem]{ulem}

\usepackage{newtxtext,newtxmath}

\title[Dynamics around spherical body with mass anomaly]{Dynamics around Non-Spherical Symmetric Bodies: I. The case of a spherical body with mass anomaly}

\author[Madeira et al.]{
G. Madeira,$^{1,2}$\thanks{E-mail: madeira@ipgp.fr}
S. M. Giuliatti Winter,$^{1}$
T. Ribeiro,$^{1}$
and O. C. Winter$^{1}$
\\
$^{1}$Grupo de Din\^amica Orbital \& Planetologia, S\~ao Paulo State University -UNESP, Av. Ariberto Pereira da Cunha, 333, Guaratinguet\'a SP, 12516-410, Brazil\\
$^{2}$Universit\'e de Paris, Institut de Physique du Globe de Paris, CNRS, F-75005 Paris, France
}

\date{Accepted 2021 December 02. Received 2021 November 10; in original form 2021 August 06}

\pubyear{2015}

\begin{document}
\label{firstpage}
\pagerange{\pageref{firstpage}--\pageref{lastpage}}
\maketitle

\begin{abstract}
The space missions designed to visit small bodies of the Solar System boosted the study of the dynamics around non-spherical bodies. In this vein, we study the dynamics around a class of objects classified by us as Non-Spherical Symmetric Bodies, including contact binaries, triaxial ellipsoids, spherical bodies with a mass anomaly, among others. In the current work, we address the results for a body with a mass anomaly. We apply the pendulum model to obtain the width of the spin-orbit resonances raised by non-asymmetric gravitational terms of the central object. The Poincaré surface of section technique is adopted to confront our analytical results and to study the system's dynamics by varying the parameters of the central object. We verify the existence of two distinct regions around an object with a mass anomaly: a chaotic inner region that extends beyond the corotation radius and a stable outer region. In the latter, we identify structures remarkably similar to those of the classical restrict and planar 3-body problem in the Poincaré surface of sections, including asymmetric periodic orbits associated with 1:1+p resonances. We apply our results to a Chariklo with a mass anomaly, obtaining that Chariklo rings are probably related to first kind periodic orbits and not with 1:3 spin-orbit resonance, as proposed in the literature. We believe that our work presents the first tools for studying mass anomaly systems.
\end{abstract}

\begin{keywords}
celestial mechanics -- Kuiper belt objects: individual: (10199) Chariklo -- minor planets, asteroids: general -- planets and satellites: dynamical evolution and stability -- planets and satellites: rings
\end{keywords}

%%%%%%%%%%%%%%%%%%%%%%%%%%%%%%%%%%%%%%%%%%%%%%%%%%
%%%%%%%%%%%%%%%%% BODY OF PAPER %%%%%%%%%%%%%%%%%%
\section{Introduction}
In the last three decades, the acquisition of data on the shape of small heliocentric bodies, by ground and space-based observations \citep{Hudson1995,Hanuvs2013,Hanuvs2017} and by space-mission explorations -- such as OSIRIS-REx and Hayabusa spacecraft \citep{Yoshikawa2015,Lauretta2017} -- fostered the study of the dynamics around these bodies. This class of objects, which involves asteroids, trans-Neptunian objects, Centaurs, and comets, characteristically have diameters of less than 1000~km  \citep{Jorda2016}. Due to their small sizes, these bodies do not have enough mass to reach hydrostatic equilibrium, showing irregular and asymmetric shapes. 

The development of space-missions was a strong motivation for the search of equilibrium regions around irregular bodies, as accomplished, e.g., by \cite{Scheeres2000} which obtained stable orbits around the asteroid 433 Eros for the spacecraft NEAR-Shoemaker \citep{Prockter2002}. Some other works with such purpose are \cite{Yu2012,Shang2015,Wang2016,Winter2020,Moura2020}. The discovery of satellites and rings around this class of objects were also justifications for the interest in the stability of irregular bodies systems \citep{Chapman1995,Merline2002,Braga2014,Ortiz2017}. 

When investigating the motion around irregular bodies, it is essential to consider the gravitational field generated by their odd shape. One method used for this is to approximate the irregular shape to a symmetric one -- such as a MacLaurin spheroid or a triaxial ellipsoid -- which allows studying the system theoretically or through low-cost simulations. Another course of action is to decompose the irregular body into a set of regular polyhedra \citep[Polyhedron Shape Model,][]{Werner1994} or mass points \citep[Mascon Model,][]{Geissler1996}. Despite the high level of accuracy, this methodology has a higher computational cost.

In the current and subsequent works, we study the dynamics around a class of objects classified by us as Non-Spherical Symmetric Bodies (NSSBs): contact binaries, triaxial ellipsoids with uniform density, and spherical bodies with a mass anomaly. The motion around NSSBs has already been studied in some articles, such as \cite{Lages2017} which analysed the stability around contact binaries through a generalized Kepler map technique \citep{Meiss1992,Shevchenko2011}, obtaining chaotic gravitational zones around the central body, similar to those found for symmetrical elongated bodies \citep{Mysen2006,Mysen2007}. Their results are appliable to the asteroids 243~Ida and 25143~Itokawa \citep{Lages2017}. 

\cite{Lages2018} also use the Kepler map technique to study the chaotic region around cometary nuclei of dumb-bell shape, obtaining that such region is responsible for engulfing most of the Hill sphere of Comet 1P/Halley. \cite{Amarante2020} studied the dynamics around 486958~Arrokoth, an object similar to a contact binary, using a Polyhedron Shape Model and found an unstable zone in the equatorial region of the asteroid. \cite{Rollin2021} obtain that the particles in the equatorial plane of 486958~Arrokoth are lost due to the chaotic diffusion of the orbits, which results in collisions or particle ejection. Interestingly, \cite{Rollin2021} also obtain theoretical dumb-bell-shaped objects with certain combinations of mass and spin period that host, not a complete chaotic zone, but a chaotic ring.

The dynamics around triaxial ellipsoids were previously studied by \cite{Scheeres1994,Vantieghem2014}, and in particular by \cite{Winter2019} which analysed the motion around 136108~Haumea, an ellipsoidal-shape object. This dwarf planet is particularly interesting due to its complex system that includes a pair of satellites, Hi’iaka and Namaka, and a ring \citep{Ragozzine2009,Ortiz2017}. The non-asymmetric terms of the gravitational field of the NSSBs create strong resonances between the orbital period of the ring particles and the spin of the central body. \cite{Ortiz2017} propose that the Haumea ring would be associated with the 1:3 resonance. However, \cite{Winter2019} using Poincar\'e surface of sections showed that this resonance is doubled, generating a large chaotic region in the resonance separatrix. Consequently, the ring is not associated with the 1:3 resonance but probably with first kind periodic orbits. 

10199~Chariklo is another irregular body with a complex system involving a pair of narrow rings and possibly small satellites \citep{Braga2014,Berard2017}. The shape of Chariklo is still not well known. Observational data suggest triaxial and Jacobi ellipsoid shapes for the object \citep{Leiva2017}. \cite{Sicardy2020} discuss the possibility of Chariklo to be a sphere with topographical features of a few kilometres, i.e., an object with a mass anomaly. Assuming a spherical Chariklo with a mass anomaly, \cite{Sicardy2019} and \cite{Sicardy2020b} show that particles inside the corotation radius migrate onto the body, and the outer material is pushed beyond the 1:2 resonance.

Here, we apply some well-known techniques to study the dynamics around a spherical body with a mass anomaly. Relations for the width and location of the spin-orbit resonances, a.k.a., sectoral resonances are presented. The dependence of resonances on the central body parameters are analysed. Poincaré surface of section technique is also applied to the system for analysing the stability of the particles. We advance the reader the existence of a chaotic region near the object with a mass anomaly. This region extension is measured, and an adjusted equation is obtained as a function of the system parameters.

In Section~\ref{dynamicalsystem}, we present the disturbing function of our case of interest. In Section~\ref{theory}, we follow the prescription of the pendulum model developed by \cite{Winter1997a} and \cite{Murray1999} for the restricted planar 3-body problem (RP3BP) to obtain an analytical recipe for the location and width of the spin-orbit resonances. Section~\ref{secPSS} presents the Poincaré Surface of Section technique \citep{Henon1965a,Henon1965b,Henon1966a,Henon1966b,Henon1969,Jefferys1971}. In Section~\ref{overview}, we identify, through numerical simulations, stable regions and give an overview of the system. In Section~\ref{resoperi}, we use the Poincaré surface of section technique to confront our analytical model and study the spin-orbit resonances in details. We apply our results to Chariklo in Section~\ref{chariklosection}, exploring the dynamics around the object, in particular in the region of the rings. We address our final comments in Section~\ref{secdiscu}.

\section{Dynamical System} \label{dynamicalsystem}
\begin{figure}
\includegraphics[width=1.\columnwidth]{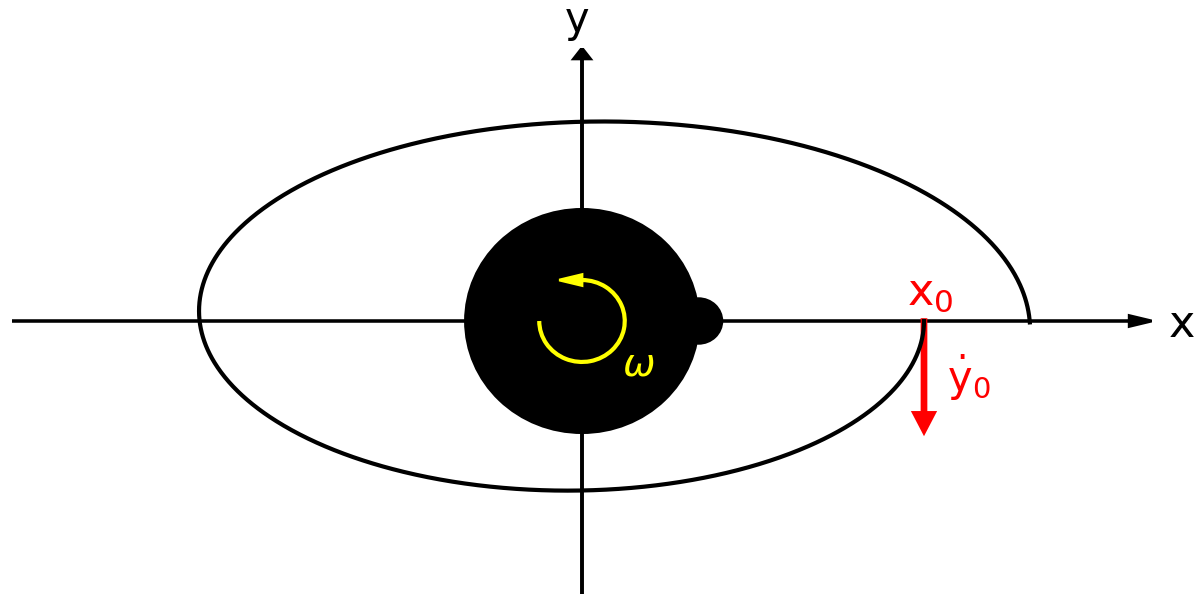}
\caption{Schematic diagram of the trajectory of a particle around a spherical object with a mass anomaly at its equator. The trajectory is fixed in the rotating frame with the central body's angular velocity $\omega$. $x_0$ is the initial position of the particle, and the red arrow indicates the initial velocity. \label{fig:system}}
\end{figure}
In the present work, we analyse the dynamics of particles orbiting a hypothetical spherical object of mass $M$ and radius $R$, with a mass anomaly $m_a$ at its equator (Figure~\ref{fig:system}). We assume the object with a uniform mass distribution, where the masses $M$ and $m_a$ have the same bulk density ($\rho=1~{\rm g/cm^3}$). The object is also assumed to rotate with constant angular velocity $\omega$ ($\omega=2\pi/T$, where $T$ is the rotation period) without wobbling motion. For simplicity, we will express our physical quantities in the following units: $GM=1$, while $R=1$ is the distance between the system centre and the mass anomaly. We also define as a unit the Keplerian frequency of the mass anomaly, scaled by the density $\rho$ of the object:
\begin{equation}
\omega_k=\sqrt{\frac{GM}{R^3}}=\sqrt{\frac{4\pi G\rho}{3}}=1.   
\end{equation}
Two dimensionless parameters will define our dynamic system: the normalized mass anomaly $\mu=m_a/M$ and the rotating rate $\lambda=\omega/\omega_k$.

Equations of motion in a frame $Oxy$ rotating with the same period as the central body's spin are given by \citep{Scheeres1996}
\begin{equation}
   \ddot{x}-2\lambda\dot{y}=\lambda^2x+U_x
\end{equation}
and
\begin{equation}
    \ddot{y}+2\lambda\dot{x}=\lambda^2y+U_y,
\end{equation}
where $U_x$ and $U_y$ stand for the partial derivatives of the gravitational potential. 

The potential acting on a particle with position-vector $\vec{r}=x\hat{x}+y\hat{y}$ ($r=|\vec{r}|$) in the rotating frame is obtained by adding the gravitational potential of the spherical portion of the object -- at the centre of the system -- with the gravitational potential of the mass anomaly, located at $\vec{R}=\hat{x}$ \citep{Sicardy2019}:
\begin{equation}
U(r)=-\frac{1}{r}-\frac{\mu}{|\vec{r}-\hat{x}|}+\lambda^2\mu(\vec{r}\cdot\hat{x}). \label{potential}
\end{equation}
Note that the potential given in Equation~\ref{potential} differs from that acting on a particle in the RP3BP \citep[][]{Murray1999} by the rotating parameter $\lambda^2$ in the indirect term. While the secondary mass in RP3BP surrounds the central body with Keplerian velocity $\omega_k$, here the mass anomaly rotates with angular velocity $\lambda\omega_k$. We introduced the rotating parameter to correct this difference.

Similar to the dynamics of a particle in the RP3BP with a disturbing internal body, we obtain the expansion of the potential $U$ for the lowest order terms in eccentricity ($e$) as:
\begin{equation}
U=-\frac{1}{r}-\sum_{j=0}^\infty\sum_{m=-\infty}^\infty \mu e^j\left[\alpha F_jb^{(m-j)}_{1/2}(\alpha)+\frac{\lambda^2}{\alpha}f_j\delta_{|m|,1}\right]\cos{\phi},  \label{potentialexp}
\end{equation}
where $\alpha=1/a<1$ ($a$=semi-major axis of the particle), $b_{1/2}^{(m)}$ is the Laplace coefficient, $f_j$ and $F_j$ are linear operators \citep[Table~\ref{operatorF},][]{Murray1999,Ellis2000}, $\delta_{|m|,1}$ is the Kronecker delta and $\phi$ is a characteristic angle of the system relating the rotation of the central object with the longitudes of the particle. The characteristic angle associated with the sectoral resonances is presented in Section~\ref{theory}.

\begin{table*}
\centering
\caption{The linear operators $f_j$ and $F_j$ for $j\leq 5$. The derivative operator D is given by D=d/d$\alpha$.\label{operatorF}}
\begin{tabular}{lrl} \hline \hline
j & $f_j$ & $F_j$ \\ \hline
1 & ${\rm -\frac{1}{2}}$ & ${\rm \frac{1}{2}\left[\left(-1+2m\right)+\alpha D\right]}$  \\ 
2 & ${\rm -\frac{3}{8}}$ & ${\rm \frac{1}{8}\left[\left(2-7m+4m^2\right)+\left(-2+4m\right)\alpha D+\alpha^2D\right]}$  \\
3 & ${\rm -\frac{1}{3}}$ & ${\rm \frac{1}{48}\left[\left(-6+29m-30m^2+8m^3\right)+\left(6-21m+12m^2\right)\alpha D+\left(-3+6m\right)\alpha^2D+\alpha^3D^3\right]}$  \\
\multirow{2}{*}{4} & \multirow{2}{*}{${\rm -\frac{125}{384}}$} &  ${\rm \frac{1}{384}\left[\left(24-146m+211m^2-104m^3+16m^4\right)+\left(-24+116m-120m^2+32m^3\right)\alpha D+\left(12-42m+24m^2\right)\alpha^2D+\right. }$ \\
 &  & ${\rm \left.+\left(-4+8m\right)\alpha^3D^3+\alpha^4D^4\right]}$ \\
\multirow{2}{*}{5} & \multirow{2}{*}{${\rm -\frac{27}{80}}$} & ${\rm \frac{1}{3840}\left[\left(-120+874m-1595m^2+1110m^3-320m^4+32m^5\right)+\left(120-730m+1055m^2-520m^3+80m^4\right)\alpha D+\right.}$   \\ 
 & & ${\rm \left.\left(-60-290m-300m^2+80m^3\right)\alpha^2D+\left(20-70m+40m^2\right)\alpha^3D^3+\left(5+10m\right)\alpha^4D^4+\alpha^5D^5\right]}$ \\
\hline 
\end{tabular}
\end{table*}

In conservative systems, such as those analysed in this work, the Jacobi constant $C_J$ is a conserved quantity used to obtain the Poincar\'e surface of sections. It is expressed here in the units $R^2\omega_k^2$ and is given by
\citep{Scheeres1996}
\begin{equation}
C_J=\lambda^2(x^2+y^2)+2U(x,y)-\dot{x}^2-\dot{y}^2.    
\label{eq:Cj}
\end{equation}

\section{Sectoral resonances} \label{theory}
At the planar limit, a pair of fundamental frequencies describe the motion of a particle: the synodic and radial epicyclic frequencies. The first, $n-\omega$ ($n$= angular frequency of the particle), corresponds to the frequency of the particle's return to a fixed position on the rotating frame. The second, $\kappa=n-\dot{\varpi}$, is the frequency of the particle's return to its pericentre, being $\dot{\varpi}$ the derivative of the particle's longitude of pericentre. If these frequencies are commensurable, the particle is in a sectoral resonance -- spin-orbit resonance -- with the central body. Once in resonance, the orbital evolution of the particle will be modelled by the energy balance provided by the resonant configuration. Sectoral resonances with real non-spherical bodies were studied in \cite{Borderes2018} and \cite{Winter2019} for the asteroid 4179~Toutatis and the dwarf planet Haumea, respectively. 

A particle at the centre of a $m$:$(m-j)$ resonance satisfies the resonance condition \citep{Sicardy2019}
\begin{equation}
m\omega-(m-j)n-j\dot{\varpi}=0, \label{phidot0}  
\end{equation}
where $m$ and $j$ are integers responsible for giving the commensurability of the frequencies. For $j=0$, the particle is in corotation resonance, while for $j=m$, we have the apsidal resonances. Both cases are out of the scope of this work \citep[for details, see][]{Sicardy2019} and here we will focus on resonances with $j\geq1$, where the numerical value of $j$ gives the order of the resonance.

When a particle is in a $m$:$(m-j)$ resonance, the characteristic angle $\phi$ -- also called resonant angle -- librates with an amplitude lower than $360^{\circ}$. The angle is given by
\begin{equation}
\phi=m\omega t-(m-j)\lambda_p-j\varpi, \label{phi}
\end{equation}
where $\lambda_p$ is the mean longitude of the particle. For simplicity, we ignore variations in the mean longitude of epoch.

\subsection{Resonance Location} \label{resonancelocation}
The angular and radial epicyclic frequencies are given by \citep{Chandrasekhar1942}
\begin{equation}
n^2=\frac{1}{r}\frac{dU_0}{dr}   \label{fn0}
\end{equation}
and
\begin{equation}
\kappa^2=\frac{1}{r^3}\frac{d(r^4n^2)}{dr}, \label{fk0}
\end{equation}
where $U_0$ is the axisymmetric part of the gravitational potential ($j=m=0$).

From equation~\ref{potentialexp}, we obtain:
\begin{equation}
U_0=-\frac{1}{r}-\frac{\mu}{2r}b_{1/2}^{(0)}\left(\alpha\right).   
\end{equation}

Expanding the Laplace coefficient up to second order in $\alpha$,
\begin{equation}
 \frac{1}{2}b_{1/2}^{(0)}\left(\alpha\right)=1+\frac{1}{4}\alpha^2,
\end{equation}
we obtain the axisymmetric part of the gravitational potential for the spherical body with a mass anomaly:
\begin{equation}
U_0=-\frac{1}{r}\left(1+\mu+\frac{\mu}{4}\alpha^2\right).
\end{equation}

Keeping the lowest order terms in $\mu$ in eqs.~\ref{fn0} and \ref{fk0}, we obtain
\begin{equation}
n^2=\frac{1}{r^3}\left(1+\mu+\frac{3\mu}{4}\alpha^2\right) \label{fn}   
\end{equation}
and
\begin{equation}
\kappa^2=\frac{1}{r^3}\left(1+\mu-\frac{3\mu}{4}\alpha^2\right).    \label{fk}
\end{equation}

The location of the resonances can be obtained by numerical methods, such as the Newton-Raphson method \citep[see][]{Press1989,Renner2006}, by applying eqs.~\ref{fn} and \ref{fk} in the resonance condition (eq.~\ref{phidot0}). Table~\ref{tab:location} shows the location of the resonances in the ranges $-4\leq m\leq 4$ and $j\leq 5$ (up to fifth-order resonances). The central body is a Chariklo-type body with $\mu=10^{-3}$ and $\lambda=0.471$ \citep[$M=6.3\times 10^{18}$~kg and $T=7.004$~hr,][]{Leiva2017}, defined as our reference object. We assume $\mu=10^{-3}$ as a reference value because it is small enough for the centre of the system to be approximately the physical centre of the spherical portion and large enough for the effects of the mass anomaly to be observed.

\begin{table*}
\centering
\caption{The location of the $m$:$(m-j)$ resonances in the ranges $-4\leq m\leq 4$ and $j\leq 5$. We assumed a central body with parameters based on the centaur Chariklo, with $\lambda=0.471$ and $\mu=10^{-3}$ (reference object). The resonances marked ``inside'' occur within the physical radius of the central body and, therefore, do not exist in the considered system. The resonances marked as ``apsidal'' are out of the scope of this work.  \label{tab:location} }
\begin{tabular}{cccccccccc} \hline \hline
 j & m $\rightarrow$  & -4  & -3  & -2  & -1  & 1               & 2       & 3 & 4  \\ \hline 
\multirow{2}{*}{1} & resonance   & 5:6 & 4:5  & 3:4 & 2:3 & 1:2 & 1:0             & 2:1     & 3:2   \\[0.1cm]
                     & $a/R$ &  1.909 & 1.993 & 2.156 & 2.612 & apsidal   & inside     & 1.256 &  1.358  \\ \hline
\multirow{2}{*}{2} & resonance   & 5:7 & 4:6  & 3:5 & 2:4 & 1:3 & 1:-1            & 2:0     & 3:1  \\[0.1cm]
                     & $a/R$ &  2.156 & 2.313 & 2.612 & 3.423 & inside & apsidal & inside & inside  \\ \hline 
\multirow{2}{*}{3} & resonance   & 5:8 & 4:7  & 3:6 & 2:5 & 1:4 & 1:-2            & 2:-1     & 3:0  \\[0.1cm]
                     & $a/R$ &  2.389 & 2.612 & 3.031 & 4.146 & inside & inside & apsidal & inside \\ \hline   
\multirow{2}{*}{4} & resonance   & 4:8  & 3:7 & 2:6 & 1:5 & 1:-3            & 2:-2     & 3:-1 & 4:0 \\[0.1cm]
                     & $a/R$ & 2.612 & 2.895 & 3.423 & 4.811 & inside & inside & inside & apsidal \\ \hline 
\multirow{2}{*}{5} & resonance   & 4:9  & 3:8 & 2:7 & 1:6 & 1:-4            & 2:-3     & 3:-2 & 4:-1 \\[0.1cm]
                     & $a/R$ & 2.825 & 3.164 & 3.793 & 5.433 & inside & inside & inside & inside \\ \hline 
\end{tabular}
\end{table*}

\subsection{Resonance Width}
In this subsection, we follow the classical approach of the pendulum model, presented in \cite{Winter1997a} and \cite{Murray1999}, to obtain the resonance width for our case of interest. A particle is in a $m$:$(m-j)$ sectoral resonance when its resonant angle $\phi$ librates, which means that the particle oscillates in the rotating frame around the central position of the resonance (eq.~\ref{phidot0}). We can evaluate the maximum amplitude of a resonant particle through the temporal variations of $\phi$:
\begin{equation}
 \dot{\phi}=m\omega-(m-j)n-j\dot{\varpi} \label{phidot}   
\end{equation}
and
\begin{equation}
\ddot{\phi}=-(m-j)\dot{n}-j\ddot{\varpi}.   \label{phiddot}
\end{equation}
Considering only the lowest order terms in eccentricity ($e$), we obtain using the Lagrange's equations \citep{Murray1999}:
\begin{equation}
\dot{n}=-3nC_r(m-j)e^j\sin{\phi}   \label{ndot}
\end{equation}
and
\begin{equation}
\dot{\varpi}=je^{j-2}C_r\cos{\phi},  \label{varpidot}
\end{equation}
where
\begin{equation}
C_r=\mu \frac{n}{\alpha}\left[\alpha F_jb_{1/2}^{(m-j)}+\frac{\lambda^2}{\alpha}f_j\delta_{|m|,1}\right]. \label{CR}
\end{equation}

From equation~\ref{varpidot}, we obtain that the second derivative of $\varpi$ is
\begin{equation}
\ddot{\varpi}=j(j-2)e^{j-3}\dot{e}C_r\cos{\phi}-je^{j-2}C_r\sin{\phi}\dot{\phi},  
\end{equation}
where the time variation of eccentricity ($\dot{e}$) obtained through Lagrange's equations is $\dot{e}=-je^{j-1}C_r\sin{\phi}$.

It can be shown that 
\begin{equation}
\ddot{\varpi}=j^2e^{2(j-2)}C_r^2\sin{j\phi}-je^{j-2}C_r(m\omega-(m-j)n)\sin{\phi}.   
\end{equation}
Therefore, 
\begin{equation}
\begin{split}
\ddot{\phi}=&-j^3e^{2(j-2)}C_r^2\sin{2\phi}+3nC_r(m-j)^2e^j\sin{\phi}+\\ +&j^2e^{j-2}C_r(m\omega-(m-j)n)\sin{\phi}. \label{phiddotb}  
\end{split}
\end{equation}

By inspection, we can evaluate the contribution of each term of equation~\ref{phiddotb}. The $C_r$ function is proportional to $\mu$, a value lower than one. In fact, for high values of mass anomaly ($\mu\gtrsim10^{-2}$), we can not assume the centre of mass of the system as the physical centre of the spherical object, and equation~\ref{potentialexp} is no longer applied -- this range of $\mu$ defines another NSSB, the contact binary. Since $\mu<<1$, the term that depends on $C_r$ will dominate those dependent on $C_r^2$, in principle. For first-order resonances ($j=1$), the first and third terms in equation~\ref{phiddotb} are proportional to $1/e^2$ and $1/e$, respectively -- $e$ is a small value -- and dominate over the second term, proportional to $e$.

\subsubsection{Second and higher-order resonances}
For second and higher-order resonances, the eccentricity exponents in equation~\ref{phiddotb} are positive, and we can approximate the equation to
\begin{equation}
\ddot{\phi}+\omega_0^2\sin{\phi}=0, \label{eq8}   
\end{equation}
where $\omega_0^2=3n|C_r|(m-j)^2e^j$. To obtain this result, we have assumed $m\omega-(m-j)n\approx0$ since the particle is in resonance.

From equation~\ref{eq8}, we can see that a resonant particle is confined in a pendulum motion around an equilibrium position of the resonance. The number of equilibrium positions of a $m$:$(m-j)$ sectoral resonance is $j$. Analogous to the simple pendulum problem, the particle reduced energy in the rotating frame is
\begin{equation}
E=\frac{\dot{\phi}^2}{2}+2\omega_0^2\sin^2\frac{\phi}{2}.    
\end{equation}

The maximum possible energy of the pendulum ($\dot{\phi}=0$~deg and $\phi=90$~deg) defines the separatrix between libration and circulation of the resonant angle. That is, the separatrix corresponds to the boundary between bounded and unbounded motions. The energy of such trajectory is $E=2\omega_0^2$, and the temporal variation of the resonant angle is $\dot{\phi}=\pm2\omega_0\cos(\phi/2)$. 

Relating $\phi$ and $n$:
\begin{equation}
dn=\frac{\dot{n}}{\dot{\phi}}d\phi=\pm\sqrt{3n|C_r|e^j}\sin{\frac{\phi}{2}}d\phi,    
\end{equation}
we obtain, by integration, the range of angular frequency in which a particle is in a $m$:$(m-j)$ sectoral resonance:
\begin{equation}
n=n_0\pm\sqrt{12n|C_r|e^j}\cos{\frac{\phi}{2}} \label{2orplusn},   
\end{equation}
where $n_0$ is the central angular frequency of the resonance.

Therefore, a particle is in a second or higher-order resonance if its semi-major axis meets the relation:
\begin{equation}
a=a_0\pm\left(\frac{16}{3}\frac{|C_r|}{n}e^j\right)^{1/2}a_0, \label{2orplus}  
\end{equation}
where $a_0$ is the central semi-major axis of the resonance (Section~\ref{resonancelocation}).

\subsubsection{First-order resonances}
For $m$:$(m-1)$ resonances, none of the terms in equation~\ref{phiddotb} can be disregarded, requiring a different solution than the one obtained. As \textit{ansatz}, we assume a solution similar to eq.~\ref{2orplusn}, $n=n_0+k\cos(\phi/2)$, where $k$ is an as-yet-unknown constant. By integrating equation~\ref{phiddotb}, we obtain the kinetic energy of the system
\begin{equation}
\begin{split}
\frac{1}{2}\dot{\phi}^2=&\int \ddot{\phi}d\phi=\frac{C_r^2}{e^2}\left(2\cos^2{\frac{\phi}{2}+\cos^2{\phi}}\right)+\\
-&6nC_r(m-1)^2e\cos^2{\frac{\phi}{2}}+\frac{4}{3}\frac{C_r}{e}(m-1)k\cos^3{\frac{\phi}{2}},  \label{e1} 
\end{split}
\end{equation}
where the constant arising from the integration was determined considering $\phi=0$~deg and $\phi=180$~deg.

Applying $n=n_0+k\cos(\phi/2)$ to equation~\ref{phidot} and assuming that the particle is exactly at the centre of the resonance ($\phi=0$~deg and $\phi=180$~deg), we find that $m\omega-(m-1)n_0=-C_r/e$. From equation~\ref{phidot}, we get
\begin{equation}
\begin{split}
\frac{1}{2}\dot{\phi}^2=&\frac{1}{2}\frac{C_r^2}{e^2}(1+\cos{\phi})^2+\frac{1}{2}(m-1)^2k^2\cos^2{\frac{\phi}{2}}+\\
+&\frac{C_r}{e}(1+\cos{\phi})(m-1)k\cos{\frac{\phi}{2}}.  \label{e2}
\end{split}
\end{equation}
Taking equations~\ref{e1} and~\ref{e2} as equal and assuming $\phi=0$~deg:
\begin{equation}
(m-1)^2k^2+\frac{4}{3}\frac{C_r}{e}(m-1)k+12nC_r(m-1)^2e=0.    
\end{equation}

Therefore, the boundaries of the  angular frequency and semi-major axis in which a particle is in a first-order resonance are, respectively:
\begin{equation}
n=n_0\pm\sqrt{12|C_r|ne}\left(1+\frac{1}{27(m-1)^2e^3}\frac{|C_r|}{n}\right)^{1/2}-\frac{|C_r|}{3(m-1)e}    
\end{equation}
and
\begin{equation}
\begin{split}
a=&a_0\pm\left(\frac{16}{3}\frac{|C_r|}{n}e^j\right)^{1/2}\left(1+\frac{1}{27(m-1)^2e^3}\frac{|C_r|}{n}\right)^{1/2}a_0\\
+&\frac{2}{9(m-1)e}\frac{|C_r|}{n}a_0.    
\end{split}
\end{equation}

\section{Poincar\'e Surfaces of Section} \label{secPSS}
Poincar\'e surface of section technique is usually applied in studies of the RP3BP \citep{Henon1965a,Henon1965b,Henon1966a,Henon1966b,Henon1969,Jefferys1971,Winter1994a,Winter1994b}, to analyse the dynamics of the third body, providing information such as the location and size of stable and chaotic regions, including the mean motion resonance regions. In RP3BP, the problem is considered in a rotating system where the primary and secondary bodies are fixed, and only the third body describes a free motion. Some works have also adopted the Poincar\'e surface of section to study dynamical systems composed of two bodies, with a non-spherical central object. \cite{Scheeres1996} applied this technique to find periodic orbits around the asteroid 4769~Castalia. This technique was also applied by \cite{Borderes2018} and \cite{Winter2019} to study the region around Toutatis and Haumea, respectively. 

This work also applies the Poincar\'e surface of section to a two-body problem composed of a  massive central body and a massless particle. Instead of the orbital motion between the primary and secondary bodies, the rotation of the central body gives the motion of the rotating frame. The Poincar\'e surface of section applied to the two-body problem with a mass anomaly provides information about stability and resonances. However, in this case, there are spin-orbit resonances instead of mean motion resonances.

\begin{figure}
\centering
\includegraphics[width=0.8\columnwidth]{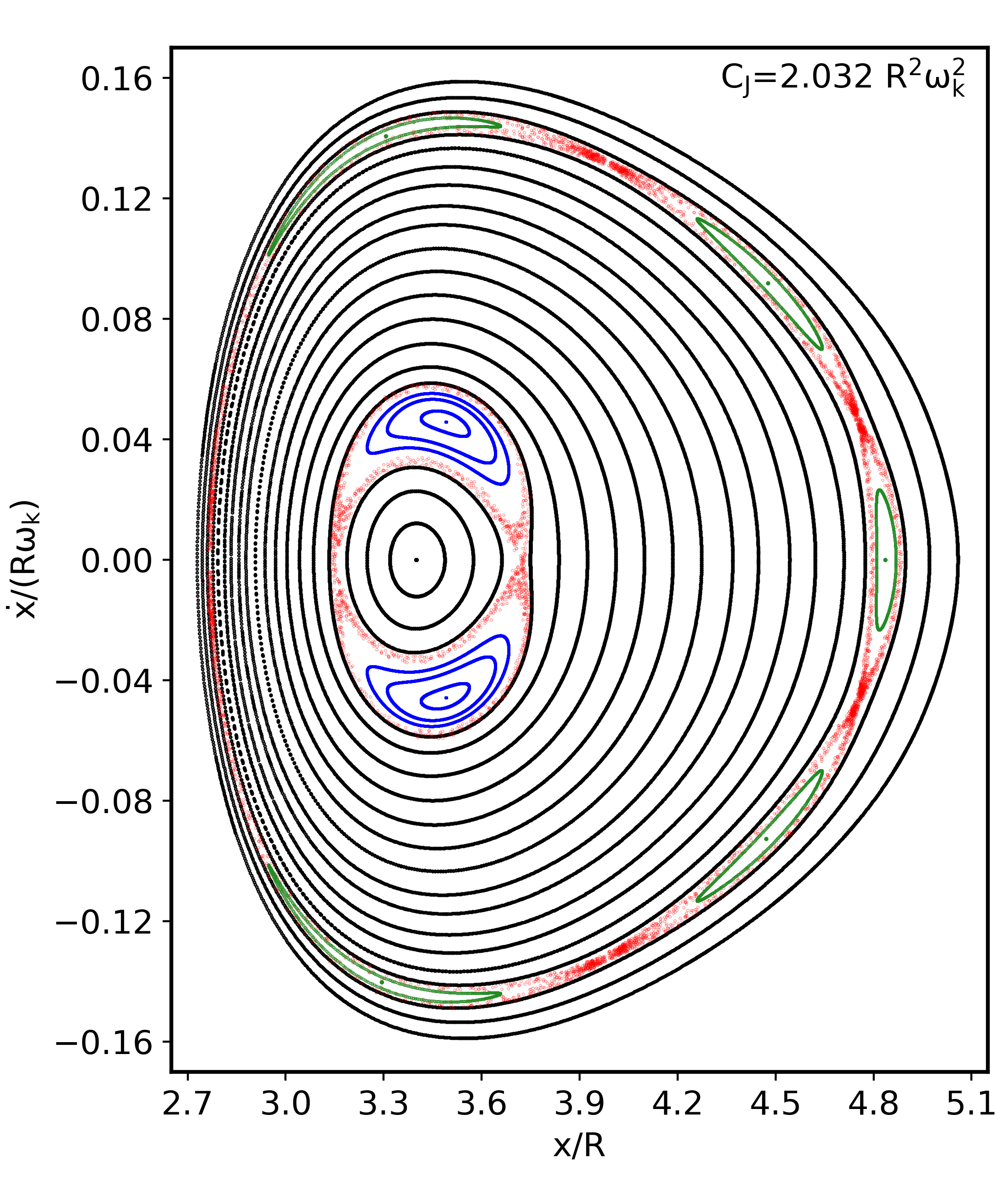}
\caption{Poincar\'e  surface of section  for ${\rm C_J}=2.032~{\rm R^2\omega_k^2}$ around an object with $\mu=10^{-3}$ and $\lambda=0.471$. The black islands are quasi-periodic orbits associated with the periodic orbit of first kind. Blue islands are associated with the 1:3 resonance and the green ones with the 2:7 resonance. The red points are chaotic orbits that cross the phase plane irregularly. \label{fig:poincare}}
\end{figure}

Poincar\'e surface of sections are maps generated in the phase space through the intersection points of the particle orbits with a fixed section in the system. These maps are generated for fixed values of the Jacobi constant (equation~\ref{eq:Cj}). In Figure~\ref{fig:poincare}, we see an example of this map for a system composed of a massive central body with a mass anomaly. The Poincar\'e surface of section was defined in the plane $y=0$ around our reference object and for the fixed value of the Jacobi constant ${\rm C_J}=2.032~{\rm R^2\omega_k^2}$. We distributed the initial conditions on the $x$-axis. 

In Figure~\ref{fig:poincare}, the different sets of closed curves, called stability islands, delimit the stable regions of the system. Each stability island is formed by a single quasi-periodic orbit that is named because it does not have a defined orbital period. At the centre of the stability islands, we have periodic orbits. The latter crosses the Poincar\'e surface of section always at the same points and can be classified into two kinds \citep{Poincare1895}: those not associated with resonances are the first kind, and those associated with resonance are the second kind.

The point in the centre of all black closed curves is a first kind periodic orbit. In contrast, the points in the centres of the blue and green islands are the second kind orbits associated with the 1:3 and 2:7 resonances, respectively (Fig.~\ref{fig:poincare}). A single stability island identifies periodic orbits of first kind, while one or more stability islands can identify the orbits of second one. The number of islands for the second kind orbits is related to the order of the resonance \citep{Winter1997b}. For example, the pair of blue islands in Figure~\ref{fig:poincare} is formed by quasi-periodic orbits that librate around the periodic orbit associated with the 1:3 resonance, a second-order resonance. In the same vein, each particle in the 2:7 resonance -- a fifth-order resonance -- generates five islands on the surface of section.

A set of first and second kind orbits belonging to the same resonance usually appears in a continuous Jacobi constant range and defines a family of orbits. The families present evolution in structure and position in the Poincar\'e surface of section during the variation of the Jacobi constant. In addition to the stability region delimited by the islands, there are also unstable regions filled with scattered red points in the figure, created by chaotic orbits. These chaotic regions are seen around the stability islands associated with the periodic orbits of 1:3 and 2:7 resonances. They are associated with the resonance separatrix and do not enter the stable regions, as we can see in Figure~\ref{fig:poincare}. A stable region bounded by quasi-periodic orbits (black curves) separates the two chaotic regions. 

In the following sections, we use Poincar\'e surface of section to explore the stability around bodies with a mass anomaly by varying the central body parameters.

\section{System overview} \label{overview}
We studied the dynamics around our object by simulating a set of particles with pericentre distance q and eccentricity in the ranges $1<q/R\leq q_f$ and $0\leq e\leq 0.5$, respectively ($\Delta e=0.05$ and $\Delta q/R=0.01$). $q_f$ is a given value of $q$ for which all particles survive. The particles were simulated for 10,000~orbits. We assumed the parameters $\lambda$ and $\mu$ in the ranges $0.01\leq \lambda\leq1$ and $10^{-6}\leq\mu\leq5\times 10^{-3}$, respectively. Except for the near-Earth asteroids, we have that the vast majority of small heliocentric bodies have $\omega>\omega_k$ \citep{Warner2009}, justifying the fact that we do not focus on cases with $\lambda>1$. 

\begin{figure}
\includegraphics[width=\columnwidth]{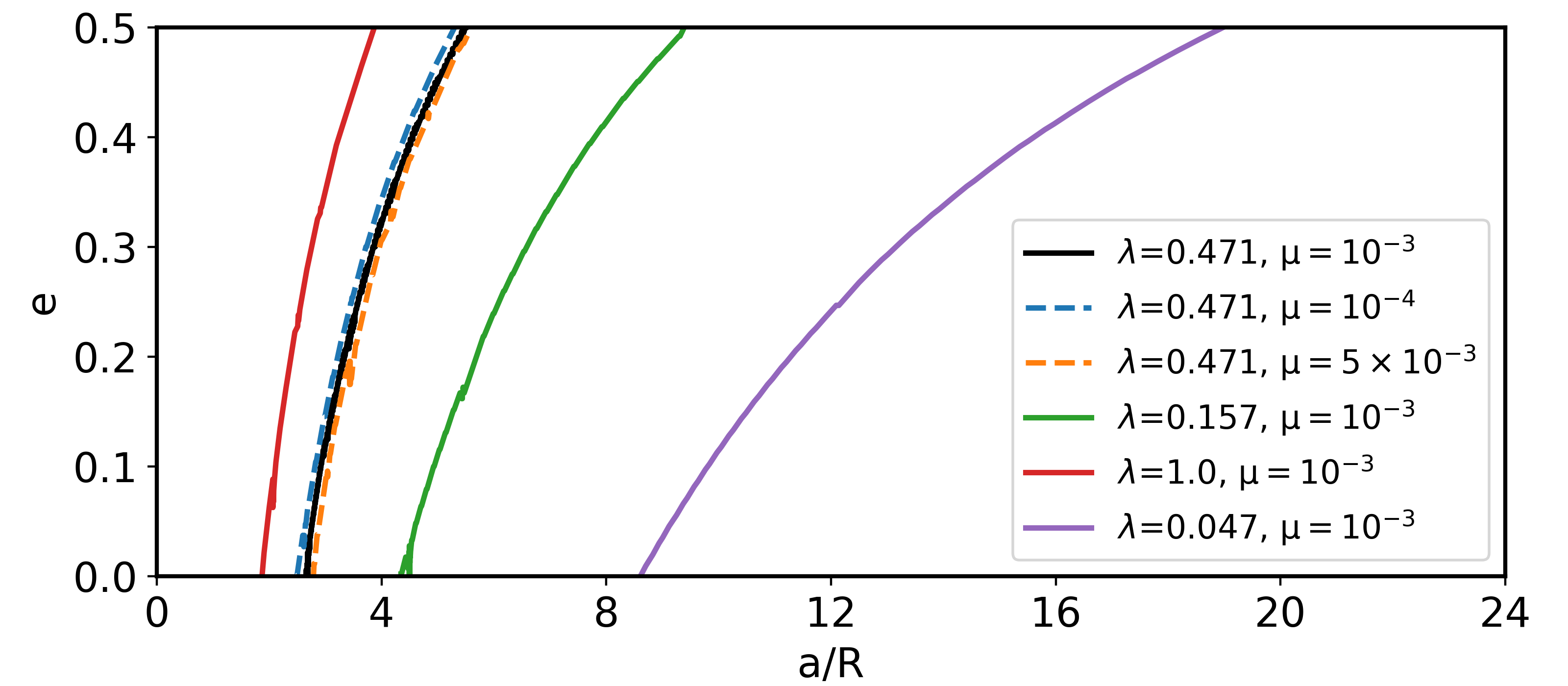}
\caption{Boundary curves between the chaotic (on the left) and stable (on the right) regions. The solid black line corresponds to the reference object, while the coloured solid and dashed lines are the cases in which we varied the parameters $\lambda$ and $\mu$, respectively. \label{fig:trans}}
\end{figure}

We verified in all numerical simulations the existence of a chaotic region just outside the central body in which particles collide or are ejected from the system. Beyond the chaotic region, there is a stable region, and the boundaries between them are shown in Figure~\ref{fig:trans}. Particles with semi-major axis and eccentricity in the region bounded by the curve (on the left side of the figure) will be lost, while those outside the boundary will survive for at least 10,000 orbits. The solid black line correspond to our reference object, while the solid coloured and dashed lines provide the boundary curves for systems where we vary $\lambda$ and $\mu$, respectively.

The successive close-encounters of the particle with the mass anomaly are responsible for exchanges of energy and angular momentum, resulting in the variation of the particles' orbital elements. Particles with sufficiently small semi-major axis show orbital evolution with chaotic diffusive character \citep[for details, see][]{Rollin2021}. In general, eccentricities in the chaotic region tend to increase, resulting in occasional collisions or until the orbit becomes hyperbolic.

\begin{figure}
\centering
\subfigure[]{\includegraphics[width=0.85\columnwidth]{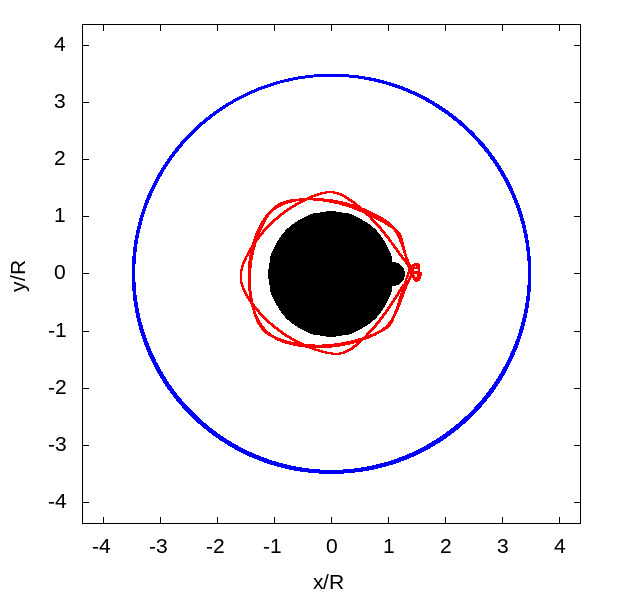}}
\subfigure[]{\includegraphics[width=\columnwidth]{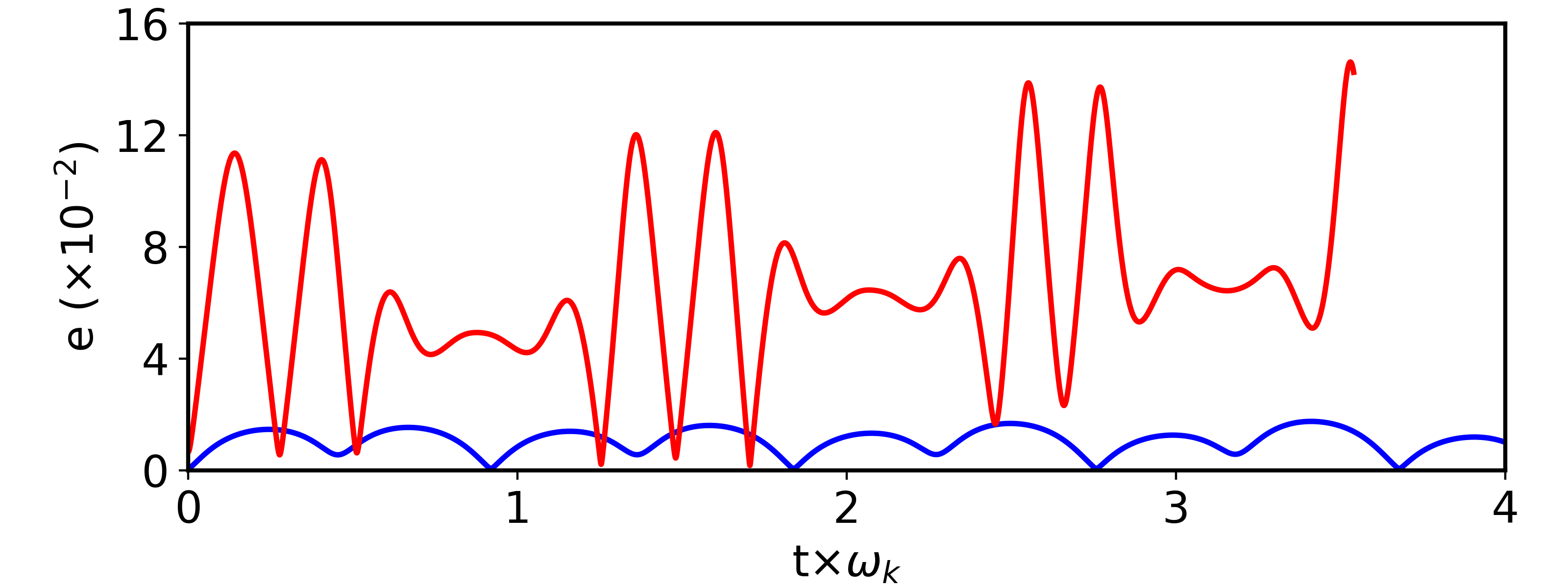}}
\caption{a) Trajectory in the rotating frame and b) temporal evolution of the eccentricity. The innermost particle (red line) is at $1.74~R$ and the outermost one (blue line) at $3.48~R$, and both are initially in circular orbits. The parameters of the central body are $\lambda=0.471$ and $\mu=5\times 10^{-3}$.  \label{fig:rc}}
\end{figure}

Figure~\ref{fig:rc} shows the trajectory in the rotating frame (a) and eccentricity (b) of a pair of particles initially in circular orbits around a central body with parameters $\lambda=0.471$ and $\mu=5\times 10^{-3}$. The semi-major axis of the innermost (red line) and outermost (blue line) particles are $1.74~R$ and $3.48~R$, respectively. We observe that the eccentricity shows a secular increase for the innermost particle, reaching values up to $0.15$. The particle collides with the central body after about $3.5\omega_k^{-1}$ or $\sim12$ spin periods. The eccentricity shows periodic variations for the outermost particle, and the particle remains stable around the central body.

The boundary curves are robust against the final simulation period and are preserved when we extend the simulations to 100,000 orbits. \cite{Lages2017} analyse through the Lyapunov exponent the stability of particles around a contact binary, obtaining a chaotic region coherent with ours. Our boundary curve is also coherent with the region where the particles are lost in the numerical simulations for a Chariklo with a mass anomaly performed by \cite{Sicardy2019}.

The boundary between chaotic and stable regions has only a slight dependence on the relative mass anomaly. Although the increase of $\mu$ generates only a small swell of the chaotic region, it produces larger increments in eccentricity and the particles are lost more quickly. The extension of the chaotic region is mainly affected by the parameter $\lambda$. Decreasing in 10 times the rotating rate, we obtain that the chaotic region is more than doubled, a result also obtained by \cite{Lages2017}.

In order to crudely evaluate the extension of the chaotic region, we calculate for a set of systems the semi-major axis at which a particle in circular orbit will survive for up to 10,000 orbits -- the threshold semi-major axis ($a_t$). The curve fitted from the numerical results is given by  
\begin{equation}
\frac{a_{t}}{R}=\left[1.298-0.007\mathcal{M}+0.006\mathcal{M}^2+0.674\lambda^{-0.75}\right], \label{trs}
\end{equation}
where $\mathcal{M}=-\log{\mu}$. Physically, we can interpret the threshold semi-major axis as the minimum semi-major axis, beyond which rings and satellites can exist around a body with a mass anomaly. 

\begin{figure}
\includegraphics[width=\columnwidth]{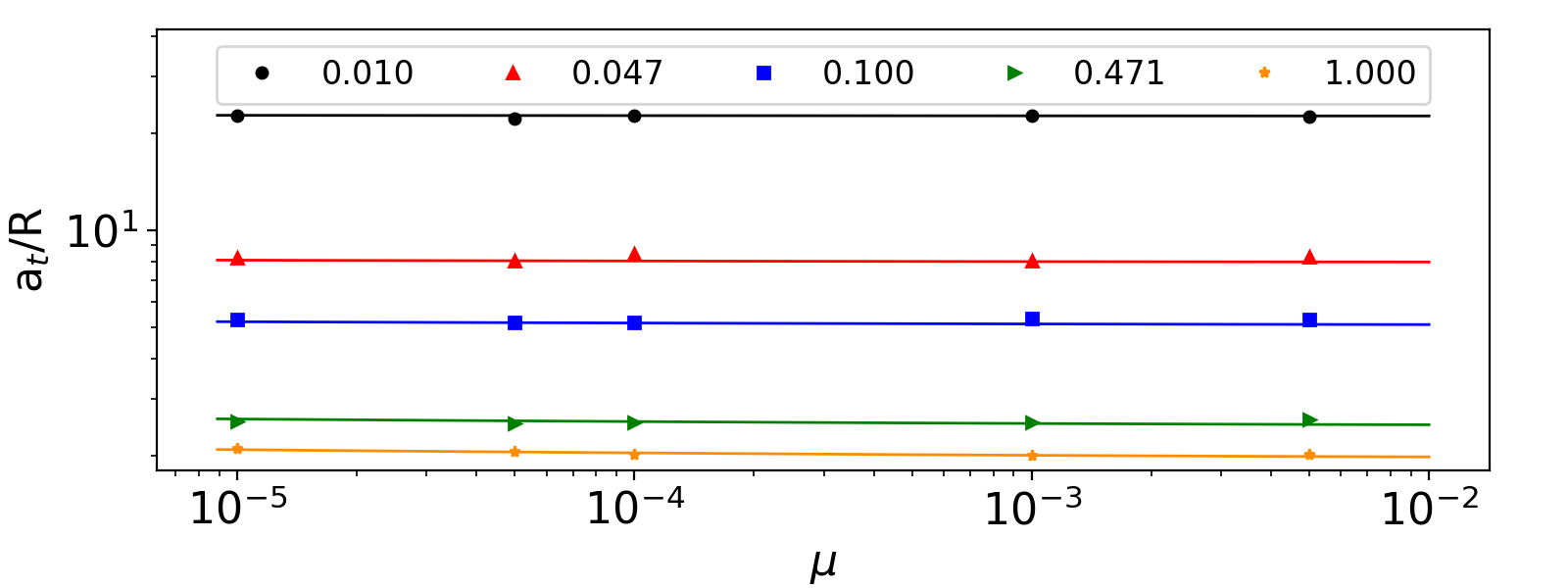}
\caption{Threshold semi-major axis obtained in selected numerical simulations (markers) and through equation~\ref{trs} (solid lines). The $x$-axis gives the normalized mass anomalies, and the different colours and markers give the rotating rates.  \label{fig:fit}}
\end{figure}

Figure~\ref{fig:fit} shows the threshold semi-major axis obtained in the numerical simulations. The $x$-axis gives the normalized mass anomaly of each simulation, and different colours and markers show the different rotating rates. The solid lines correspond to the curves given by equation~\ref{trs} (the colour of the lines matches the colour of the markers for the same $\lambda$).

\begin{figure*}
\centering
\subfigure[$\mu=10^{-4}$]{\includegraphics[width=1.5\columnwidth]{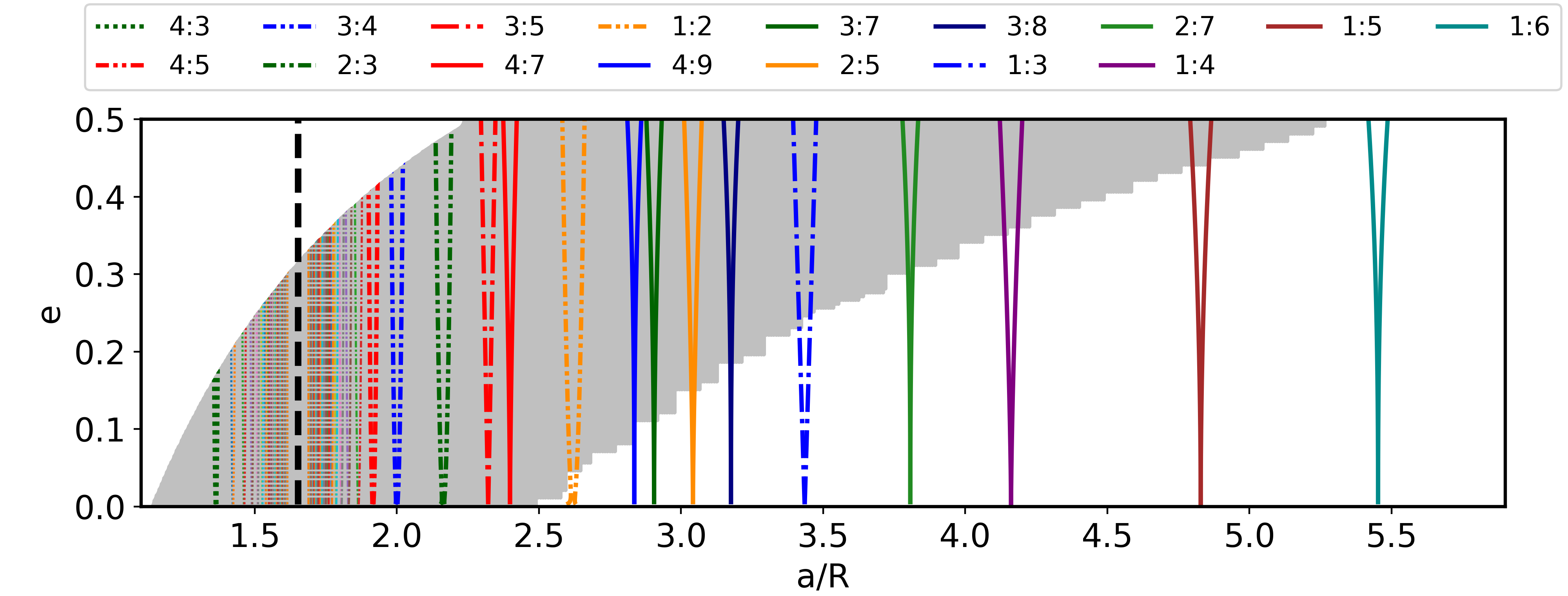}}
\centering
\subfigure[$\mu=10^{-3}$]{\includegraphics[width=1.5\columnwidth]{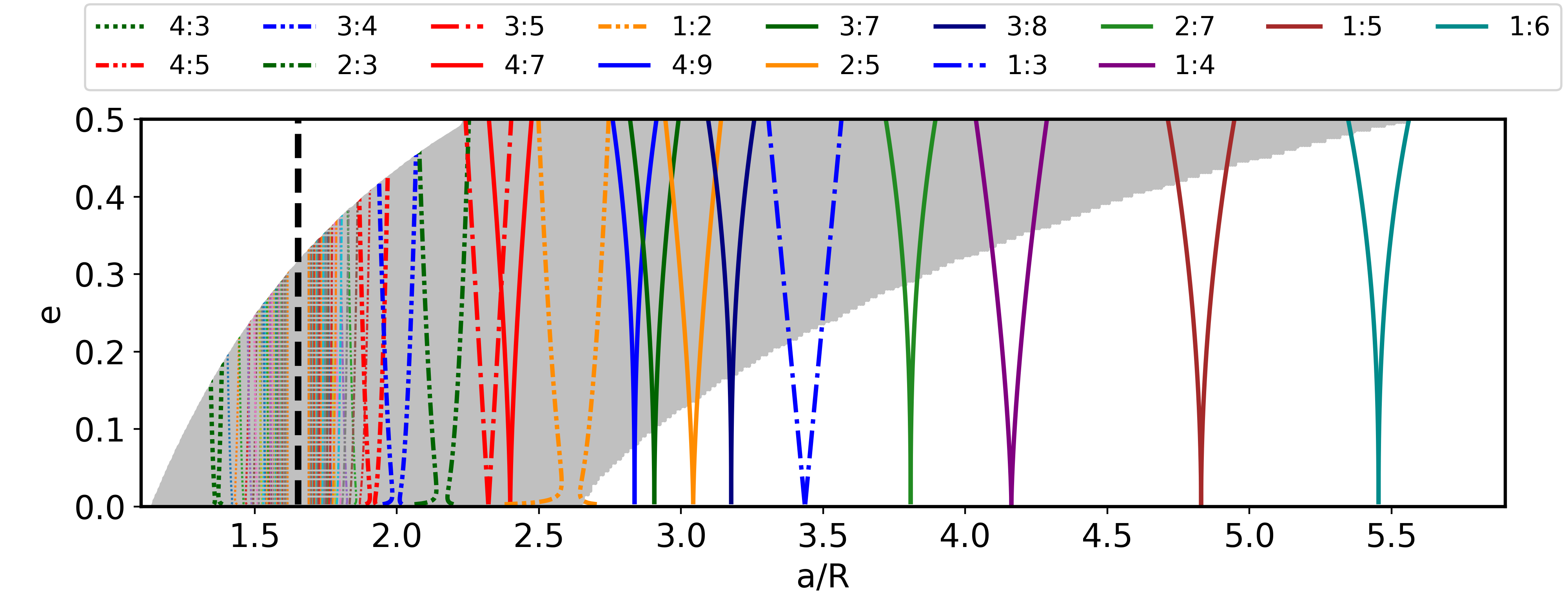}}
\centering
\subfigure[$\mu=5\times 10^{-3}$]{\includegraphics[width=1.5\columnwidth]{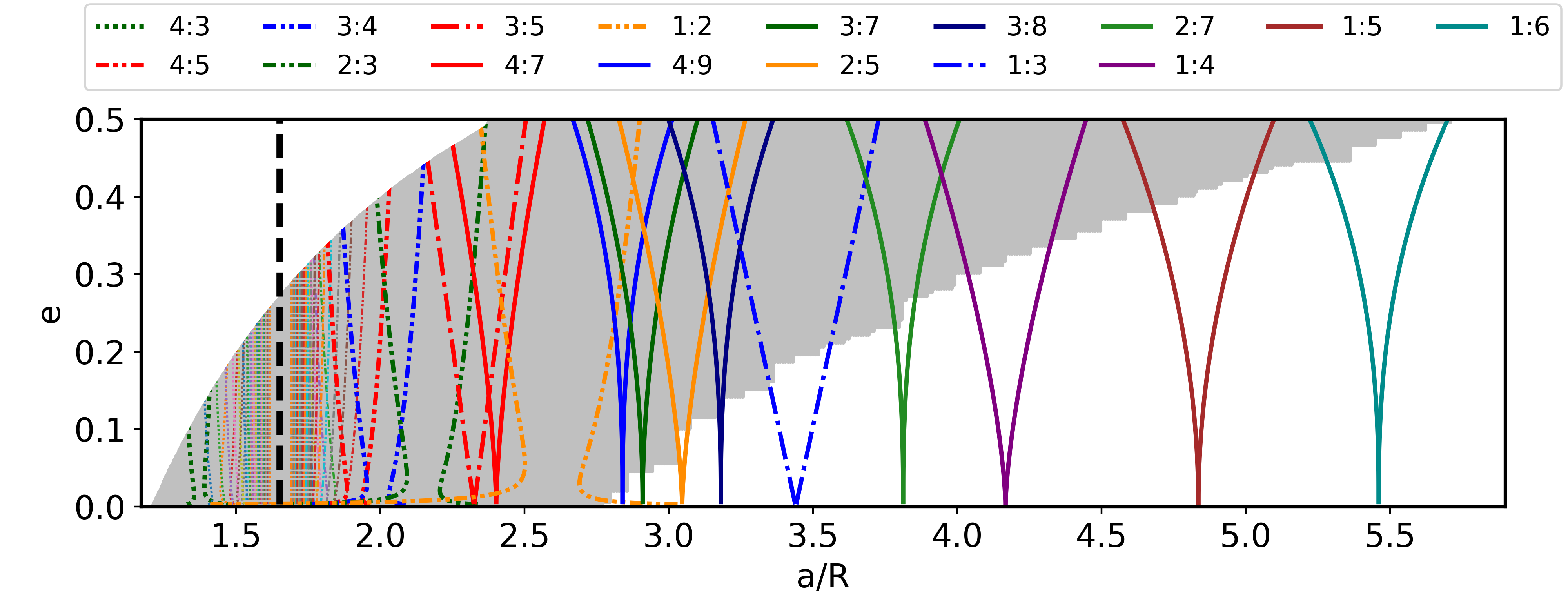}}
\caption{Semi-major axis ($a/R$) versus eccentricity ($e$) for systems, with $\lambda=0.471$ and a) $\mu=10^{-4}$, b) $\mu=10^{-3}$, and c) $\mu=5\times 10^{-3}$. Particles with initial $a/R$ and initial $e$ in the left white region have pericentre within the central body and collide. Particles in the grey area collide with the central body or are ejected, and those in the right white one remain in the system for more than 10,000 orbits. The dashed black lines correspond to the corotation radius, and the coloured lines provide the theoretical boundaries of the resonances. Coloured lines not referenced on the label and close to the corotation radius correspond to first order resonances with $|m|>4$. \label{fig:mu}}
\end{figure*}

\begin{figure*}
\centering
\subfigure[$\lambda=0.047$]{\includegraphics[width=1.5\columnwidth]{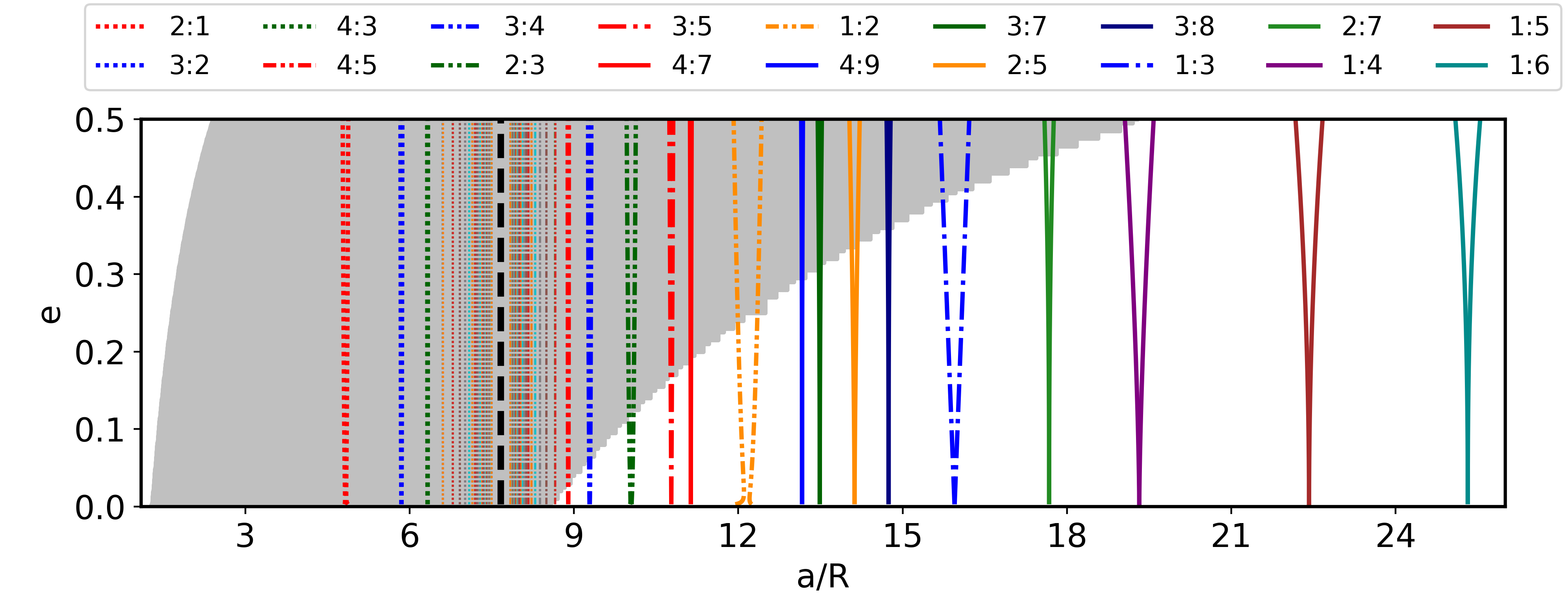}}
\centering
\subfigure[$\lambda=0.471$]{\includegraphics[width=1.5\columnwidth]{Figures/q047_mu1e-3.png}}
\centering
\subfigure[$\lambda=1.000$]{\includegraphics[width=1.5\columnwidth]{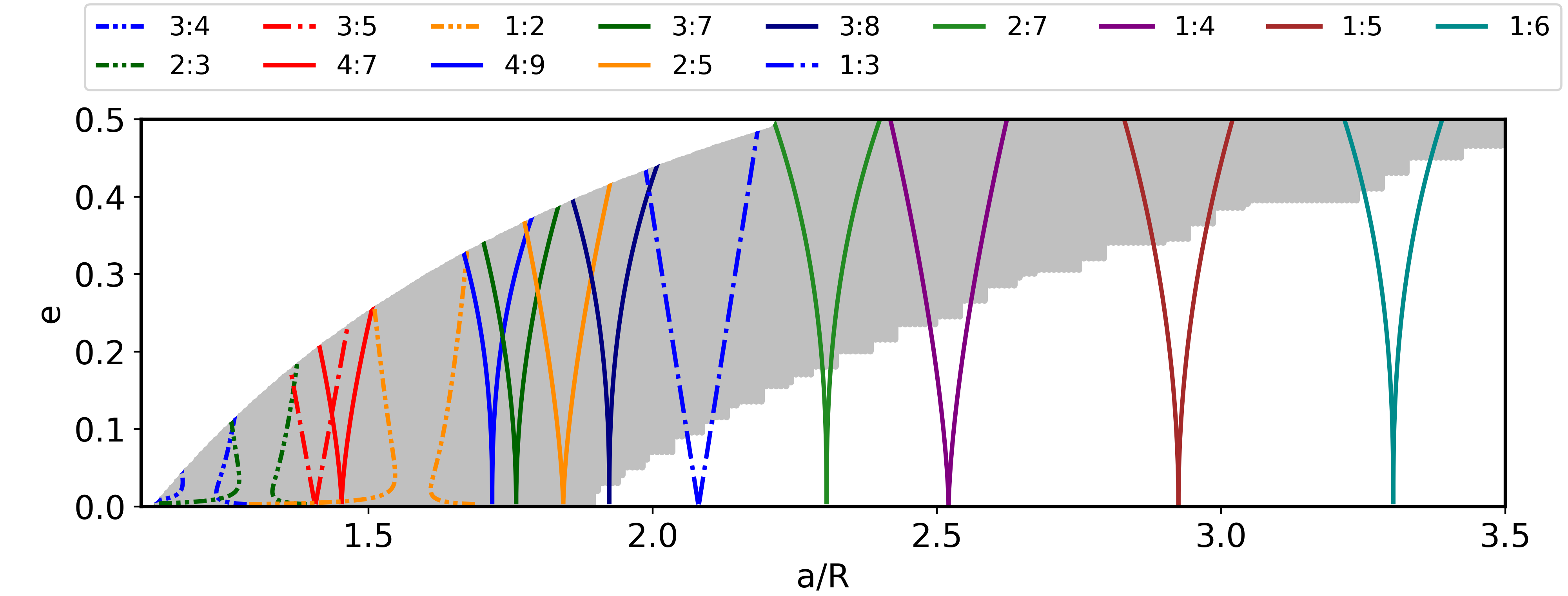}}
\caption{Semi-major axis ($a/R$) versus eccentricity ($e$) of systems with $\mu=10^{-3}$ and rotating rate $\lambda$ given in the caption of each panel. Particles with initial conditions in the white region on the left have pericentre within the central body and collide, while those in the grey area show chaotic behaviour. The white region on the right is the stable region. The dashed black line provides the corotation radius, and the coloured lines give the theoretical boundaries of the resonances. Coloured lines not referenced on the label and close to the corotation radius correspond to first order resonances with $|m|>4$. \label{fig:lambda}}
\end{figure*}

Figure~\ref{fig:mu} shows the position and width of the sectoral resonances, obtained theoretically (Section~\ref{theory}), where each panel corresponds to a different normalized mass anomaly, while the rotating rate is fixed as $\lambda=0.471$. The vertical black line in each panel corresponds to the corotation radius $a_c$ of the system ($a_c=\lambda^{-2/3}$), and the coloured lines give the boundaries of the resonances. The white region on the left provides the initial conditions of particles with pericentre within the central body, and the white area on the right is the stable region. The grey area places the chaotic region.

Since the sectoral resonances are spin-orbit resonances, we have that $\mu$ has a minor effect on their locations, as seen in the figure. However, the resonance width will depend on $\mu$, as an increase in the mass anomaly will enhance the gravitational perturbation felt by the particles, allowing larger regions to be connected to the resonance equilibrium points. As we increase the numerical value of $m$, the resonances approach the corotation radius. The first-order resonances with $|m|>4$ overlap for the case with $\mu=10^{-4}$ (fig~\ref{fig:mu}a). For $\mu=10^{-3}$ (fig~\ref{fig:mu}b), we see that additional first-order resonances, such as 4:5 and 3:4, overlap for high eccentricities, while for $\mu=5\times10^{-3}$ (fig~\ref{fig:mu}c) the overlap is intensified, covering the 2:3 resonance.

The overlap of first-order resonances is generally responsible for eliminating stable regions associated with the resonance \citep{Wisom1980,Winter1997a,Winter1997b}. Thus, they should contribute to the chaotic behaviour verified in the systems. It is not by chance that the region surrounding the corotation radius is always chaotic. However, while the overlap helps to carve the chaotic region around the central object, it is not the primary source of chaoticity for the system. Such a fact can be seen in \ref{fig:lambda}a, where the chaotic region covers a region where there is no overlap of first-order resonances. As already mentioned, encounters with the mass anomaly produce chaotic diffusion of the orbits, clearing an entire region that extends beyond the corotation radius. In our numerical simulations, we did not find stable particles in internal resonances.

Analogously to Figure~\ref{fig:mu}, we present in Figure~\ref{fig:lambda} the resonances and chaotic and stable regions around an object with a mass anomaly, now keeping $\mu$ constant and varying $\lambda$. As we can see in the figure, the parameter affects both the location and the width of the resonances (Eqs.~\ref{phidot0} and \ref{CR}). When $\lambda $ is increased by one order, resonances move more than four times closer to the body, while chaotic regions approach it by only a factor of two. So, by changing $\lambda$, we change which resonances will be in the stable region.

In the case shown in Figure~\ref{fig:lambda}c, the rotation frequency is equal to the Keplerian one, which places the corotation radius on the surface of the spherical portion of the body. Consequently, the internal resonances and some external ones reside within the central body, with only a few resonances in the stable region. Assuming objects with even faster rotation, we get a narrower chaotic region with fewer resonances outside the object, corresponding to unattractive cases. In the hypothetical case where the spin frequency tends to infinity, it would be not sectoral resonances or chaotic region since it falls in the case of a non-rotating spherical object with a ridge at its equator.

In Section~\ref{resoperi}, we analyse the evolution of the stable region and the external resonances using Poincar\'e surfaces of section. 

\section{Stable Region} \label{resoperi}
In Section~\ref{overview}, we have shown the existence of two distinct regions around a spherical body with a mass anomaly: a chaotic region where particles collide or are ejected, and a stable region, which will be our focus in this section. First, we compare the resonance widths obtained by numerical simulations with those predicted by the analytical model described in Section~\ref{theory}. We then analyse the motion of test particles in the vicinity of external resonances.

We put our analytical model to the test using the following methodology: i) for a given resonance, we theoretically calculate its central position (Section~\ref{resonancelocation}) and the Jacobi constant (Equation~\ref{eq:Cj}), initially assuming a circular orbit; ii) we generate  Poincar\'e surface of section of a broad region around the central position; iii) by visual inspection, we obtain the position of the stable fixed point of the resonance -- which corresponds to the real central position of the resonance -- and the limits of the widest island surrounding the point -- the width of the resonance; iv) then, we successively increase the eccentricity by $10^{-2}$ and repeat the previous steps until the islands disappear or until we reach $e=0.5$.

\begin{figure*}
\centering
\subfigure[$\lambda=0.471$ and $\mu=10^{-3}$]{\includegraphics[width=1.6\columnwidth]{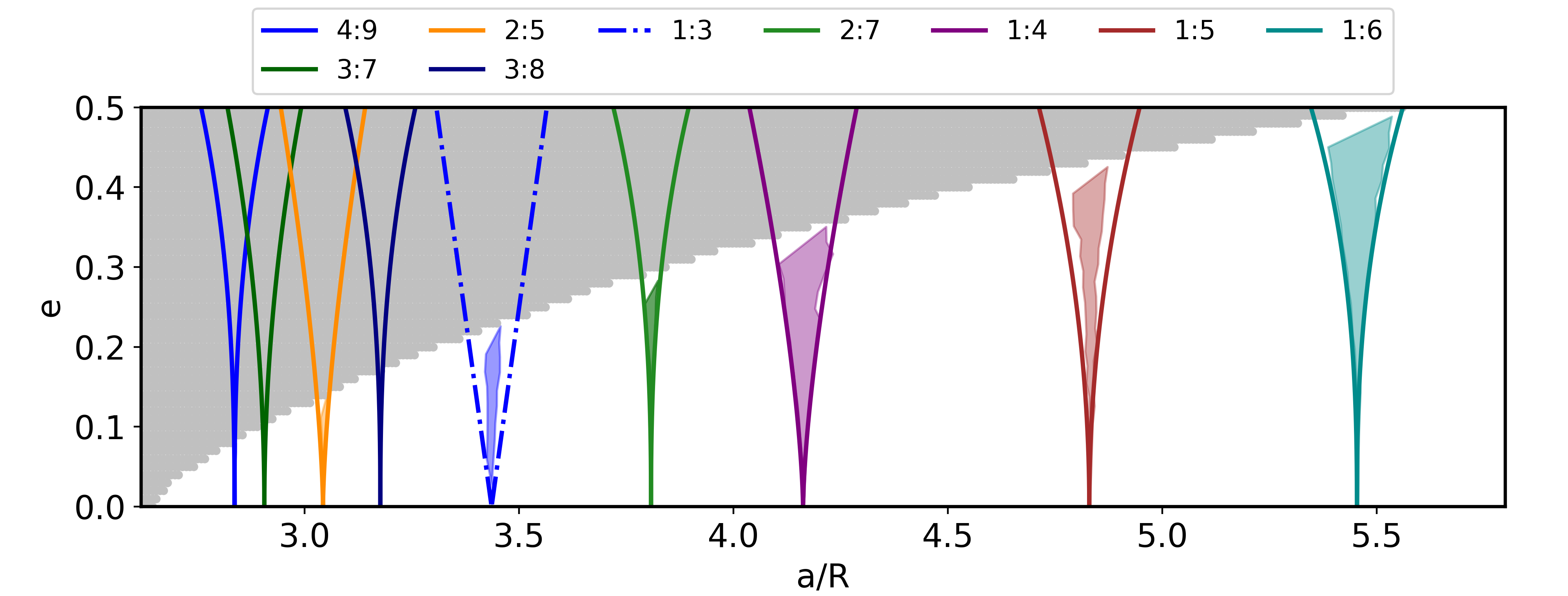}}
\centering
\subfigure[$\lambda=0.157$ and $\mu=10^{-3}$]{\includegraphics[width=1.6\columnwidth]{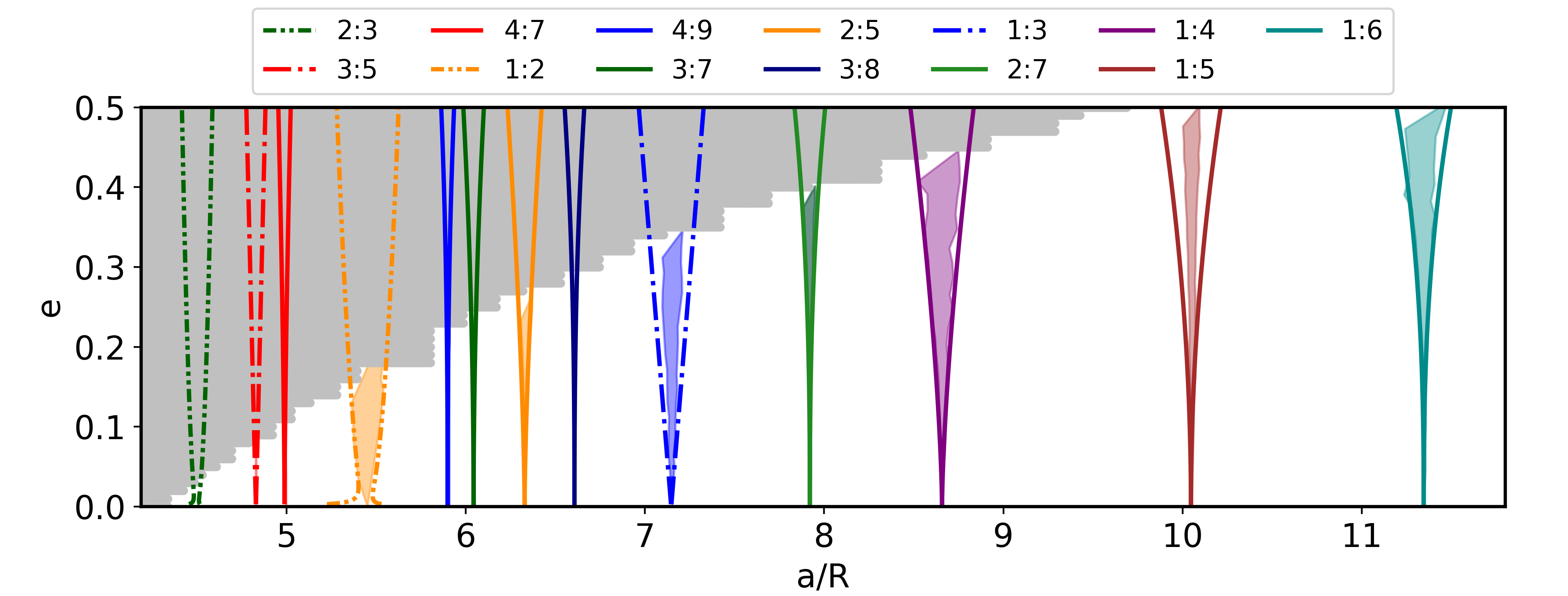}}
\caption{The width of the external sectoral resonances in the stable region for a) an object with $\lambda=0.471$ and $\mu=10^{-3}$ and b) for an object with $\lambda=0.157$ and $\mu=10^{-3}$. The solid and dashed lines give the widths predicted by the analytical model, and the coloured filled regions delimit the obtained numerically widths. The grey region corresponds to the chaotic region near the central body. \label{fig:TSSP}}
\end{figure*}

Figure~\ref{fig:TSSP} shows the resonance widths obtained theoretically and through the Poincar\'e surfaces of sections, for our reference object and an object with $\lambda=0.157$ and $\mu=10^{-3}$. We found that numerical data agree reasonably well with the analytical model, indicating that the pendulum model with necessary adaptations applies to our system. In general, we obtain that the largest divergences occur for larger eccentricities ($e\gtrsim 0.2$). It is expected, since we assumed first-order approximations in eccentricity in the development of the pendulum model.

We verify that the innermost resonances present the largest displacements in the central position for the reference case. These same displacements are verified for the case with $\lambda=0.157$ (Figure~\ref{fig:TSSP}b) in which resonances are at least twice as far from the central body. As a rule, we obtain that displacements depend on the distance $a_t/a_c$ from the resonance to the corotation radius. The central positions we obtained differ by less than 5\% from those theoretically obtained, demonstrating the robustness of the analytical method.

\begin{figure}
\centering
\subfigure[]{\includegraphics[width=0.88\columnwidth]{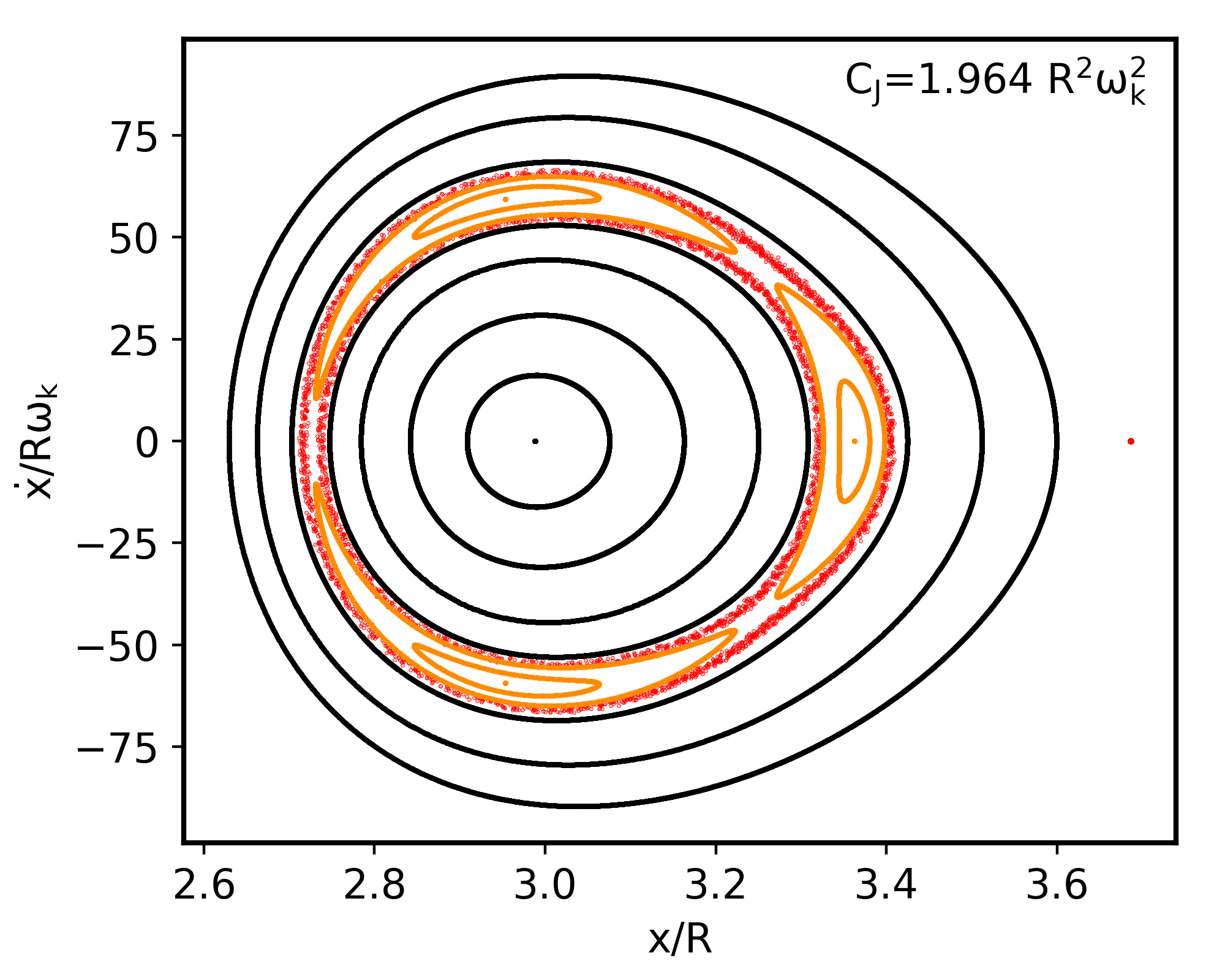}} 
\subfigure[]{\includegraphics[width=0.88\columnwidth]{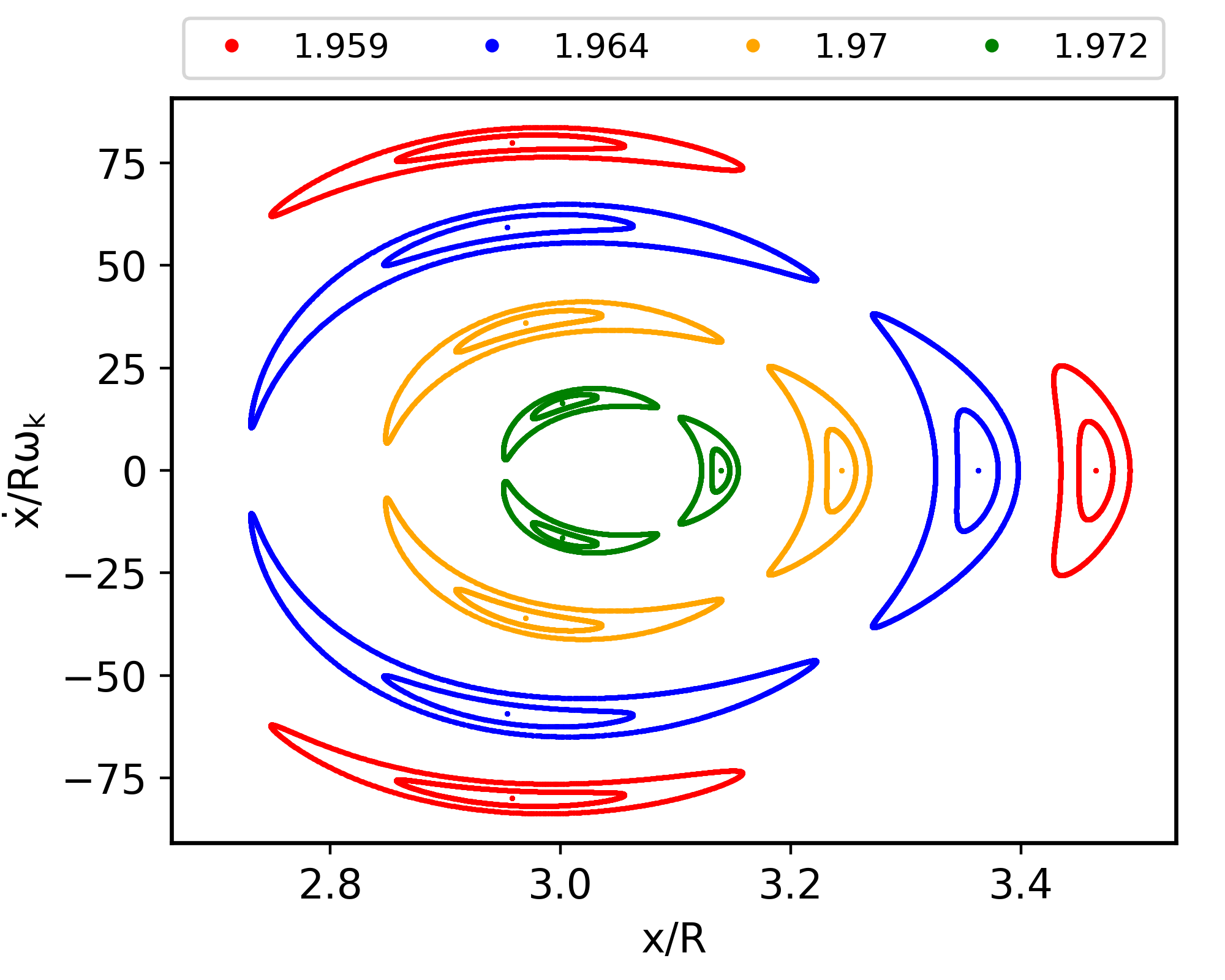}}
\centering
\subfigure[]{\includegraphics[width=0.88\columnwidth]{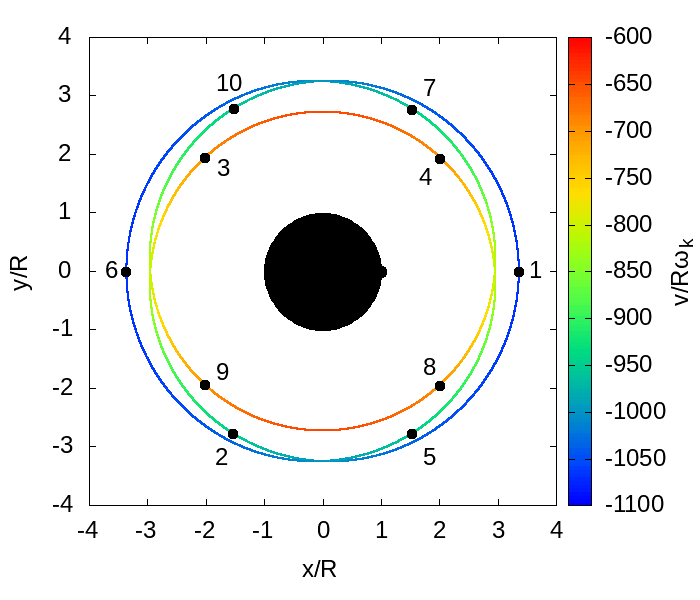}}
\caption{a) Poincar\'e surface of section for $C_J=1.964~{\rm R^2\omega_k^2}$, with $\lambda=0.471$ and $\mu=10^{-3}$. We assume initial conditions with $3.15\leq x_0/R\leq 3.84$. The black curves are the periodic and quasi-periodic orbits of first kind, and the orange curves are orbits associated with the 2:5 resonance. Red dots correspond to chaotic orbits. b) Evolution of the 2:5 resonance islands, where the colours of the dots correspond to values of $C_J$ given on the figure's label. c) Central orbit of the 2:5 resonance for $C_J=1.964~{\rm R^2\omega_k^2}$ in the rotating frame. The temporal evolution of the orbit is given by numbers and dots equally spaced in time, while the colour-coding gives the velocity in the rotating frame. \label{fig:25}}
\end{figure}

After the validity of the analytical model is attested, we turn our attention to the resonance dynamics. Figure~\ref{fig:25}a shows Poincar\'e surface of section of the region around the 2:5 resonance for the reference object and $C_J=1.964~{\rm R^2\omega_k^2}$. In the figure, we can identify four types of motion in the stable region:  periodic motion of first kind, quasi-periodic motion associated with the latter, resonant and chaotic motion -- which is not stable despite being within the region defined by us as stable. The single black dot at $x_0/R=3$ corresponds to the orbit classified as periodic of first kind. Periodic orbits in a Poincar\'e surface of section divide the $x$-axis positions into pericentric positions -- at smaller $x$ -- and apocentric positions -- at larger $x$. Seeing the right part of the figure, we have particles with higher initial eccentricity, forming black closed curves surrounding the periodic orbit. They are quasi-periodic orbits and define regions where particles remain indefinitely in stable motion without other effects.  

The orange islands correspond to orbits associated with the 2:5 resonance, where every single dot in the centre of an island is a stable fixed point of the resonance. All three orange dots in the figure correspond to a single periodic orbit of second kind. Due to energy exchanges between the central body and resonant particles, the latter can remain stable even in the presence of other effects, depending on the system conditions.

The red dots show the chaotic zone between the resonant and quasi-periodic orbits. Appendix~\ref{pssrc} shows  Poincar\'e surfaces of section of the resonances found around the reference object. As one can see, the chaotic zones at the resonances separatrices are always narrow, showing that the region we named the ``stable region'' has in fact a few very small strips of confined chaotic motion. Moving to the right in Figure~\ref{fig:25}a, there is another region with quasi-periodic orbits that extends up to $x_0/R=3.57$. After this limit, we have the chaotic region, where the red dot at $x_0/R=3.65$ corresponds to a particle that collides with the central body. 

In Figure~\ref{fig:25}b, we present the evolution of the 2:5 resonance, showing the largest stable orbit of the resonance, an intermediate one, and the central orbit, for different values of the Jacobi constant. Since the 2:5 resonance is a third-order resonance ($j=3$), each initial condition produces three distinct islands in  Poincar\'e surfaces of section. Except for the case with $C_J=1.959~{\rm R^2\omega_k^2}$, we see that the three islands shrink and get closer as the value of $C_J$ increases. The Jacobi constant and the eccentricity are inversely proportional, so the latter decreases from right to left in the figure. The resonance width decreases with the eccentricity, and the resonant orbits tend towards the periodic orbit of the first kind, explaining why the islands shrink until they disappear.

\begin{figure}
\centering
\subfigure[]{\includegraphics[width=0.88\columnwidth]{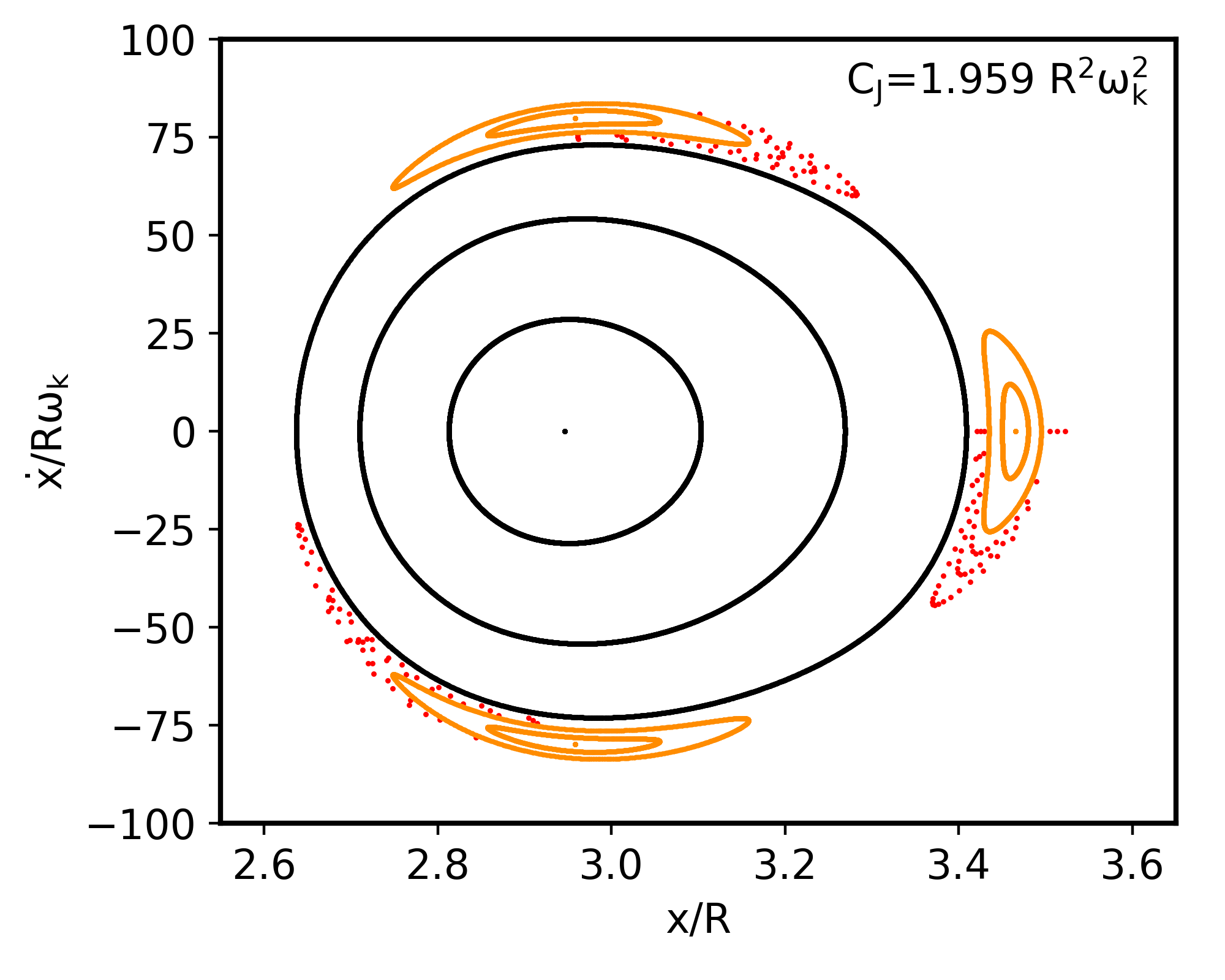}}
\subfigure[]{\includegraphics[width=0.88\columnwidth]{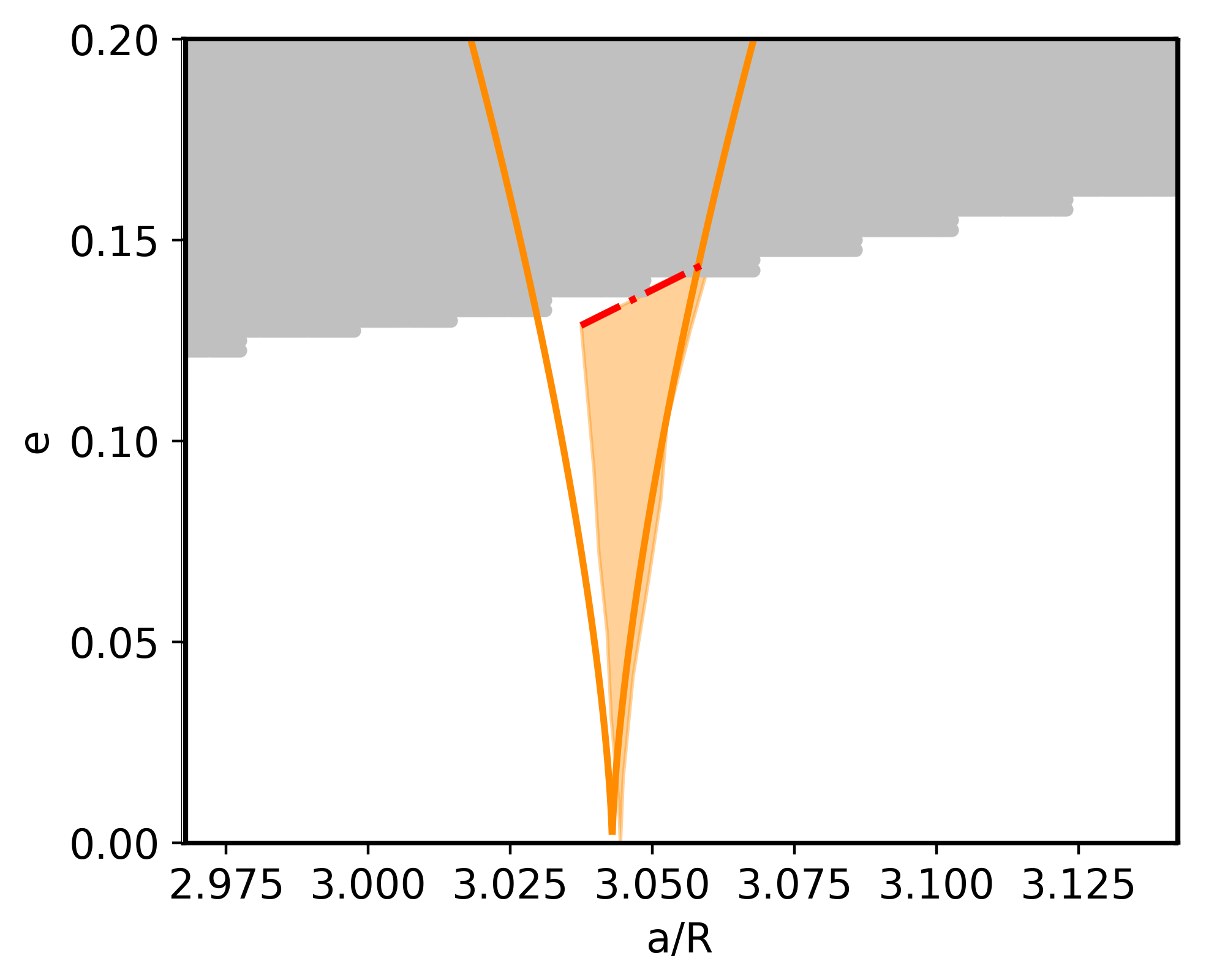}} 
\centering
\subfigure[]{\includegraphics[width=0.88\columnwidth]{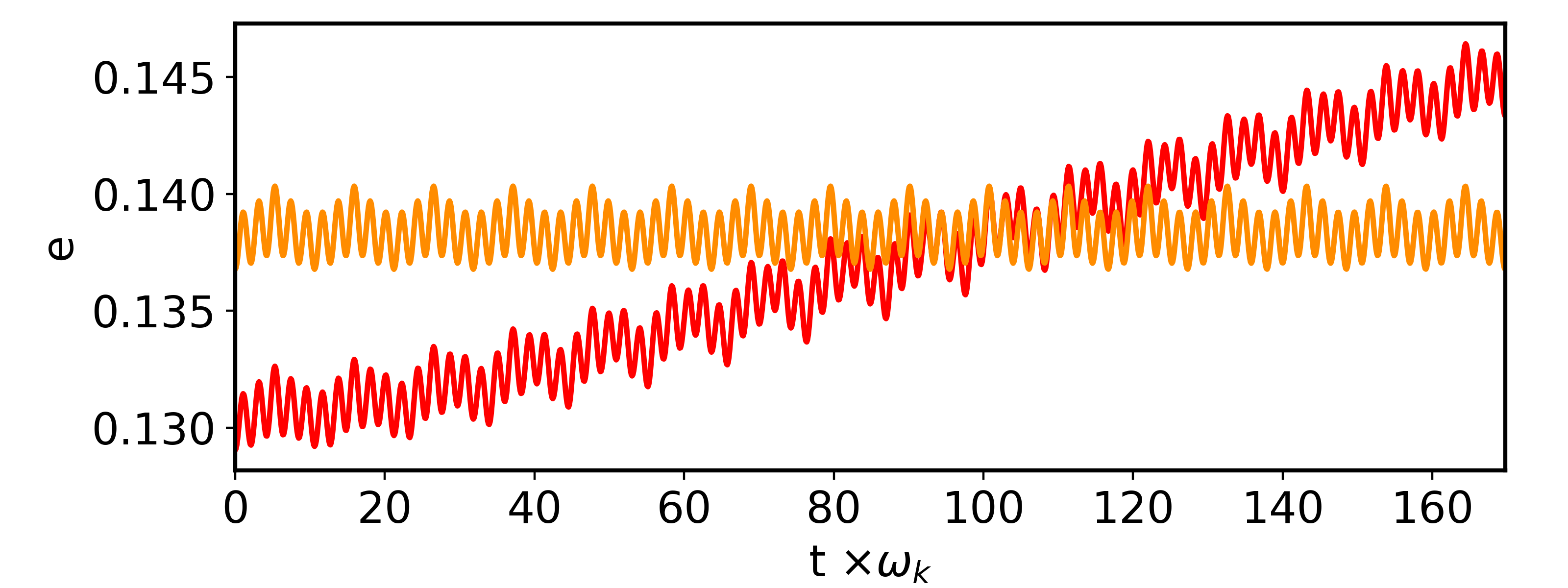}}
\caption{a) Poincar\'e surface of section for $C_J=1.959~{\rm R^2\omega_k^2}$ where the periodic/quasi-periodic orbits of first kind are in black, the 2:5 resonance islands are in orange, and the particles in the chaotic region are in red. b) Theoretical boundaries of the 2:5 resonance are shown by the solid orange lines. In contrast, the filled orange and grey regions are regions numerically obtained for the 2:5 resonance and the chaotic region, respectively. The red dashed line gives the initial conditions of the simulation with $C_J=1.959~{\rm R^2\omega_k^2}$. c) Eccentricity of a pair of particles: the one that remains in the system is orange, and the unstable one is red.\label{fig:25zoom}}
\end{figure}

To understand why the largest red island is smaller than the largest blue island (Figure~\ref{fig:25}b), we present in Figure~\ref{fig:25zoom}a the Poincar\'e surfaces of section for $C_J=1.959~{\rm R^2\omega_k^2}$, and in Figure~\ref{fig:25zoom}b, the theoretical and numerical boundaries of the 2:5 resonance. The red dashed line places the case with $C_J=1.959~{\rm R^2\omega_k^2}$ and the grey area is the chaotic region. For this value of Jacobi constant, the resonance is at the edge between the stable and chaotic regions. As a result, the particles most bounded to the resonance - closer to the stable fixed point -- remains stable (in orange), while particles closer to the resonance boundaries initially follow the pattern expected for resonant particles. However, they are showing the stickiness phenomenon behaviour mimicking the resonant behaviour, but they are lost from the system at some point.

The eccentricity of one of these less bounded particles is shown in Figure~\ref{fig:25zoom}c by the solid red line, while the eccentricity of the central resonant orbit is the solid orange line. Both particles show a periodic variation in eccentricity. However, the eccentricity of the less bounded particle also shows an increase, reaching $e\sim 0.145$. 

Figure~\ref{fig:25}c shows, in the rotating frame, the periodic orbit of the second kind seen in Figure~\ref{fig:25}a, where the colour-coding gives the velocity. Since the particle is at the stable fixed point of the resonance, the orbit is closed. Also, the orbit is retrograde ($v<0$) because the resonance is beyond the corotation radius. As the central body is symmetric,  there will always be at least one axis that divides the orbit into two symmetric parts. For example, for the orbit shown in Figure~\ref{fig:25}c, this axis corresponds to $y=0$.

\cite{Sicardy2020} discusses some additional symmetries expected for the trajectory of a particle in a $m:m-j$ sectoral resonance. The orbit is invariant by a rotation of $360/|m|$~deg, and it has a total of $|m|(j-1)$ self-crossing. For the 2:5 case ($j=3$ and $m=-2$), we see that the orbit is invariant by a rotation of 180~deg and has four self-crossing.

A peculiarity of our system is the positions of the particle pericentre and apocentre. In RP3BP, in which the disturbing body is at $x_0/R=1$ and the particle is initially at $x_0/R<1$, the gravitational effect felt by the particle is weaker (stronger) when it is on the $x$-axis at $x/R>0$ ($x/R<0$). Consequently, the particle starts at the pericentre, the apocentre being in the opposite direction. In our case, we have the opposite scenario. The orbit position where a particle feels the strongest gravitational effect is on the $x$-axis at $x/R>0$ -- where the modulus of gravitational force is the sum of the forces of the two portions of the central body. Thus, a particle initially on the $x$-axis ($x/R>0$) starts at its apocentre (minimum velocity), as we can see from the dot labelled ``1'' in  Figure~\ref{fig:25}.

\begin{figure*}
\begin{minipage}[t]{0.49\linewidth}
\vfill \vfill
\subfigure[]{\includegraphics[width=0.88\columnwidth]{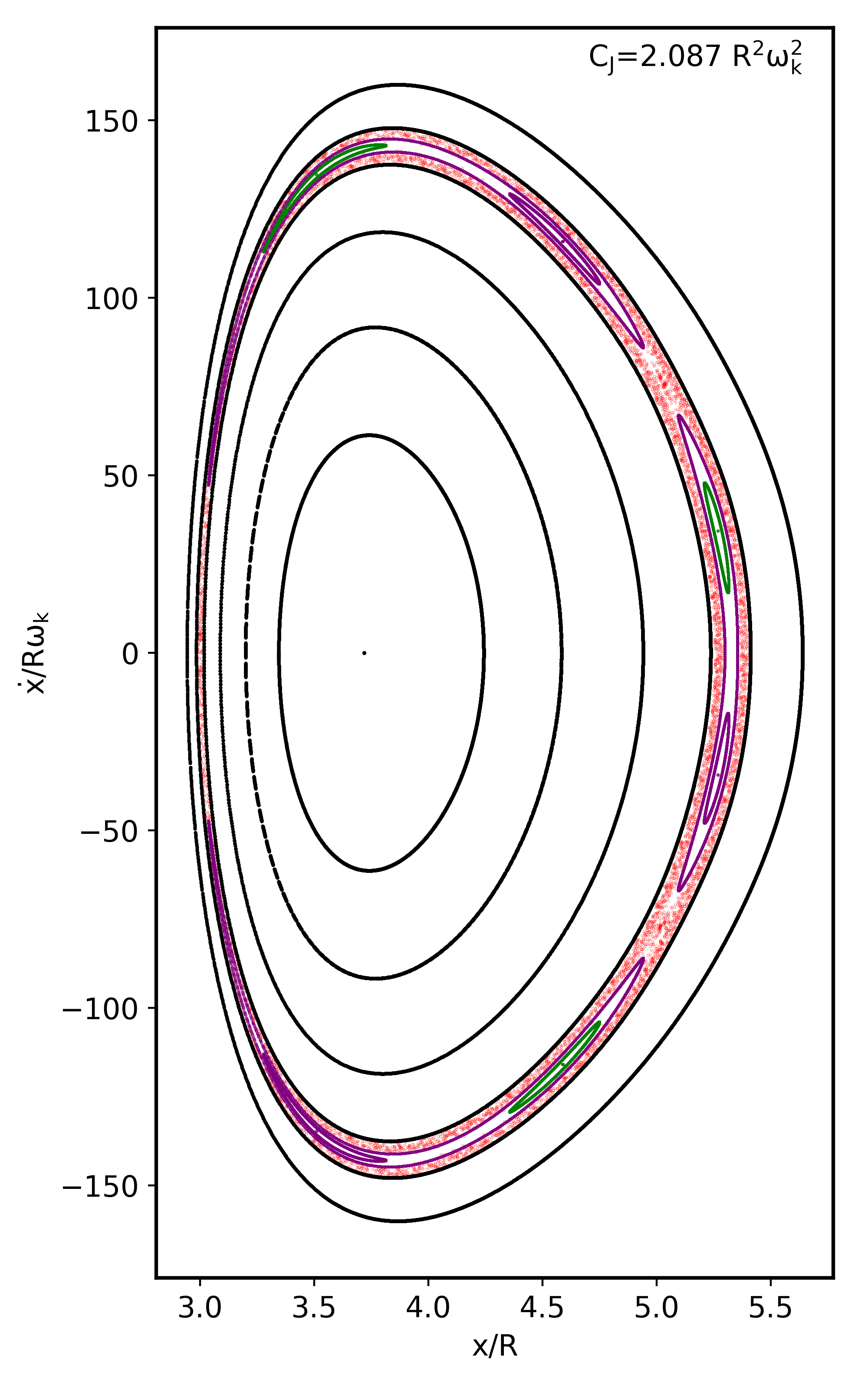}} 
\end{minipage}
\hfill
\begin{minipage}[t]{0.49\linewidth}
\subfigure[]{\includegraphics[width=0.88\columnwidth]{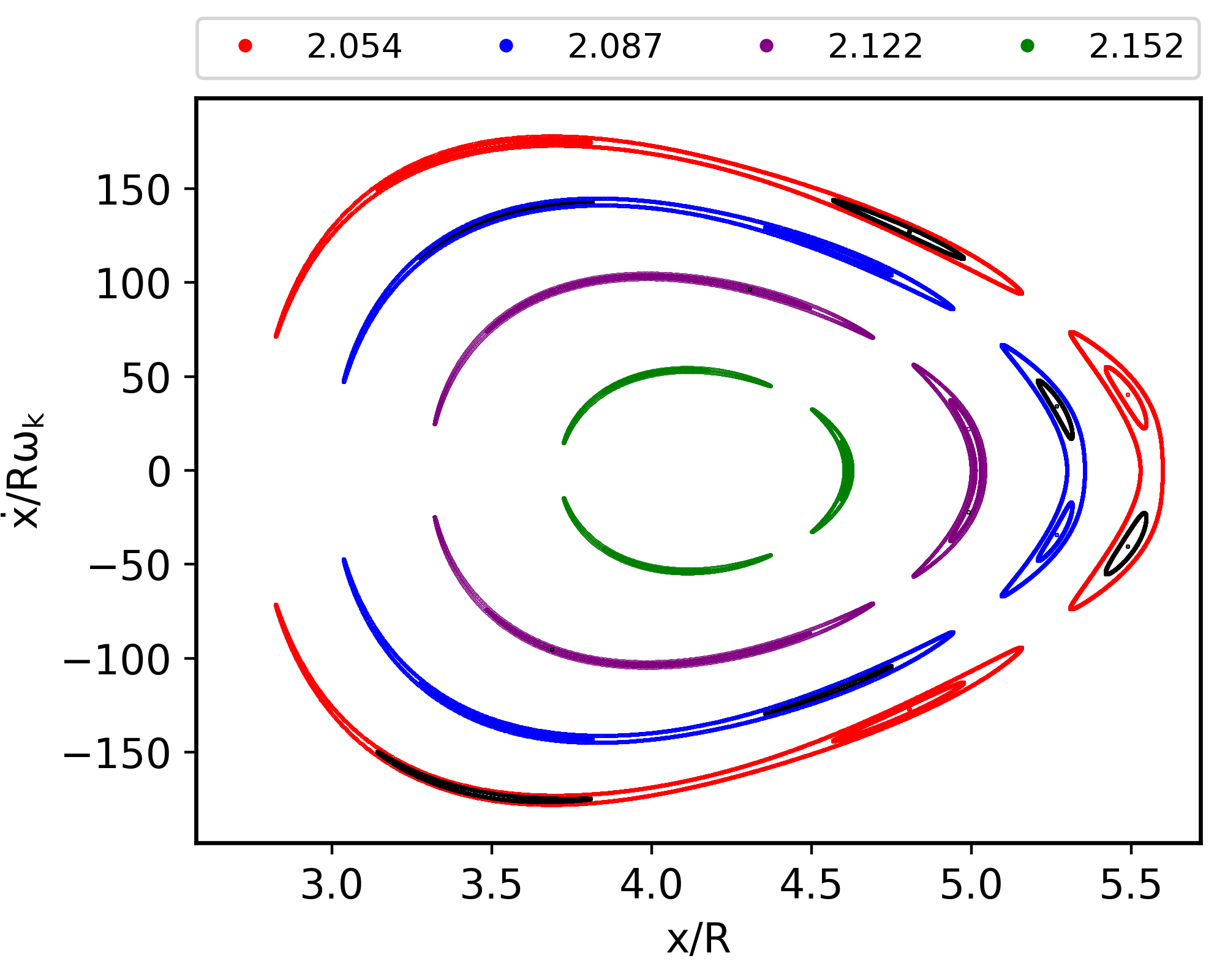}} 
\subfigure[]{\includegraphics[width=0.88\columnwidth]{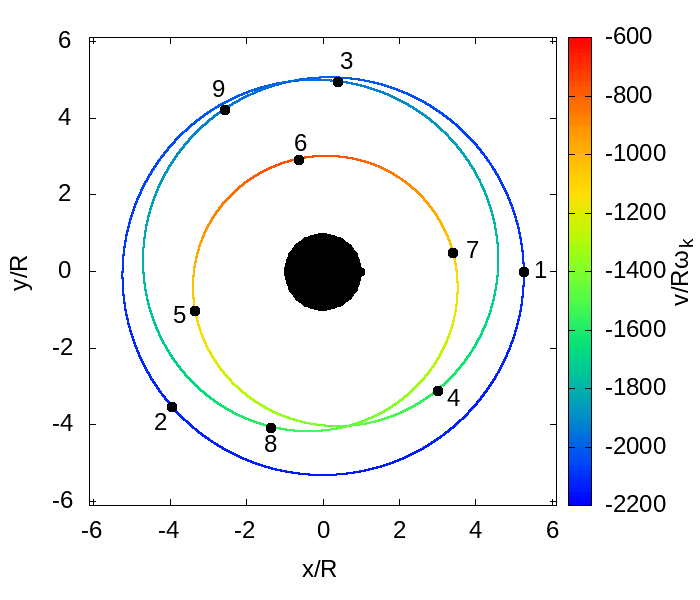}}
\end{minipage}
\caption{a) Poincar\'e surface of section for $C_J=2.087~{\rm R^2\omega_k^2}$, with $\lambda=0.471$ and $\mu=10^{-3}$. We assumed initial conditions with $3.70\leq x_0/R\leq 5.97$ and separated the distinct types of orbits by colour: the periodic/quasi-periodic orbits of first kind are in black, the 1:4 resonance orbits are in purple and green and chaotic ones in red. b) Resonance islands for different values of $C_J$. The label on the panel gives the colour of the largest island for each value of $C_J$. c) Central orbit in the rotating frame of one of the families associated with the 1:4 resonance (in green in the top panel) for $C_J=2.087~{\rm R^2\omega_k^2}$. The numbers and colours on the panel provide time evolution and velocity in the rotating frame, respectively. \label{fig:14}}
\end{figure*}

Figure~\ref{fig:14} shows, from top to bottom, the Poincar\'e surface of section for $C_J=2.087~{\rm R^2\omega_k^2}$, the whole evolution of the islands of the 1:4 resonance and the trajectory of a particle at a stable fixed point of the resonance. As shown in Figure~\ref{fig:14}a, the overview of the resonance neighbourhood is similar to the 2:5 resonance, with a narrow, chaotic region at the resonance boundaries, surrounded by a region with periodic/quasi-periodic orbits. A crucial difference, however, is obtained in the resonance islands. While the 2:5 resonance has three stable fixed points, we obtained in Figure~\ref{fig:14}a six stable points for the 1:4 resonance.

To understand the dynamics of the resonance, we colour green (Figure~\ref{fig:14}a) the trajectory of a particle near one of these points. The particle is responsible for forming three islands around three of the stable fixed points (in green in the figure). This fact leads us to conclude that the resonance is the 1:4 (of third-order) and not the 2:8 (of sixth-order) as we would obtain in the ellipsoidal problem -- which we will address in a following publication. The islands produced by the particle have the particularity of being asymmetric in relation to the $x$-axis -- we say that the particle is in asymmetric libration \citep{Beauge1994,Winter1997b}.

In Figure~\ref{fig:14}b, we highlight the Poincar\'e surface of section islands of some particles by plotting them in black, intending to show the asymmetric libration. Each island produced by a particle in asymmetric libration has a mirror image obtained from the motion of a different particle in asymmetric libration. Closer to the resonance boundaries, we also get  ``horseshoe fashion'' orbits encompassing pairs of fixed points of two different trajectories.

When we refer to asymmetric resonance or libration, we refer to the symmetry of the trajectory in Poincar\'e surface of section and not in the $xy$-plane. As already mentioned, the trajectory in the $xy$-plane of the resonant particles has a symmetry axis due to the symmetric mass distribution in the central body. For example, in Figure~\ref{fig:14}c, the axis of symmetry would correspond to the axis connecting the point ``6''  to the centre of the system.

Several works such as \cite{Message1970}, \cite{Frangakis1973a,Frangakis1973b}, \cite{Message1978}, \cite{Bruno1994}, \cite{Beauge1994} and \cite{Winter1997b} have studied asymmetric periodic orbits in the context of RP3BP, showing that these orbits are characteristics of $1:1+p$ resonances and are obtained only for particles with eccentricities above a critical value. Similar to particles in $m:m-j$ resonance with $m\neq -1$, the ones with eccentricity lower than this threshold value present symmetric libration in  Poincar\'e surface of section. We obtained these same results for the case with mass anomaly. In Figure~\ref{fig:14}b, the critical eccentricity is reached somewhere between the Jacobi constants $2.122~{\rm R^2\omega_k^2}$ and $2.152~{\rm R^2\omega_k^2}$. Carrying out a set of Poincar\'e surface of section in this interval, we obtain that the critical eccentricity for the 1:4 resonance is $e\sim0.167$ ($2.136~{\rm R^2\omega_k^2}$).

\begin{figure}
\centering
\subfigure[]{\includegraphics[width=0.88\columnwidth]{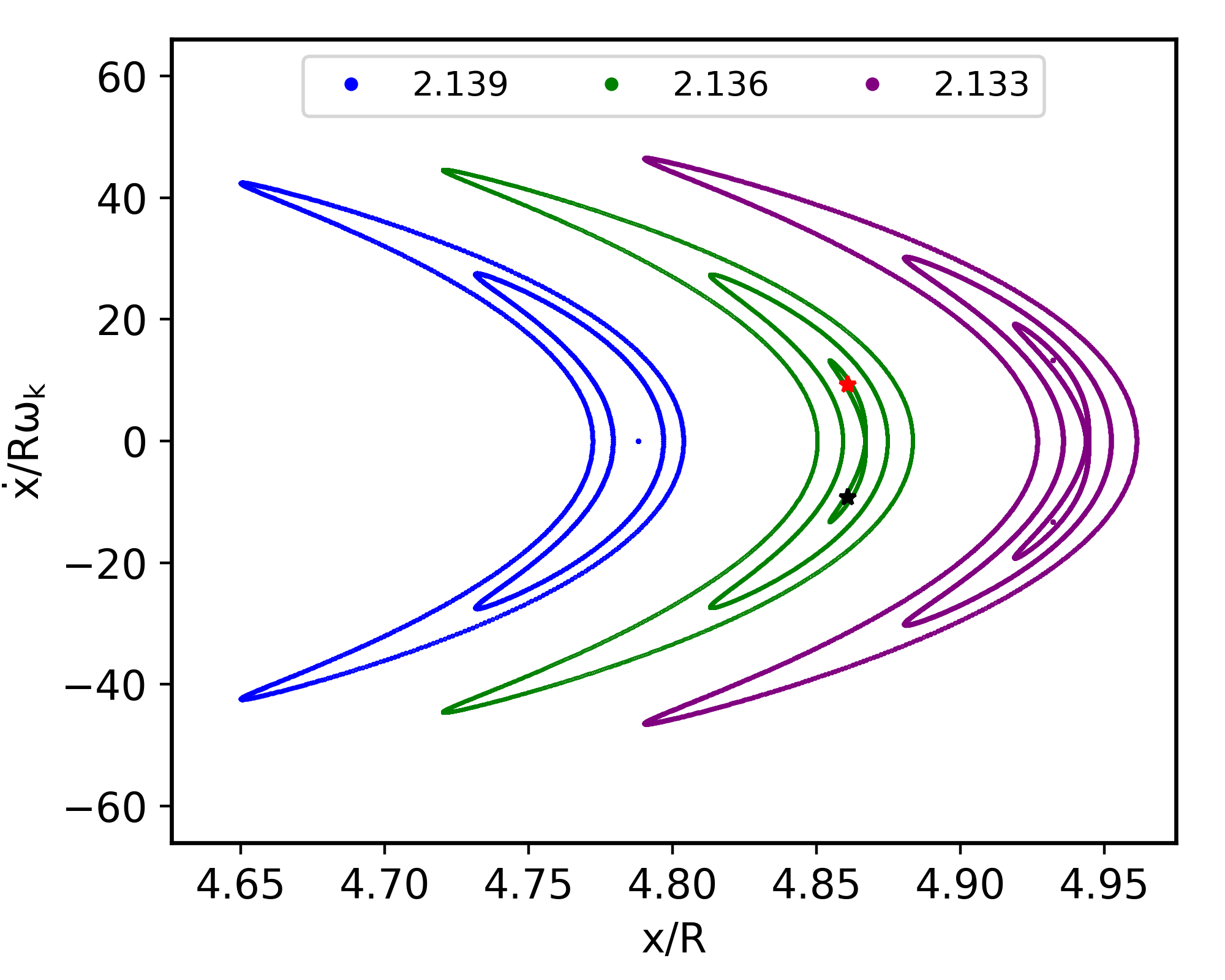}} 
\subfigure[]{\includegraphics[width=0.88\columnwidth]{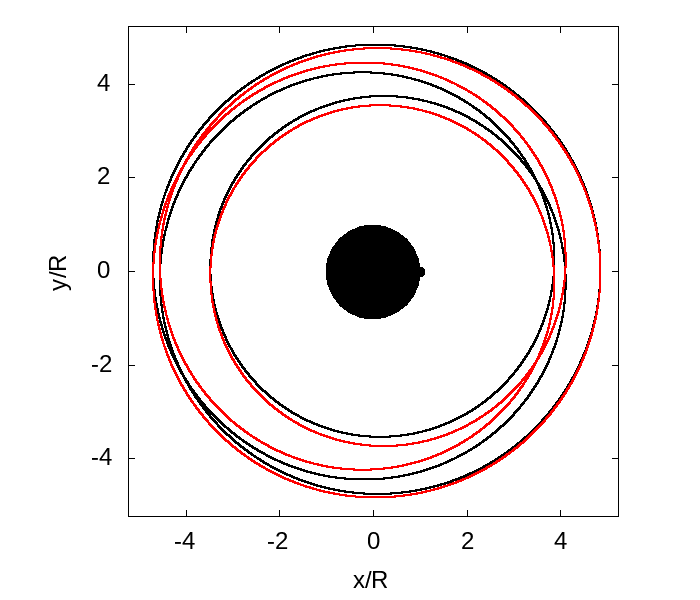}} 
\subfigure[]{\includegraphics[width=0.88\columnwidth]{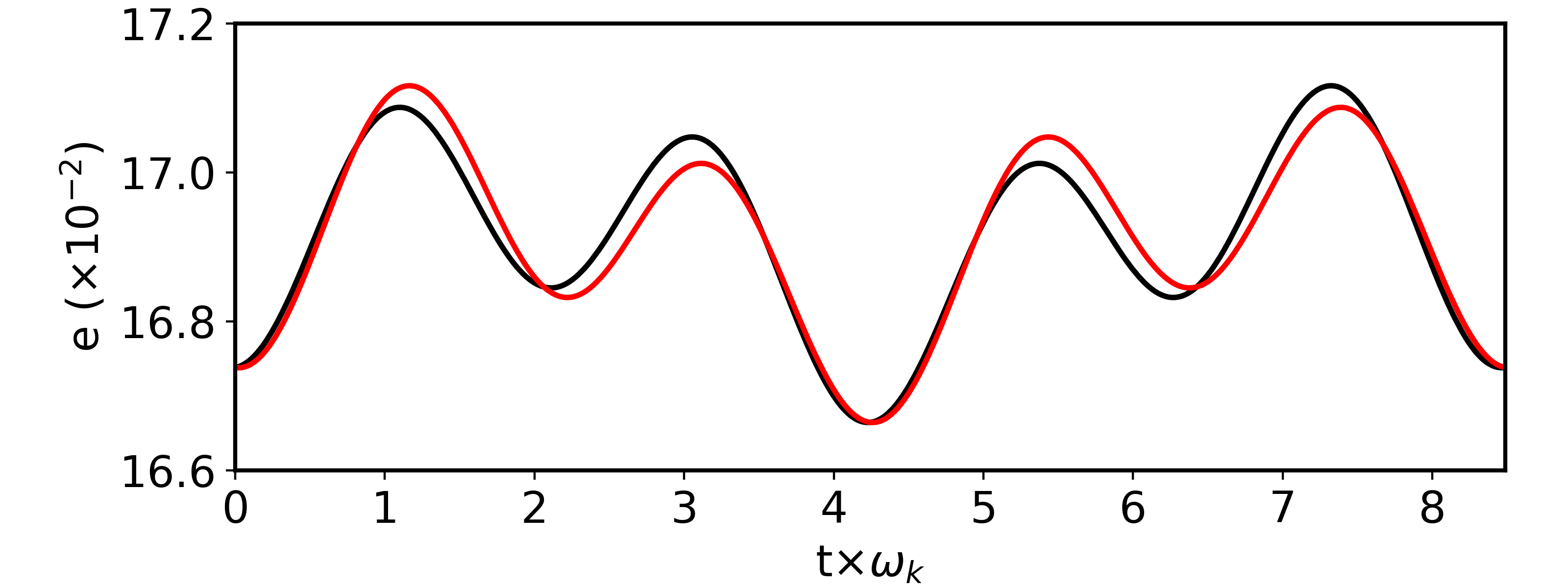}} 
\caption{a) Poincar\'e surface of section of one island of the 1:4 resonance for $C_J=2.133$,~$2.136$, and~$2.139~{\rm R^2\omega_k^2}$ (in purple, green, and blue, respectively). The black and red stars are the stable points obtained after bifurcation. b) Trajectories and c) eccentricities of the stable points given by stars in the top panel, where the colour of the solid lines coincides with the colour of the star for the same stable point.\label{fig:14bisec}}
\end{figure}

Figure~\ref{fig:14bisec}a shows one island of the 1:4 resonance for the critical eccentricity ($2.136~{\rm R^2\omega_k^2}$, in green), for $2.133~{\rm R^2\omega_k^2}$ (in purple) and $2.139~{\rm R^2\omega_k^2}$ (in blue). For the highest value of Jacobi constant (smallest eccentricity), we see a single stable point in the figure related to a single family of resonant orbits. The critical eccentricity is reached by decreasing the Jacobi constant, and the stable point bifurcates into two points (the stars in the figure). Each of the points gives rise to an independent family of resonant orbits. The $x$-axis, which previously allocated the single stable point, now allocates the unstable equilibrium point after the bifurcation, corresponding to the inflexion position of the ``horseshoe fashion'' orbits. 

Figure~\ref{fig:14bisec}b shows the trajectories of the stable points given by stars in the top panel. The orbits are mirror versions of each other. The same is obtained for eccentricities in Figure~\ref{fig:14bisec}c, in which the red curve is the mirror version of the black one with respect to time $t\approx4.2\omega_k^{-1}$ (pericentre passage time). As discussed in \cite{Bruno1994}, the bifurcation of the stable points is related to the indirect term of the disturbing function, which differs from zero only for $1:1+p$ resonances (equation~\ref{potentialexp}). 

\begin{figure}
\centering
\subfigure[]{\includegraphics[width=0.88\columnwidth]{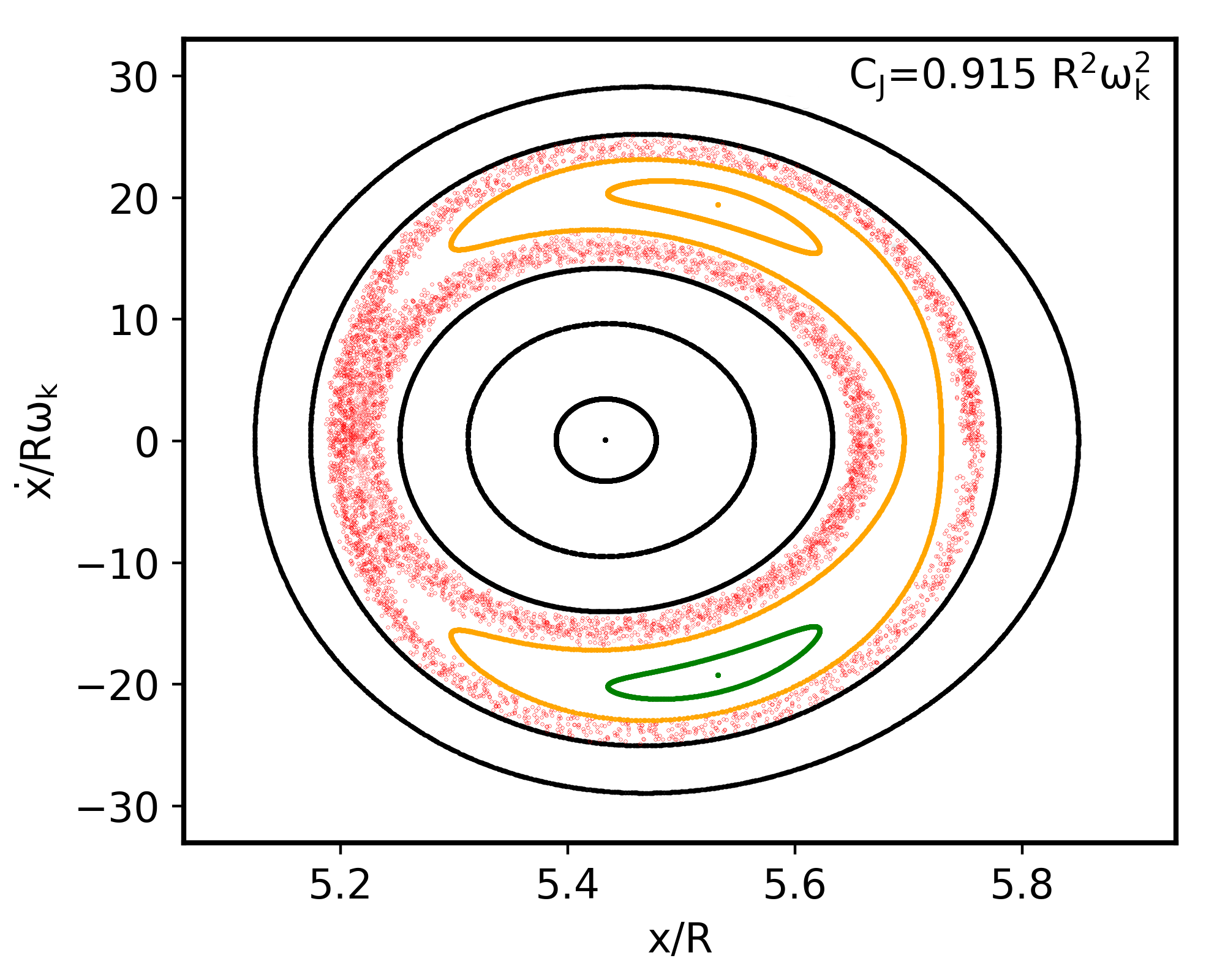}} 
\subfigure[]{\includegraphics[width=0.88\columnwidth]{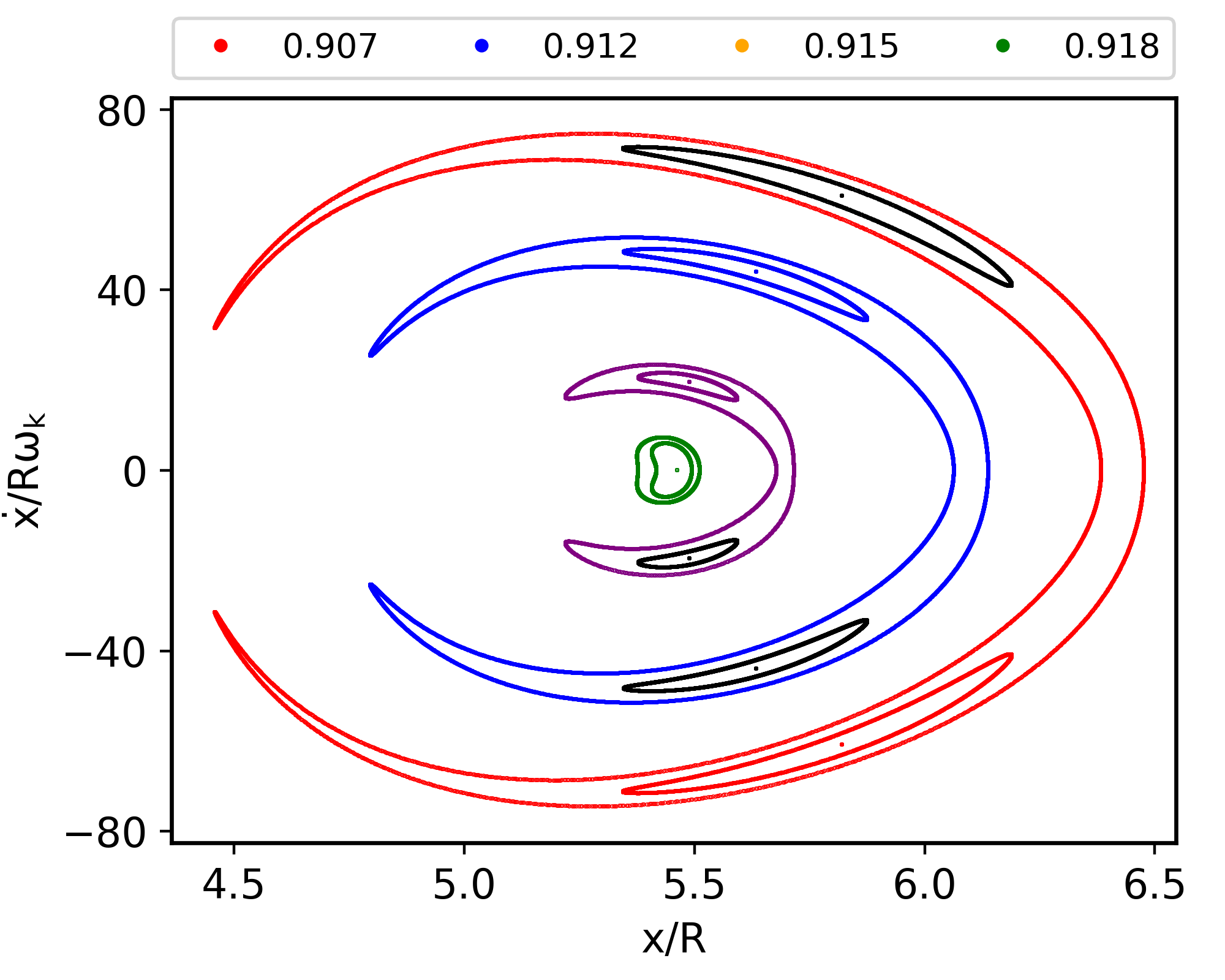}} 
\centering
\subfigure[]{\includegraphics[width=0.88\columnwidth]{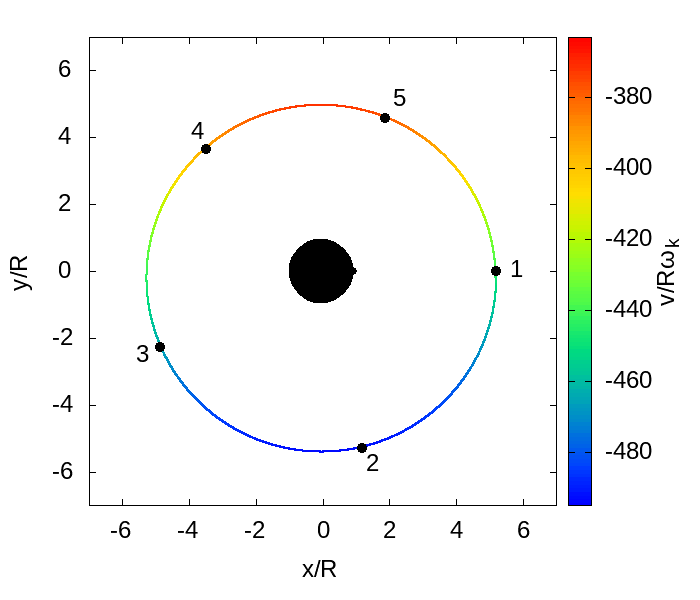}}
\caption{a) Poincar\'e surface of section for $C_J=0.915~{\rm R^2\omega_k^2}$, with $\lambda=0.157$ and $\mu=10^{-3}$. The non-resonant orbits are in black. Particles in 1:2 resonance and chaotic orbits are in orange and green and red, respectively. b) Poincar\'e surface of section for some particles in 1:2 resonance, with $C_J=0.907$, $0.912$, $0.915$, and $0.918~{\rm R^2\omega_k^2}$. Different colours of the islands involved by the same ``horseshoe fashion'' orbit correspond to different particles. c) Trajectory of a stable fixed point shown in orange in the top panel, where the colour-coding gives the velocity and the numbers and dots, the time evolution of the orbit. \label{fig:12}}
\end{figure}

We show another example of particles in asymmetric libration in Figure~\ref{fig:12}. From top to bottom, this figure shows Poincar\'e surface of section of the region of 1:2 resonance for a central body with $\lambda=0.157$ and $\mu=10^{-3}$, the whole evolution of the resonance, and the trajectory of a particle in a stable fixed point of the resonance. As in the reference case, chaotic behaviour is seen only in a narrow region in the separatrices, with a large regular region of periodic/quasi-periodic orbits of first kind around the resonances. For the 1:2 resonance, we have a low critical eccentricity ($e\sim 10^{-2}$),  with symmetric libration only in the cases where the resonance islands are tiny. 

Trajectories of particles in 1:2 resonance are the only ones without self-crossings in the rotating frame, as we can see in the bottom panel. Such fact has implications for the temporal evolution of a ring of particles, as self-crosses increase collisions between particles. In this context,  a ring with particles into the 1:2 resonance or in periodic/quasi-periodic orbits of first kind -- which do not show self-crossing either -- should have a lower rate of collisions than a ring with particles in other resonances,  disregarding other external effects. The particles shown in Figures~\ref{fig:14}c and \ref{fig:12}c do not start at the apocentre of the orbit because they are not initially with $\dot{x}=0$. A particle around a body with mass anomaly will start at its apocentre only when that condition is met.

\section{Application to the Chariklo system} \label{chariklosection}
\cite{Leiva2017} using stellar occultation data, investigated the shape of Chariklo, obtaining four distinct shapes models for the object: a sphere, a MacLaurin spheroid, a triaxial ellipsoid, and a Jacobi ellipsoid. According to \cite{Sicardy2019}, observational data suggest the presence of topographic features of typical heights of 5~km in the spherical solution. This fact places Chariklo as a possible body with a mass anomaly. In this section, we briefly study the dynamics around Chariklo, in particular in the region of the ring. The rings have orbital radii of 391~km and 405~km, with radial widths of 7~km and 3~km, respectively \citep{Berard2017}.

\begin{figure*}
\centering
\includegraphics[width=1.5\columnwidth]{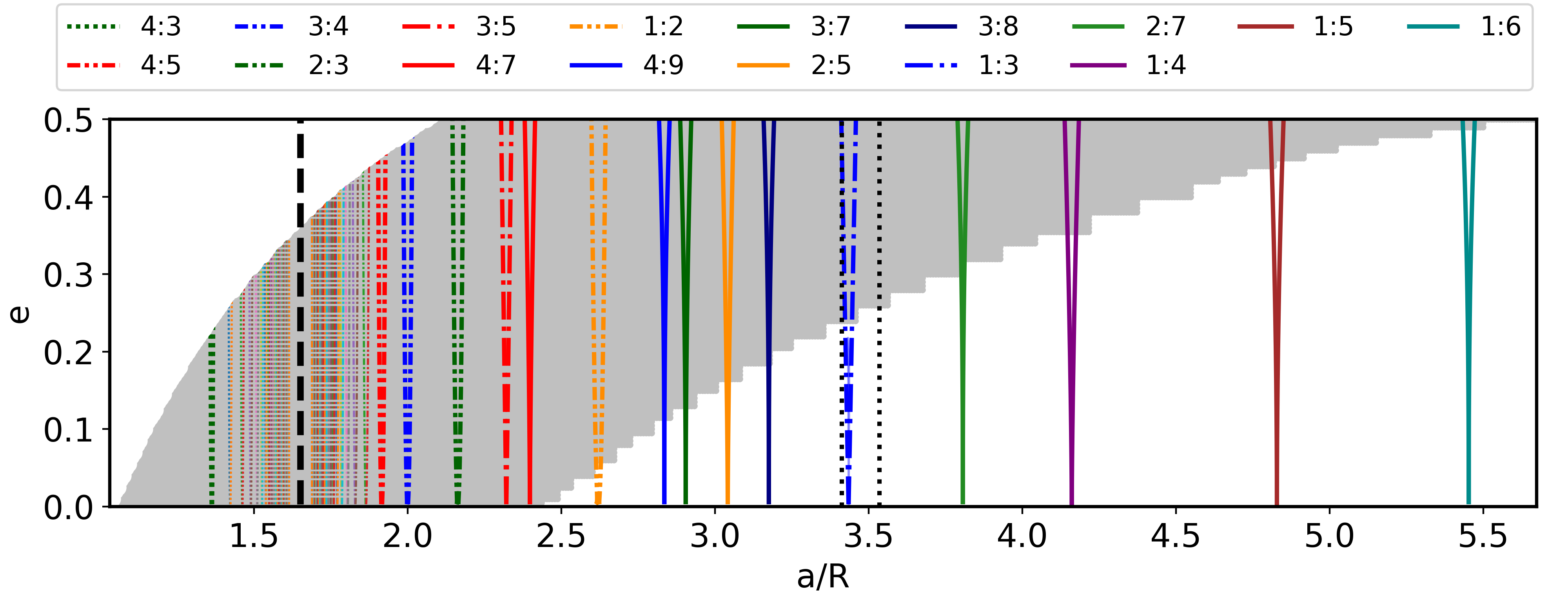}
\caption{Semi-major axis versus eccentricity for Chariklo system, where coloured lines place the sectoral resonances, and the grey area corresponds to the chaotic region. Coloured lines not referenced on the label, between $1.4-1.8$, correspond to first order resonances with $|m|>4$. A vertical dashed line at $a/R\approx1.7$ gives the corotation radius, while vertical dotted lines give the central location of the rings.\label{fig:chgeneral}}
\end{figure*}

We performed numerical simulations adopting the spherical Chariklo given by \cite{Leiva2017}, $\lambda=0.471$,  with a mass anomaly of $\mu=7\times10^{-6}=(5~{\rm km}/(2\times129~{\rm km}))^3$. Figure~\ref{fig:chgeneral} shows the width of the resonances and the location of the chaotic region. The vertical dashed line gives the corotation radius and the central location of the rings by the vertical dotted lines. We obtained a threshold semi-major axis of $a_t/R=2.5$ in the numerical simulation. This result is in good agreement with our adjusted equation (eq.~\ref{trs}) which returns $a_t/R=2.6$. The 1:2 resonance is the only first-order resonance beyond the chaotic region. The region beyond the chaotic one is essentially stable, hosting the rings and possibly moons, depending on their eccentricity.

\begin{figure}
\centering
\subfigure[]{\includegraphics[width=0.90\columnwidth]{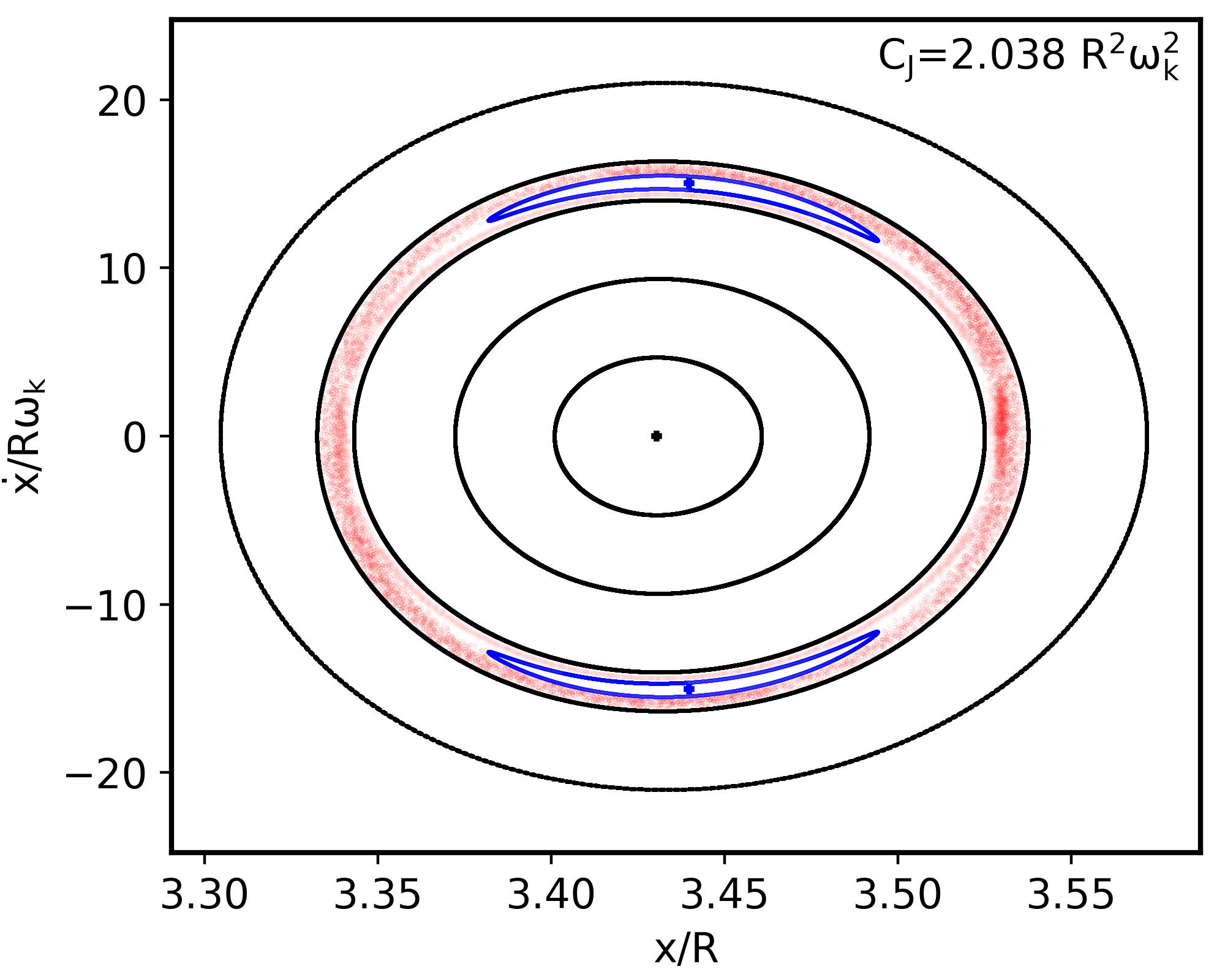}} 
\subfigure[]{\includegraphics[width=0.88\columnwidth]{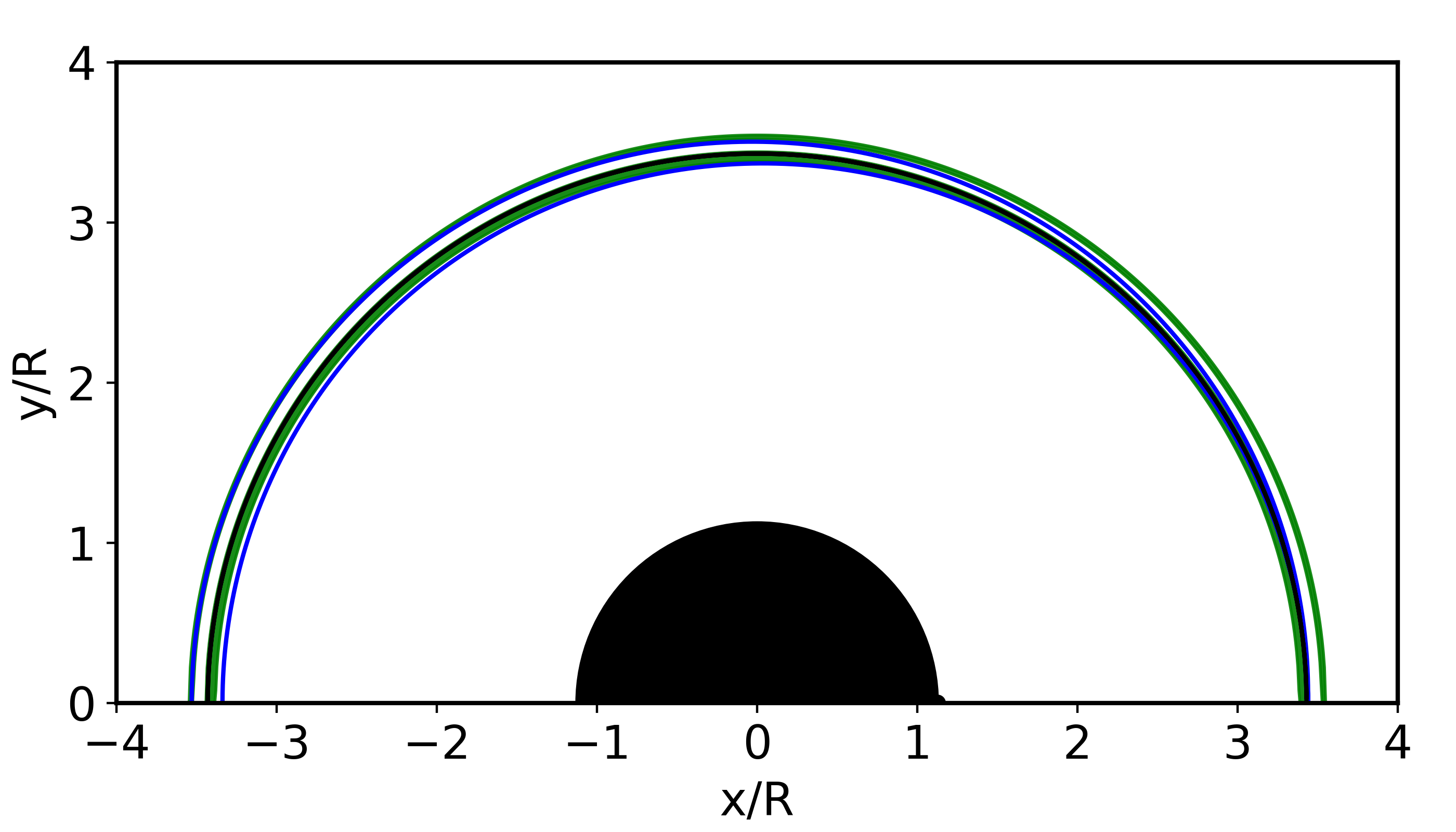}} 
\subfigure[]{\includegraphics[width=0.88\columnwidth]{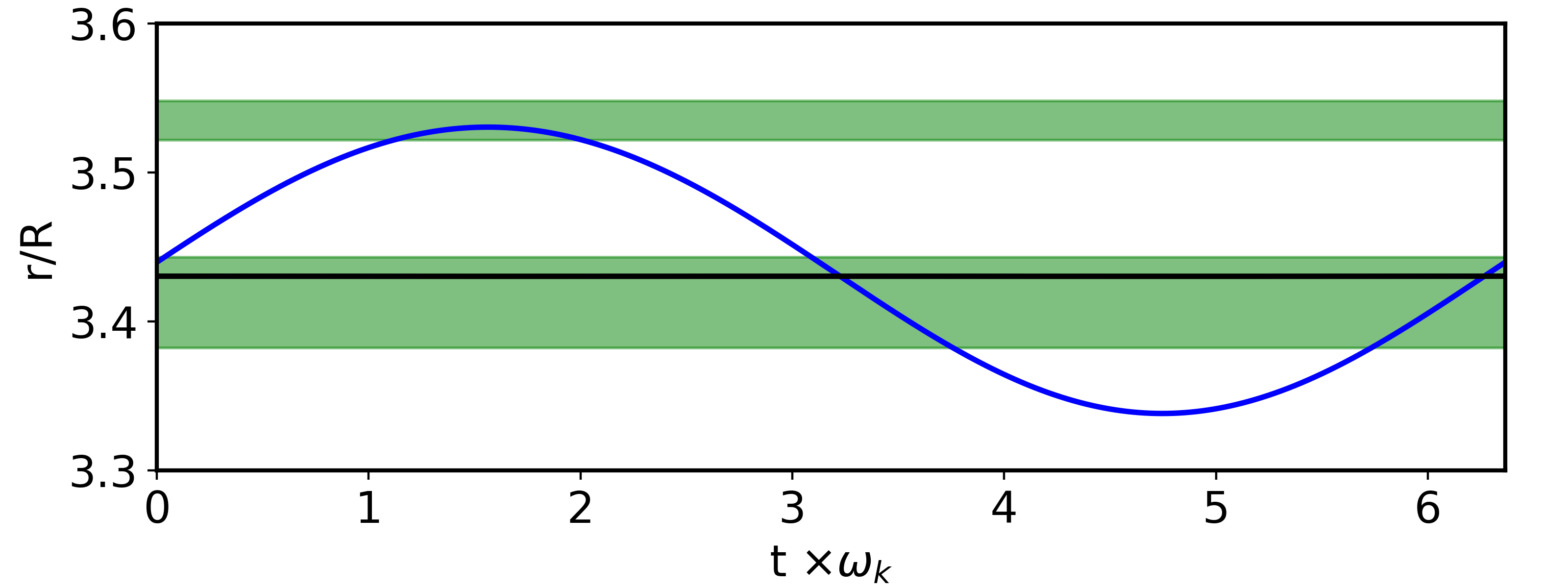}} 
\caption{a) Poincar\'e surface of section of  Chariklo system for $C_J=2.038~{\rm R^2\omega_k^2}$. We show different orbits by different colours: the non-resonance orbits are black, the 1:3 resonant orbits are blue, and the chaotic ones are red. b) motion in the rotating frame for $y>0$ and c) radial variation of periodic orbits shown in panel a). The orbits of the first and second kind are given by black and blue lines, respectively, and the green regions correspond to the positions of  Chariklo rings. \label{fig:ch13}}
\end{figure}

\cite{Leiva2017} proposes that the inner ring is associated with the 1:3 spin-orbit resonance. Therefore, we studied this resonance in detail, as it is close to both rings. Figure~\ref{fig:ch13}a shows Poincar\'e surface of section for the largest Jacobi constant obtained by us for the 1:3 resonance ($C_J=2.038~{\rm R^2\omega_k^2}$). For this value of $C_J$, the resonance has not reached the critical eccentricity, and we obtain only a single symmetric periodic orbit. A narrow, chaotic region surrounds the islands of resonance, but the whole set is surrounded by a stable region associated with periodic/quasi-periodic orbits of first kind.

Figure~\ref{fig:ch13}b shows the motion in the rotating frame of the periodic orbits given in Figure~\ref{fig:ch13}a, in which the colour of the orbits matches those given in the top panel, and the green regions correspond to the location of the rings. For clarity, we only show the portions of the orbits with $y>0$. We show the radial variation of the orbits in Figure~\ref{fig:ch13}c.

The 1:3 resonance orbit has one self-crossing at $y=0$ and a period of almost $6.4\omega_k^{-1}$. In contrast, the trajectory of the first kind follows the Chariklo shape, with a period of almost $3.2\omega_k^{-1}$. As one can see in the figure, both orbits are initially in the inner ring -- near its outer edge.  However, only the first kind of periodic orbit remains within the ring throughout the simulation, while the resonant orbit crosses the ring edges and reaches the outer ring.

The difference in radial variation is due to the different nature of the orbits. Periodic orbits of first kind correspond to nearly circular orbits, while those of second kind are intrinsically eccentric, explaining why the latter has a significantly larger radial variation. Here, when we refer to eccentricity, we are referring to osculating elements defined in the context of the classical 2-body problem. We refer the reader to the work of \cite{Ribeiro2021} for a detailed discussion regarding the orbital elements in the context of NSSBs.
\begin{figure}
\centering
\includegraphics[width=\columnwidth]{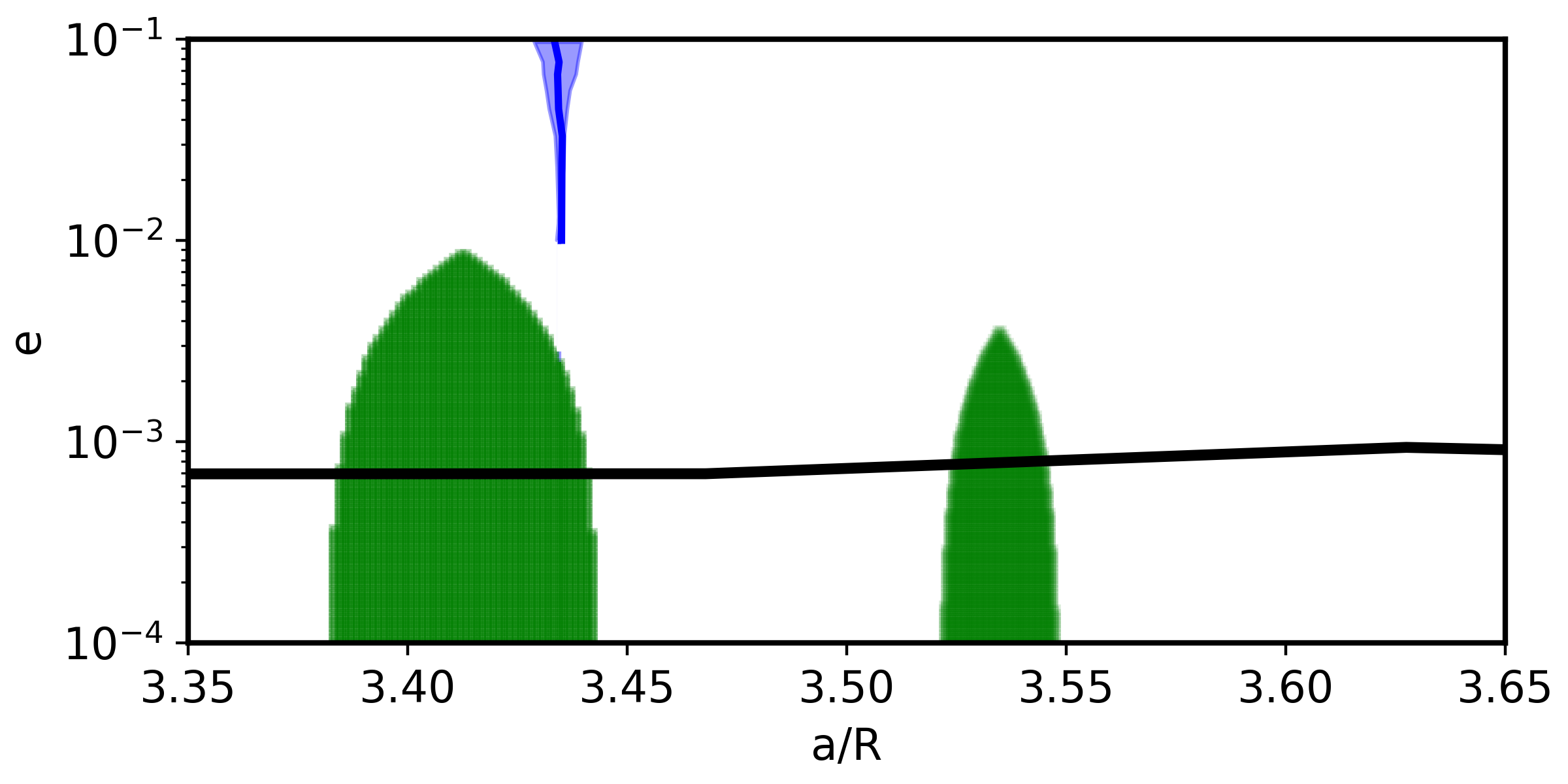}
\caption{Diagram of the semi-major axis versus eccentricity. The green regions show the range of values that corresponds to the location of the rings. The blue line shows the orbital elements obtained for the central orbit of the 1:3 resonance, and the blue filled regions give the boundaries of the resonance. The black line gives the periodic orbits of first kind. \label{fig:chzoom}}
\end{figure}

The results discussed in the last paragraph lead us to question whether the inner ring is associated with the 1:3 resonance. To verify this, we present in Figure~\ref{fig:chzoom} a diagram with the semi-major axis versus eccentricity for a range of values corresponding to the rings (in green). We also show the orbital elements obtained for the periodic orbits of first and second kinds (black and blue lines, respectively). The region filled in blue shows the 1:3 resonance boundaries obtained in the Poincar\'e surface of sections.

The largest possible eccentricity for a particle to remain within the boundaries of the inner ring is $e=9\times10^{-3}$, which is smaller than the smallest eccentricity obtained for the resonant orbits ($e=10^{-2}$). In addition, the resonant periodic orbits and the ring are displaced, indicating that the ring is not confined by such resonance.

Meanwhile, periodic orbits of first kind cover a broad region and encompass both rings. The entire region shown in Figure~\ref{fig:chzoom}, which is not associated with the 1:3 resonance (blue region), is composed of periodic/quasi-periodic orbits of first kind, including the ring region. Therefore, we conclude that Chariklo rings are associated with first kind orbits and not with the 1:3 resonance, as proposed by \cite{Leiva2017}. Similar results were obtained by \cite{Winter2019} for the Haumea ring.

\section{Conclusions and discussion} \label{secdiscu}
In this paper, we have attempted to perform a general analysis of the dynamics of particles around a spherical body with a mass anomaly. For this, we used well-known techniques of the 3-body problem study, varying the parameters of the central object. We can summarise our overall results as follows:
\begin{itemize}
    \item The pendulum model with the necessary adaptations and the Poincar\'e surface of section proved to apply to the mass anomaly problem. We verified a strong agreement between the results by comparing both techniques.  
    \item There is a chaotic region near the central object where particles collide or are ejected due to chaotic diffusion caused by successive close-encounters with the mass anomaly. \cite{Mysen2006,Mysen2007} and \cite{Lages2017} also obtains chaotic regions near the central object for elongated bodies and contact binaries, respectively. 
    \item For the set of parameters analysed by us, the chaotic region extends beyond the corotation radius. This fact indicates a lack of stable internal sectoral and corotation resonances in the mass anomaly system.
    \item Resonances location is mainly affected by the mass of the spherical portion and the spin period. In contrast, the masses of the spherical and anomalous portions of the body and spin period are responsible for determining the width of the resonances.
    \item Beyond the chaotic region, there is a region where the motion of the particles is dynamically stable. In such a region, there is chaotic behaviour only in a narrow region in the separatrices of the resonances.
    \item The behaviour of the particles in the external sectoral resonances is similar to those obtained for the mean motion resonances in the RP3BP \citep{Winter1997a,Winter1997b}. Similar to RP3BP, we verify the existence of asymmetric periodic orbits associated with $1:1+p$ resonances.
\end{itemize}

Although objects with the shape assumed in this work are unknown so far, the completely irregular shapes known for some asteroids lead us to speculate that such a class of object might exist. We emphasize that bodies with mass anomaly are perfectly reasonable outputs from a collision of a satellite that spirals towards the central body due to tidal dissipation or a collision between two objects at low velocity, with partial accretion \citep{Leinhardt2011}. 

\cite{Sicardy2019} discuss the possibility of Chariklo having a spherical shape with topographic feature with ${\rm \mu\sim10^{-5}}$, which places the Centaur as a first candidate to integrate the class of mass anomaly objects. We studied the dynamics around a Chariklo with mass anomaly and found that 1:3 resonant particles present radial variations too large for the radial extension of the inner ring. On the other hand, particles in periodic/quasi-periodic first kind orbits show radial motions that match the extension of the two rings of Chariklo. Consequently, the ring must be associated with these orbits and not with orbits of second kind as proposed by \cite{Leiva2017}. With the constant increase in data on small heliocentric bodies, we believe that objects with shapes similar to bodies with mass anomalies may soon be detected.

It is essential to point out that in the current study, we limited ourselves to analyse the dynamics of an isolated particle around a NSSB, disregarding the effects associated with the ring particles, such as collisions between them, local viscous, and self-gravity effects. We also disregard external effects that modulate the dynamics of small particles, such as solar radiation pressure and Poynting-Robertson drag.  Nevertheless, the location and width of resonances and the chaotic region are general results and should remain almost unchanged in the presence of other effects. Therefore, our work presents some tools and first general results for studies on dynamics of mass anomaly systems. 

\section*{Acknowledgements}
This study was financed in part by the Coordena\c{c}\~ao de Aperfei\c{c}oamento de Pessoal de N\' ivel Superior - Brasil (CAPES) - Finance Code 001, Funda\c{c}\~ao de Amparo \`a Pesquisa do Estado de S\~ao Paulo (FAPESP) - Proc.~2016/24561-0 and Proc.~2018/23568-6, Conselho Nacional de Desenvolvimento Cient\' ifico e Tecnol\' ogico (CNPq) - Proc.~305210/2018-1 and Proc.~313043/2020-5. Finally, we thank the anonymous reviewer for the comments that significantly improved our work.
%%%%%%%%%%%%%%%%%%%%%%%%%%%%%%%%%%%%%%%%%%%%%%%%%%
\section*{Data Availability}
The data underlying this article will be shared on reasonable request to the corresponding author.



\bibliographystyle{mnras}
\bibliography{madeira} % if your bibtex file is called example.bib

%\begin{thebibliography}{99}
%\end{thebibliography}

%%%%%%%%%%%%%%%%%%%%%%%%%%%%%%%%%%%%%%%%%%%%%%%%%%

%%%%%%%%%%%%%%%%% APPENDICES %%%%%%%%%%%%%%%%%%%%%

\appendix

\section{Poincar\'e surfaces of section for the reference object} \label{pssrc}
\begin{figure}
\centering
\subfigure[]{\includegraphics[width=0.8\columnwidth]{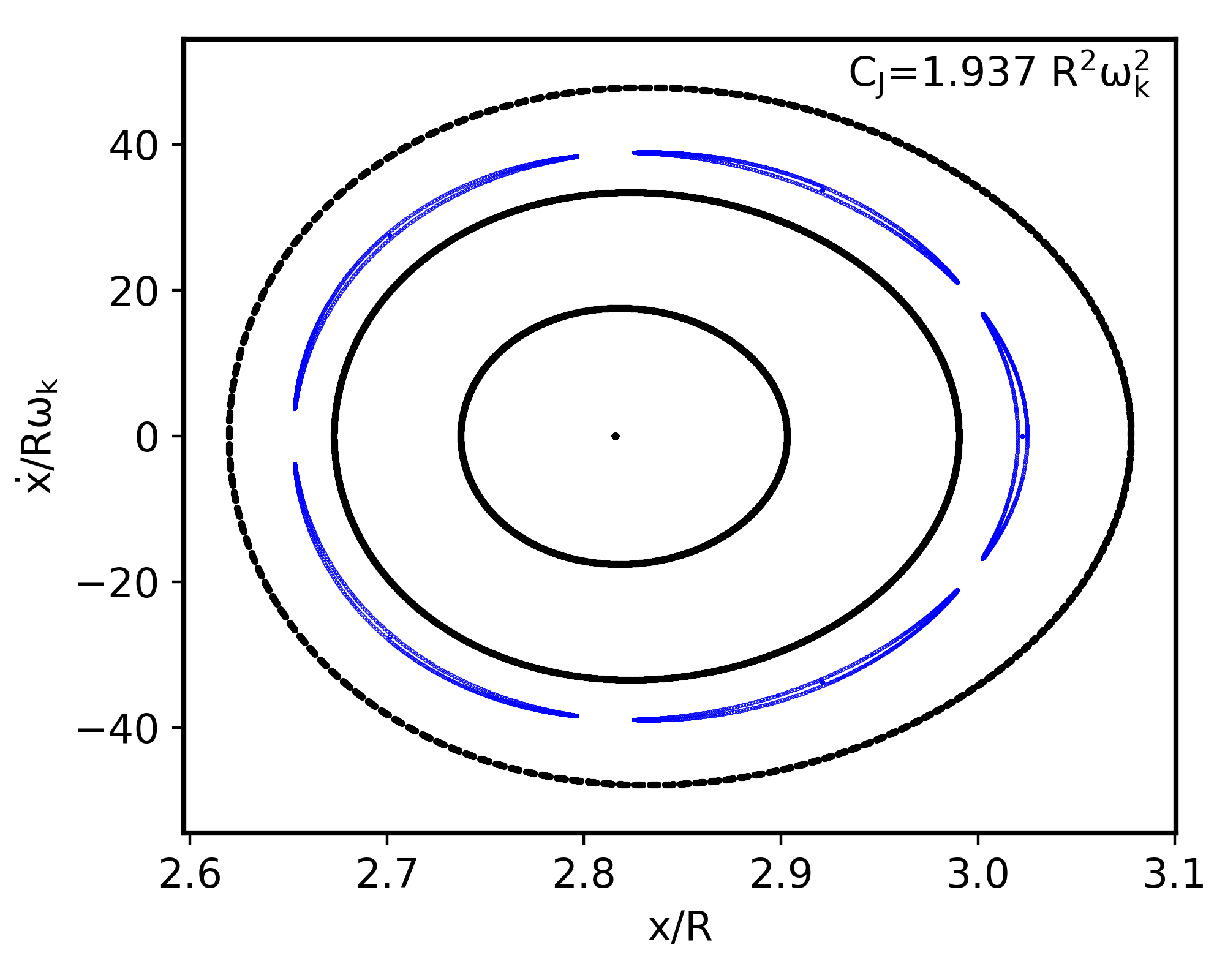}} 
\centering
\subfigure[]{\includegraphics[width=0.8\columnwidth]{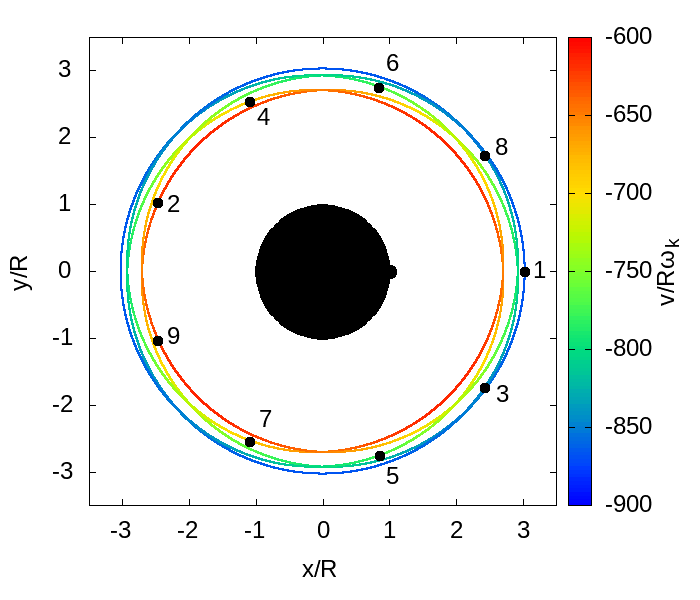}}
\caption{a) Poincar\'e surface of section for $C_J=1.937~{\rm R^2\omega^2}$ in which the black curves are the periodic and quasi-periodic orbits of first kind and the blue curves are orbits associated with the 4:9 resonance. b) Central orbit of the 4:9 resonance for $C_J=1.937~{\rm R^2\omega^2}2$ in the rotating frame. The temporal evolution of the orbit is given by numbers and dots equally spaced in time, while the colour-coding gives the velocity in the rotating frame. \label{fig:49}}
\end{figure}

\begin{figure}
\centering
\subfigure[]{\includegraphics[width=0.8\columnwidth]{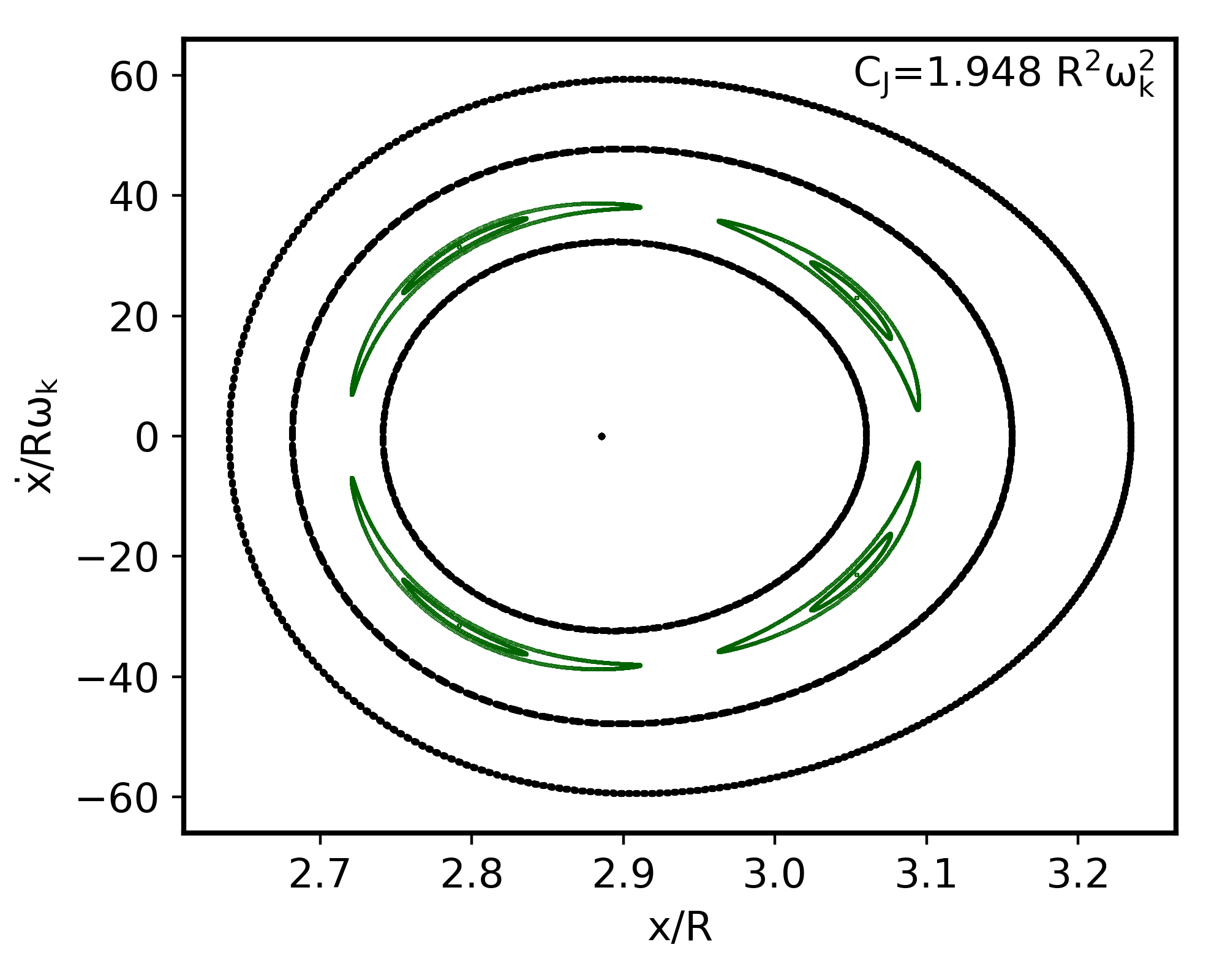}} 
\subfigure[]{\includegraphics[width=0.8\columnwidth]{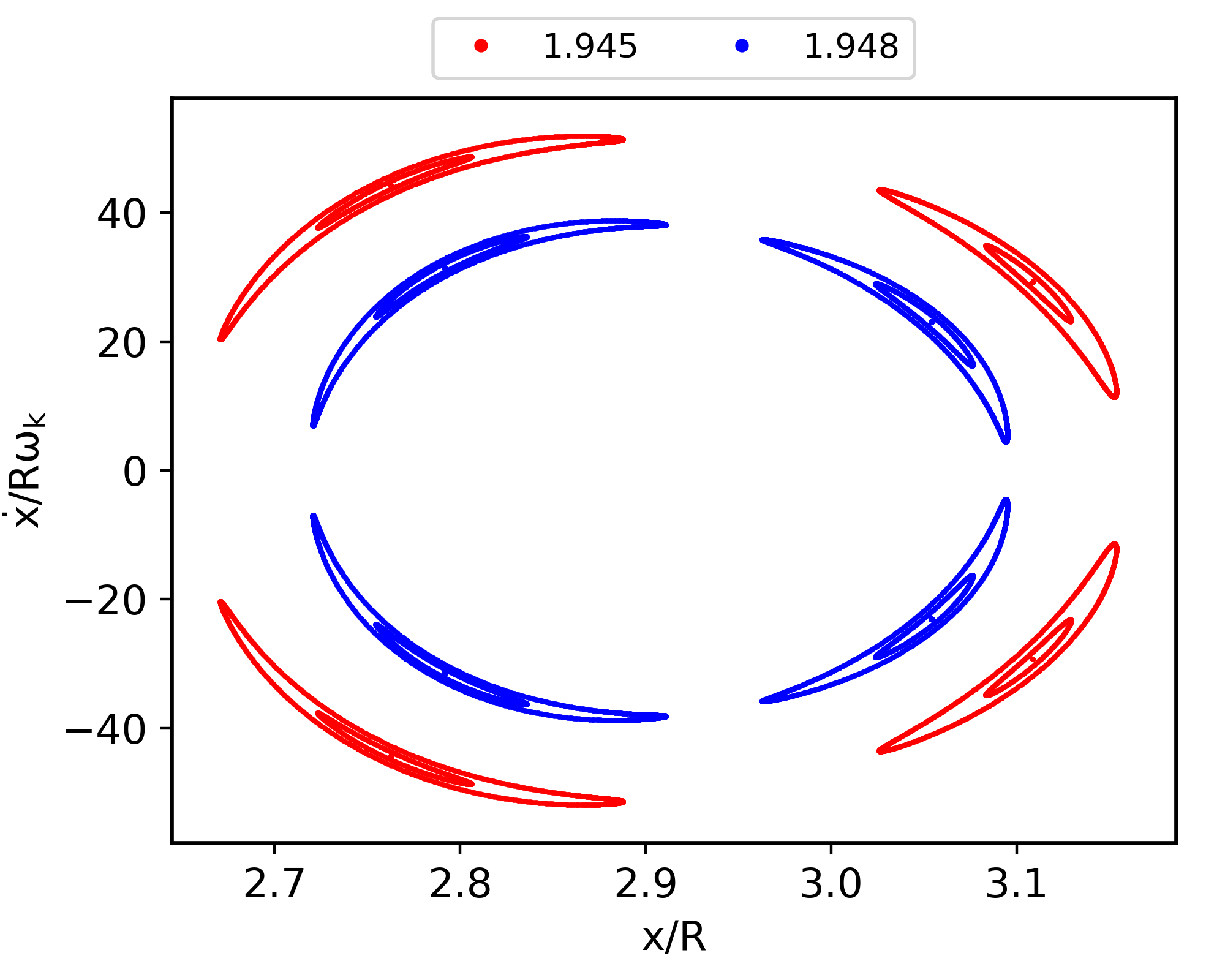}}
\subfigure[]{\includegraphics[width=0.8\columnwidth]{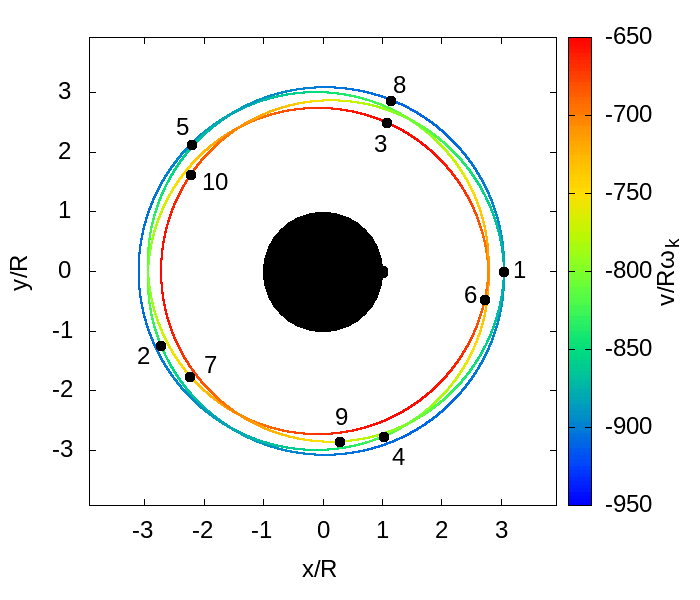}}
\caption{a) Poincar\'e surface of section for $C_J=1.948~{\rm R^2\omega^2}$ in which the periodic/quasi-periodic orbits of first kind are in black and the 3:7 resonance orbits are in green. b) Resonance islands for different values of $C_J$ given in the upper panel. c) Central orbit of the 3:7 resonance for $C_J=1.948~{\rm R^2\omega^2}$. The numbers and colours on the panel provide time evolution and the velocity in the rotating frame, respectively. \label{fig:37}}
\end{figure}

\begin{figure}
\centering
\subfigure[]{\includegraphics[width=0.8\columnwidth]{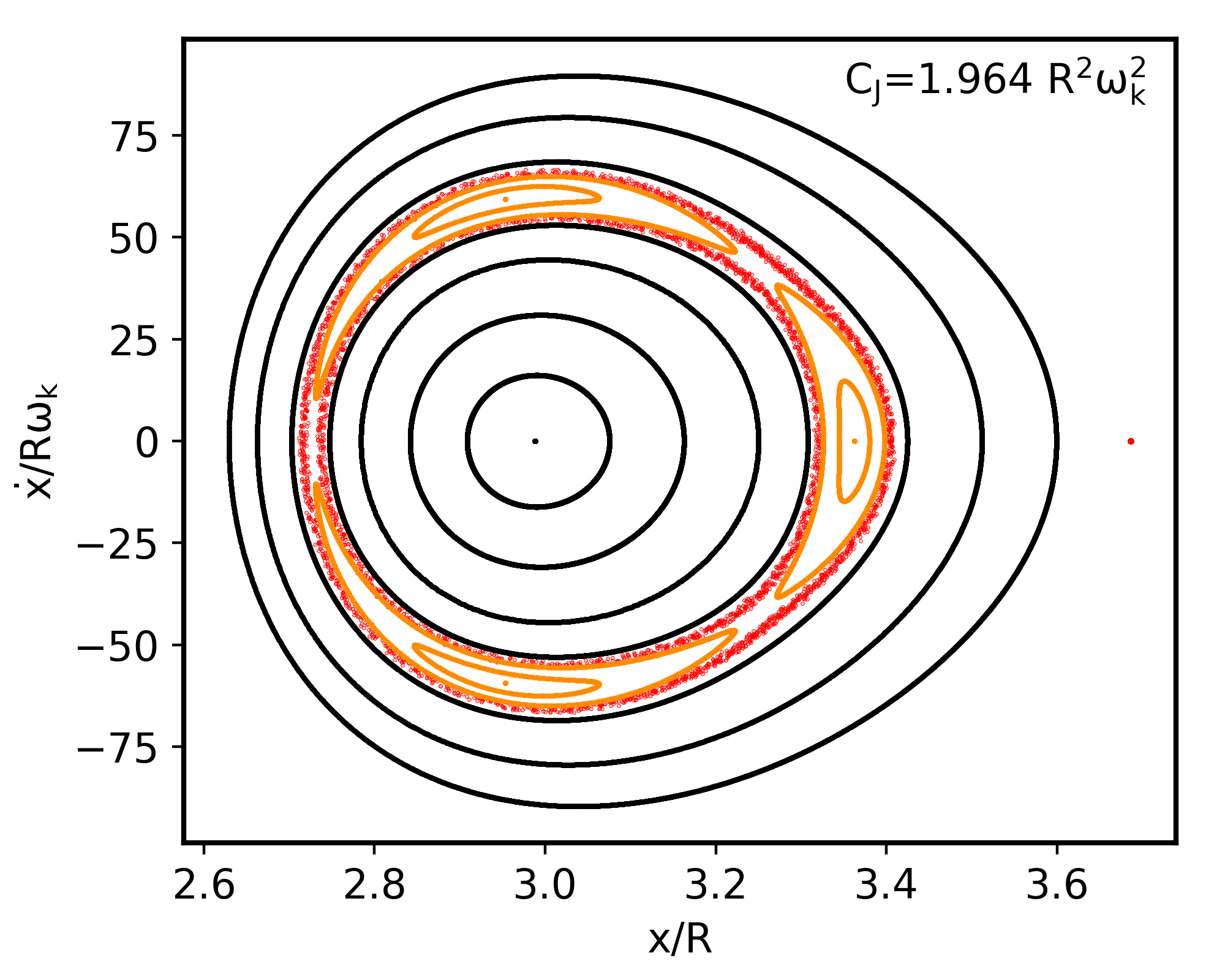}} 
\subfigure[]{\includegraphics[width=0.8\columnwidth]{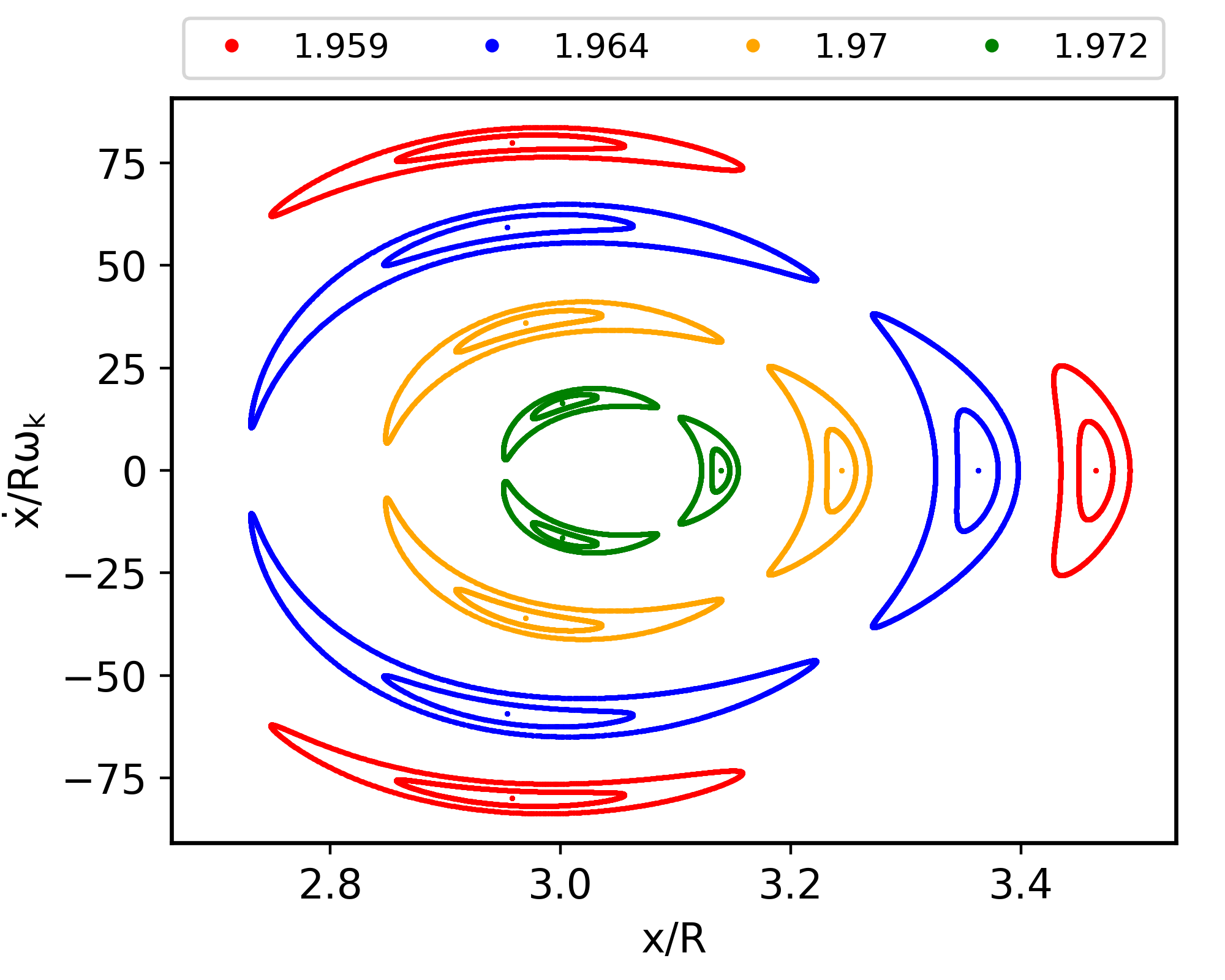}}
\centering
\subfigure[]{\includegraphics[width=0.8\columnwidth]{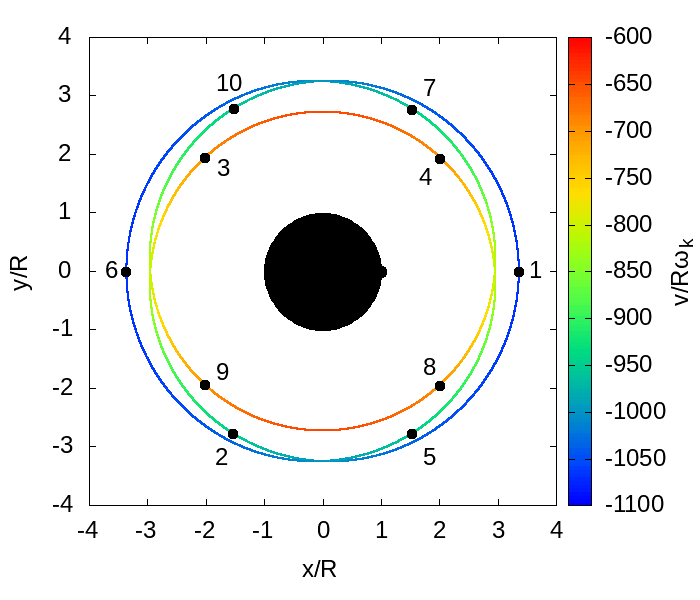}}
\caption{a) Poincar\'e surface of section for $C_J=1.964~{\rm R^2\omega^2}$. The black curves are  periodic and quasi-periodic orbits of first kind, and the orange curves are orbits associated with the 2:5 resonance. Red dots correspond to chaotic orbits. b) Evolution of the 2:5 resonance islands, where the colours of the dots correspond to the values of $C_J$ given on the label of the figure. c) Central orbit of the 2:5 resonance for $C_J=1.964~{\rm R^2\omega^2}$ in the rotating frame. The temporal evolution of the orbit is given by numbers and dots equally spaced in time, while the colour coding gives the velocity in the rotating frame. \label{fig:252}}
\end{figure}

\begin{figure}
\centering
\subfigure[]{\includegraphics[width=0.8\columnwidth]{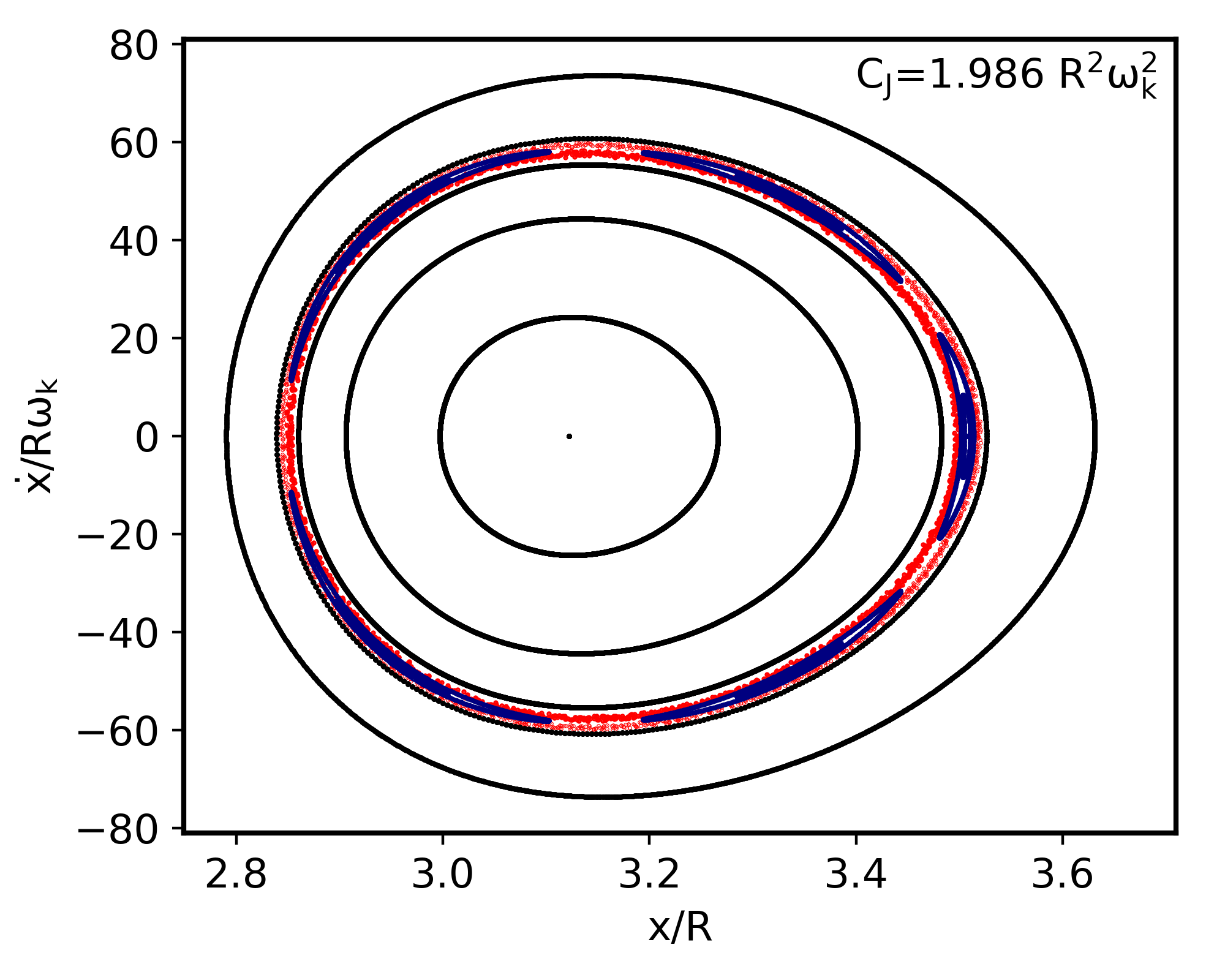}} 
\subfigure[]{\includegraphics[width=0.8\columnwidth]{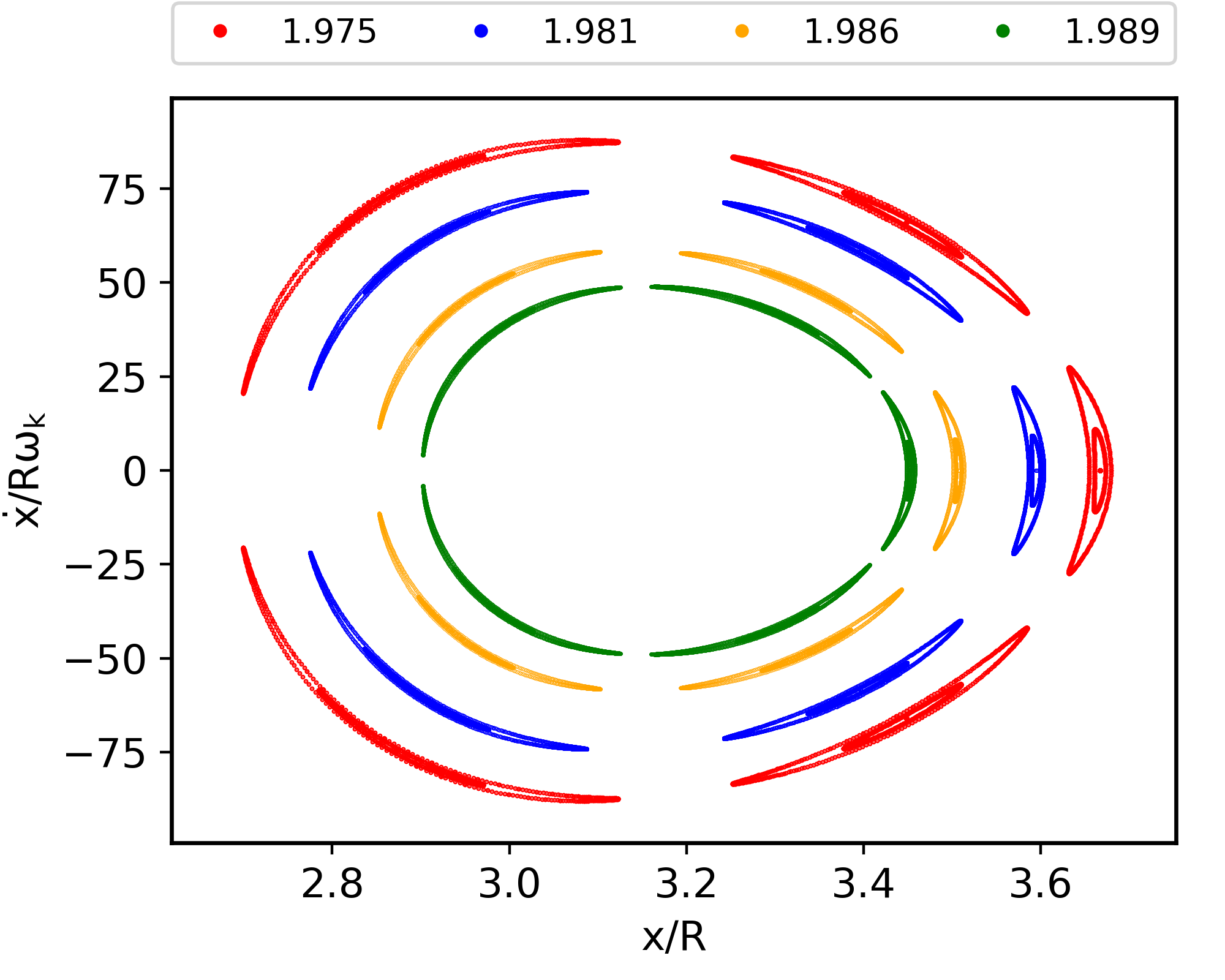}}
\centering
\subfigure[]{\includegraphics[width=0.8\columnwidth]{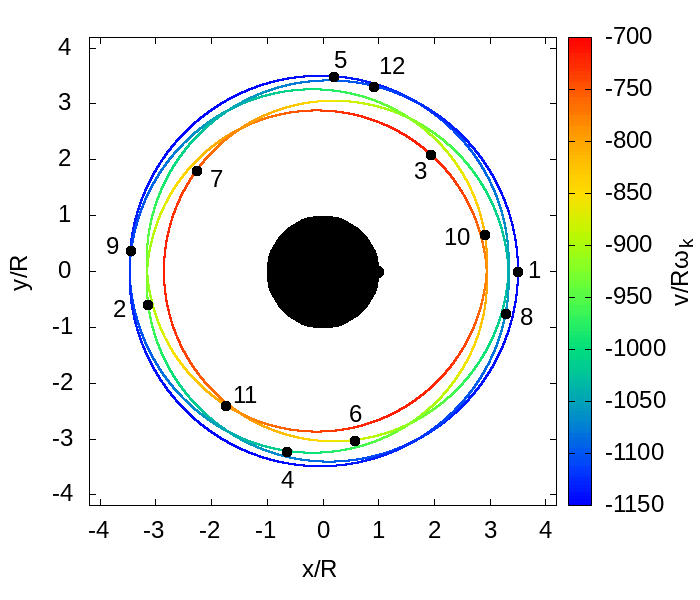}}
\caption{a) Poincar\'e surface of section for $C_J=1.986~{\rm R^2\omega^2}$. Black curves are  periodic and quasi-periodic orbits of first kind, and  blue curves are orbits associated with the 3:8 resonance. Red dots correspond to chaotic orbits. b) Evolution of the 3:8 resonance islands, where the colours of the dots correspond to  values of $C_J$ given on the label of the figure. c) Central orbit of the 3:8 resonance for $C_J=1.986~{\rm R^2\omega^2}$ in the rotating frame. The temporal evolution of the orbit is given by numbers and dots equally spaced in time, while the colour coding gives the velocity in the rotating frame. \label{fig:38}}
\end{figure}

\begin{figure}
\centering
\subfigure[]{\includegraphics[width=0.8\columnwidth]{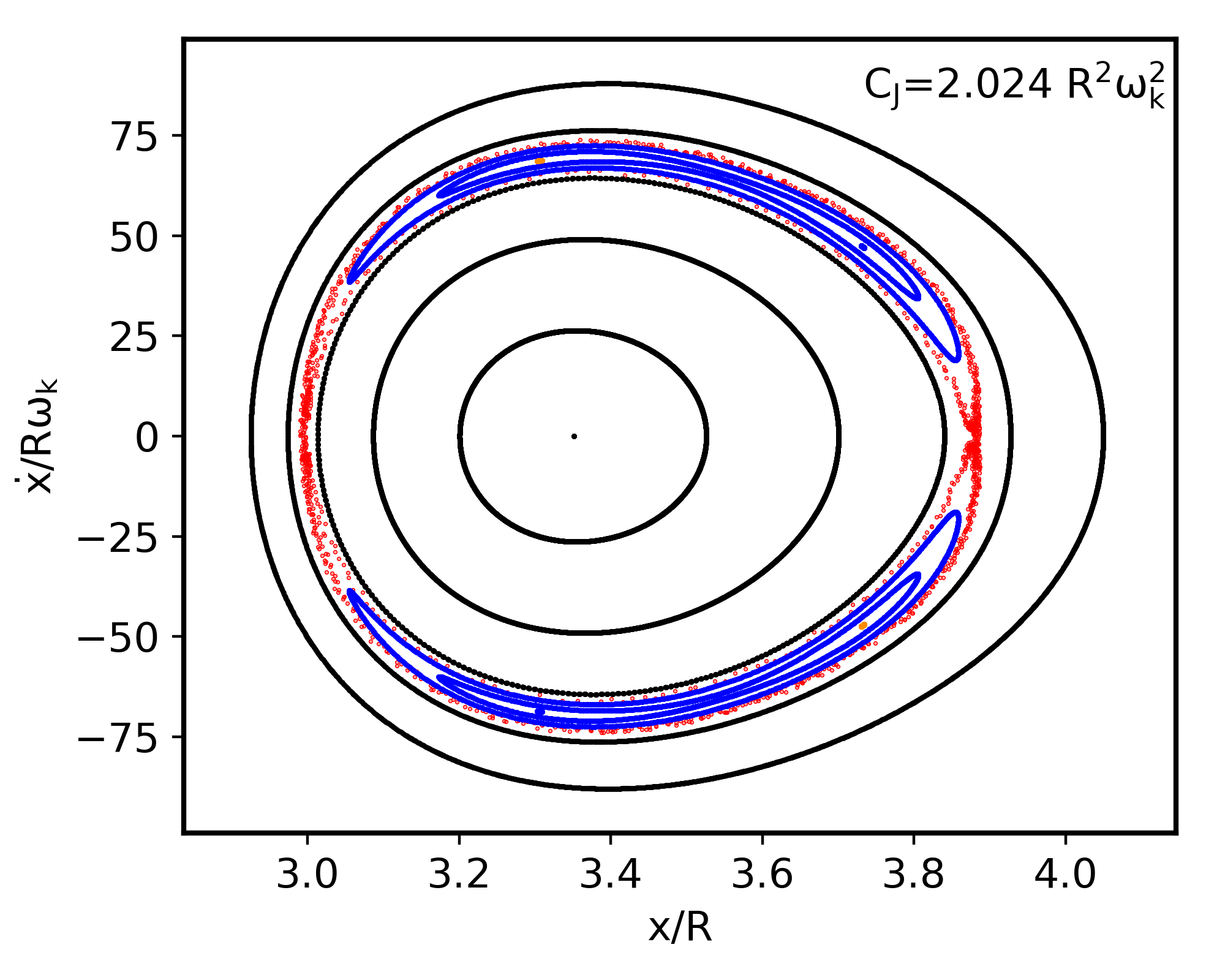}} 
\subfigure[]{\includegraphics[width=0.8\columnwidth]{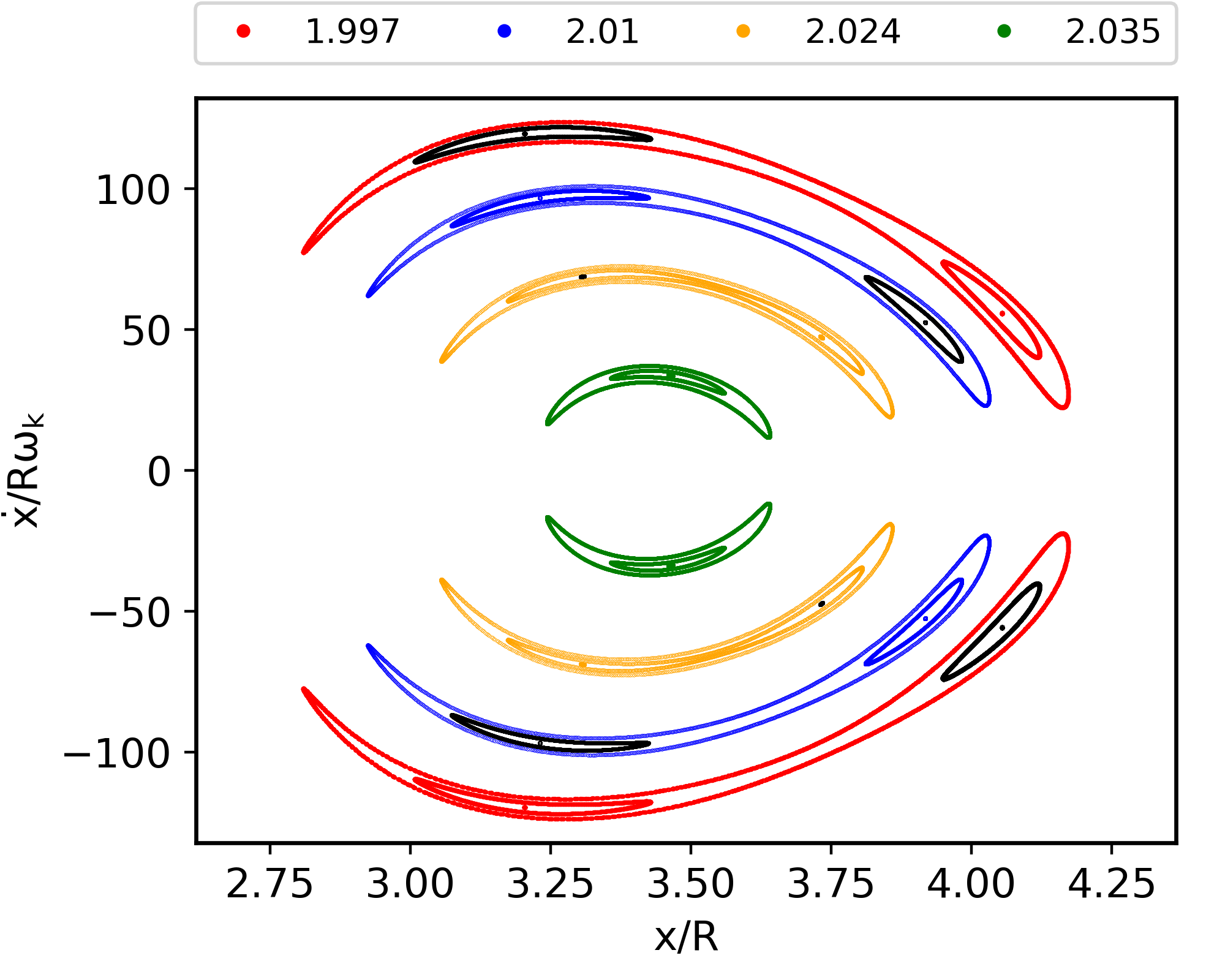}}
\centering
\subfigure[]{\includegraphics[width=0.8\columnwidth]{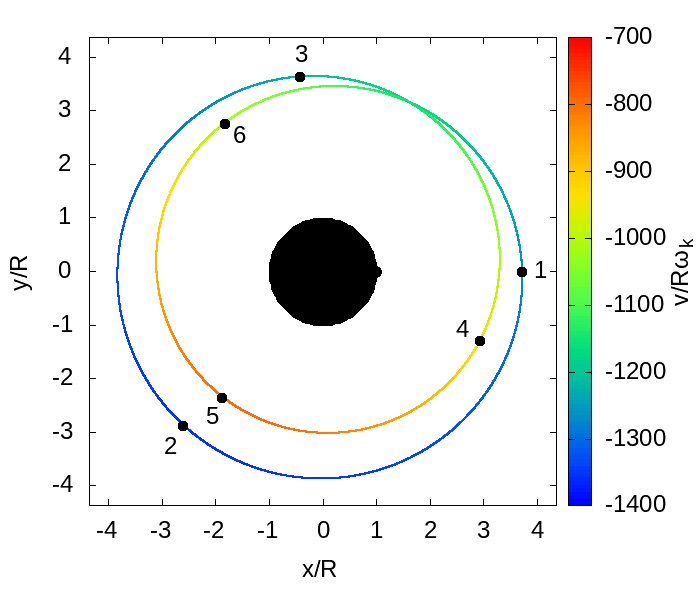}}
\caption{a) Poincar\'e surface of section for $C_J=2.024~{\rm R^2\omega^2}$. Periodic/quasi-periodic orbits of first kind are in black, the 1:3 resonance orbits are in blue and orange and chaotic ones in red. b) Resonance islands for different values of $C_J$. The label on the panel gives the colour of the largest island for each value of $C_J$. c) Central orbit in the rotating frame of one of the families associated with the 1:3 resonance (in blue in the top panel) for $C_J=2.024~{\rm R^2\omega^2}$. The numbers and colours on the panel provide time evolution and the velocity in the rotating frame, respectively.  \label{fig:13}}
\end{figure}

\begin{figure}
\centering
\subfigure[]{\includegraphics[width=0.8\columnwidth]{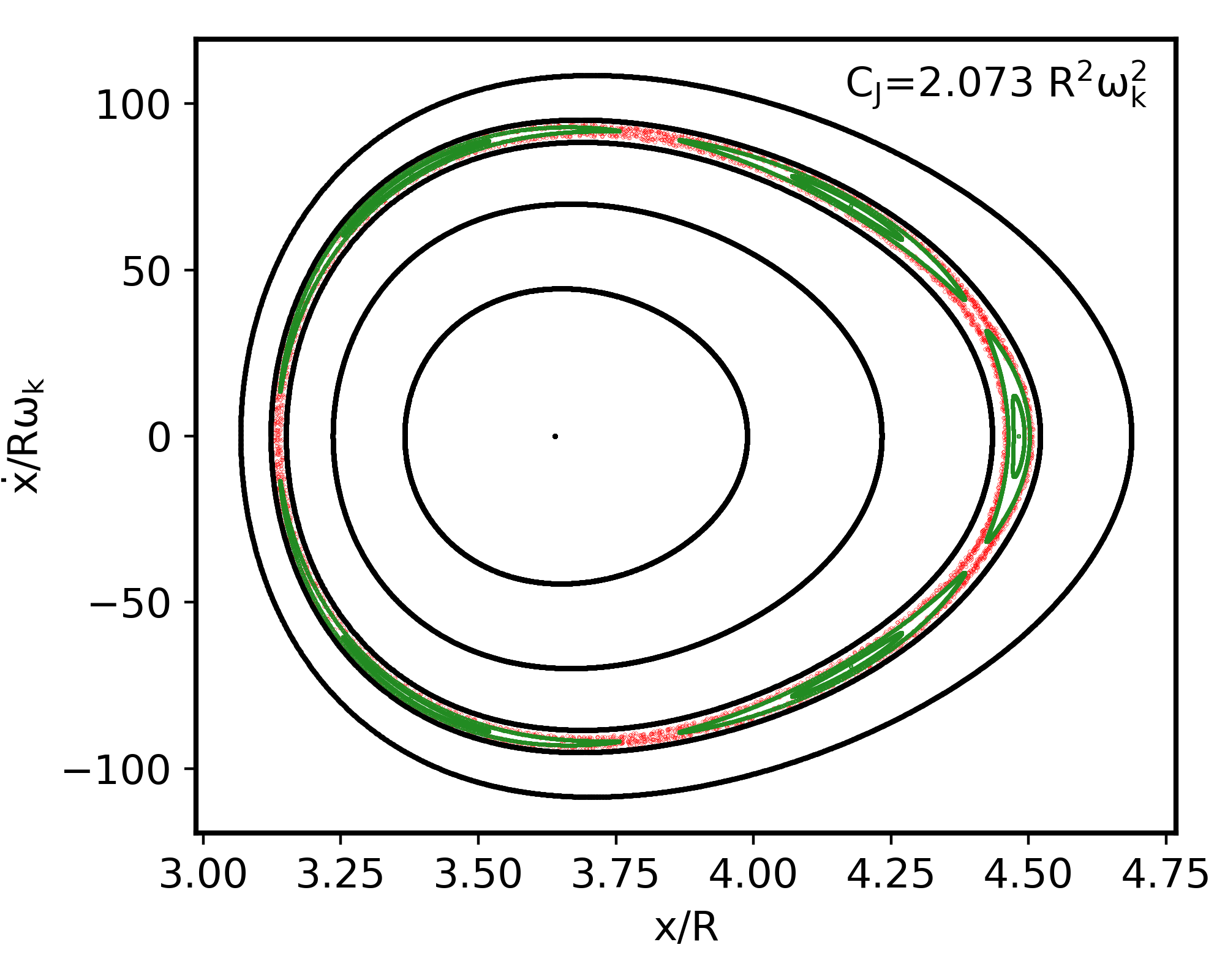}} 
\subfigure[]{\includegraphics[width=0.8\columnwidth]{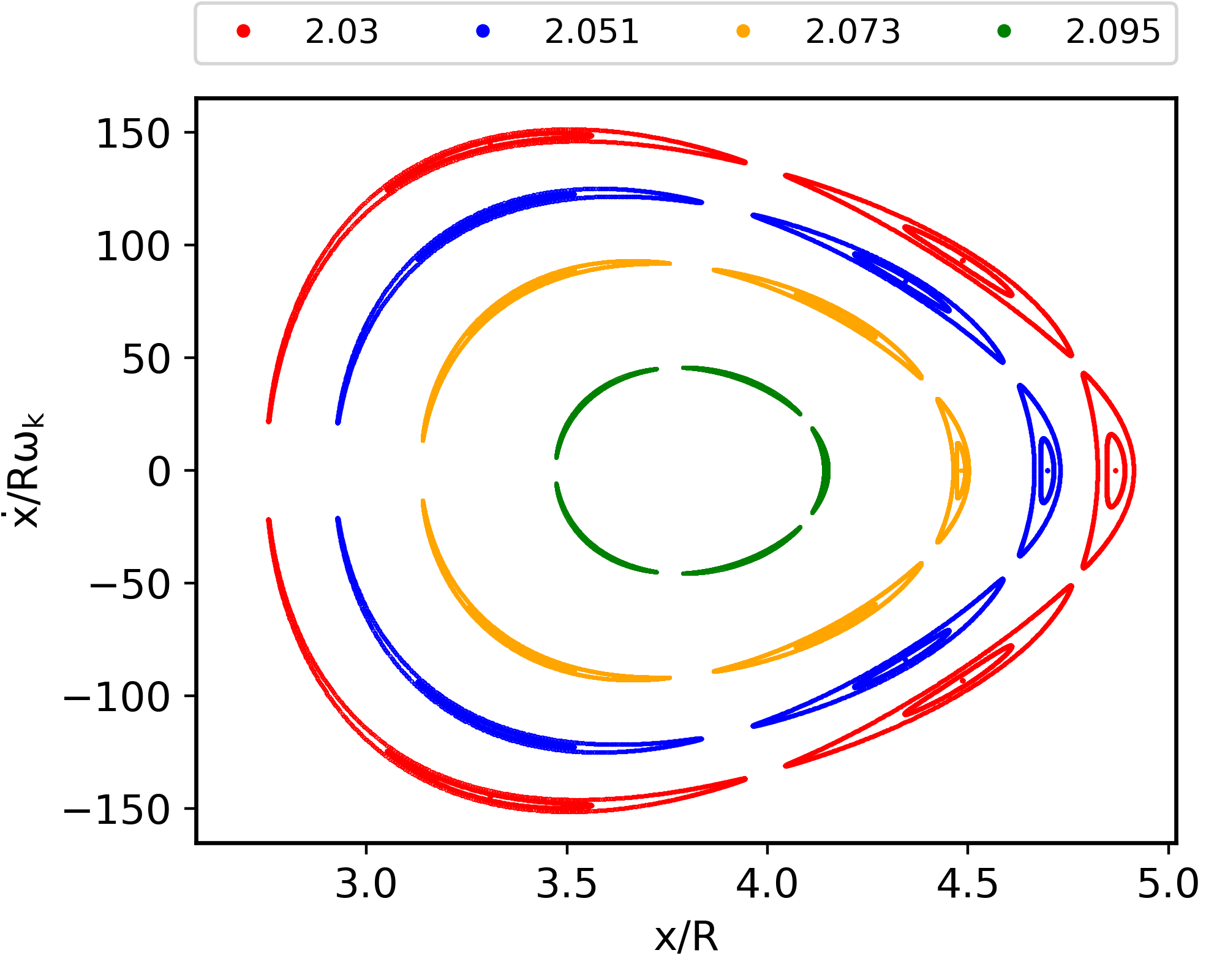}}
\centering
\subfigure[]{\includegraphics[width=0.8\columnwidth]{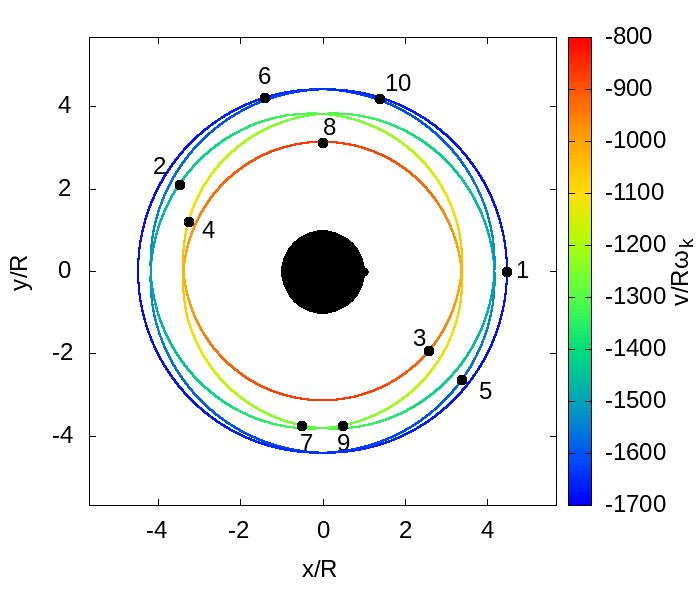}}
\caption{a) Poincar\'e surface of section for $C_J=2.073~{\rm R^2\omega_k^2}$. Black curves are  periodic and quasi-periodic orbits of first kind, and  green curves are orbits associated with the 2:7 resonance. Red dots correspond to chaotic orbits. b) Evolution of the 2:7 resonance islands, where the colours of the dots correspond to the values of $C_J$ given on the label of the figure. c) Central orbit of the 2:7 resonance for $C_J=2.073~{\rm R^2\omega_k^2}$ in the rotating frame. The temporal evolution of the orbit is given by numbers and dots equally spaced in time, while the colour coding gives the velocity in the rotating frame. \label{fig:27}}
\end{figure}

\begin{figure}
\centering
\subfigure[]{\includegraphics[width=0.8\columnwidth]{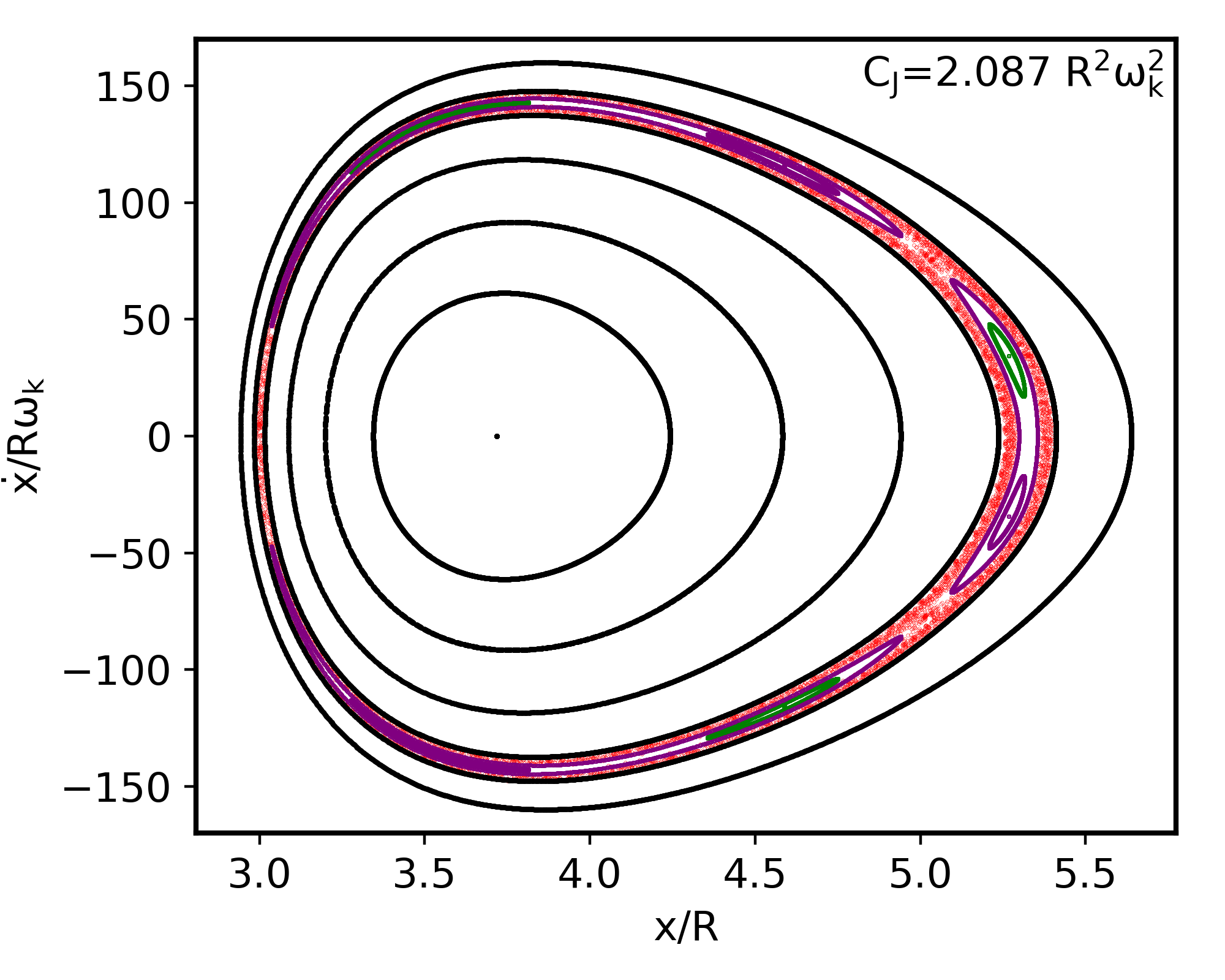}} 
\subfigure[]{\includegraphics[width=0.8\columnwidth]{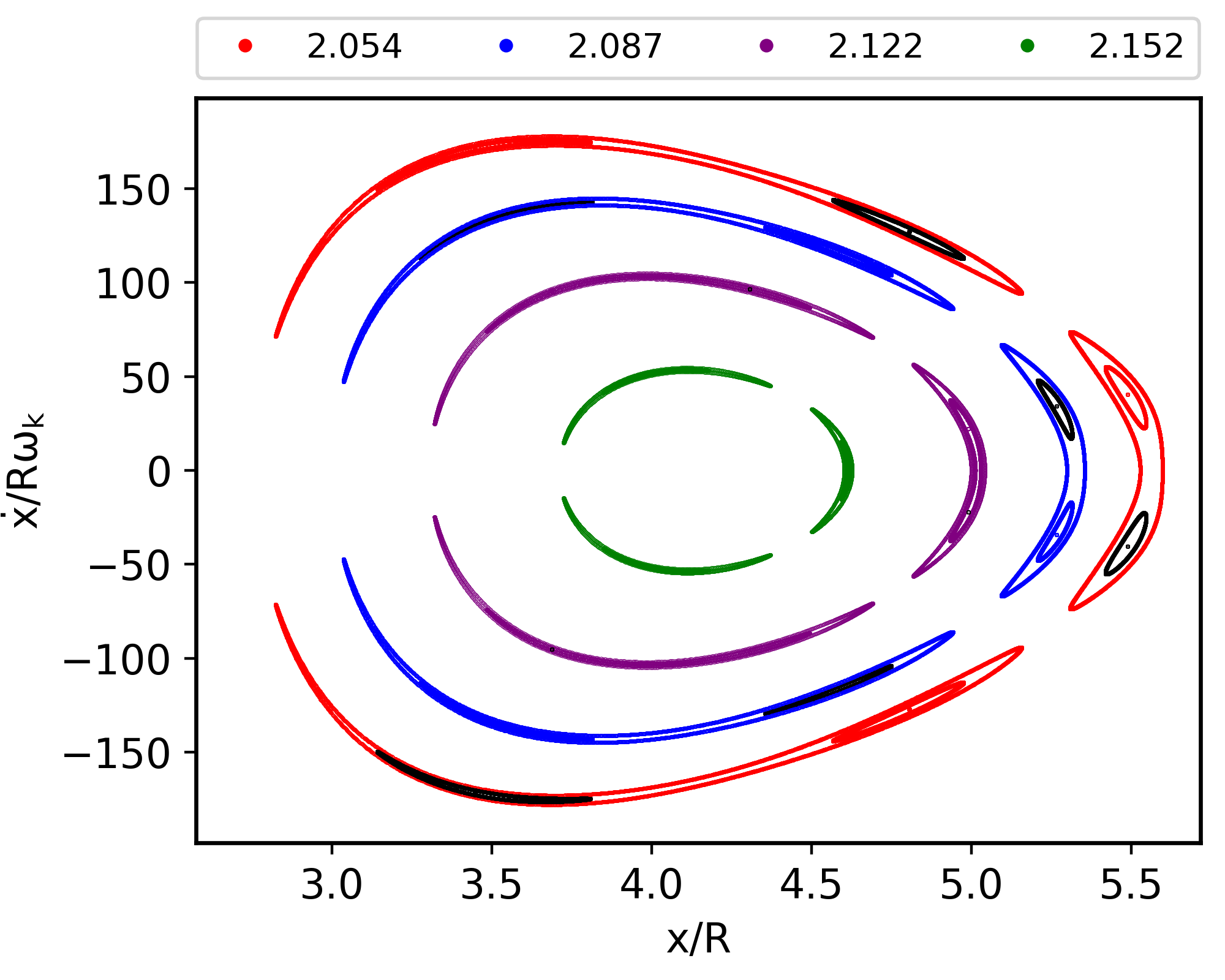}}
\centering
\subfigure[]{\includegraphics[width=0.8\columnwidth]{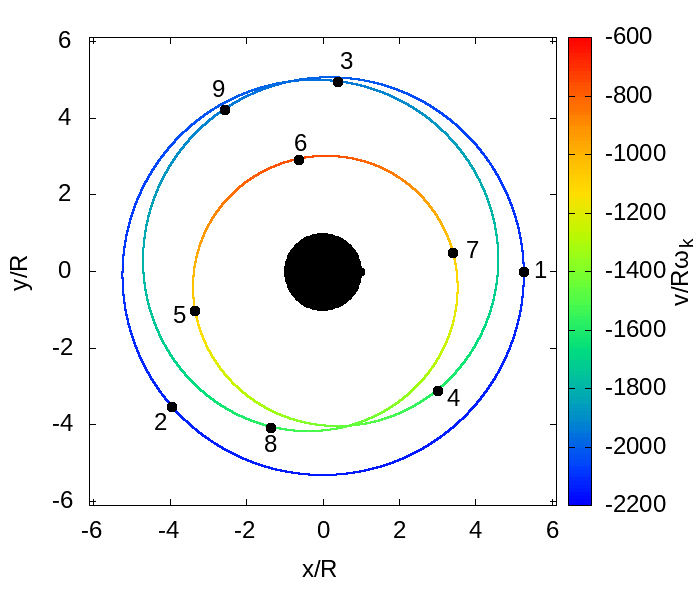}}
\caption{a) Poincar\'e surface of section for $C_J=2.087~{\rm R^2\omega_k^2}$. Periodic/quasi-periodic orbits of first kind are in black, the 1:4 resonance orbits are in purple and green and chaotic ones in red. b) Resonance islands for different values of $C_J$. The label on the panel gives the colour of the largest island for each value of $C_J$. c) Central orbit in the rotating frame of one of the families associated with the 1:4 resonance (in green in the top panel) for $C_J=2.087~{\rm R^2\omega_k^2}$. The numbers and colours on the panel provide time evolution and  velocity in the rotating frame, respectively. \label{fig:142}}
\end{figure}

\begin{figure}
\centering
\subfigure[]{\includegraphics[width=0.8\columnwidth]{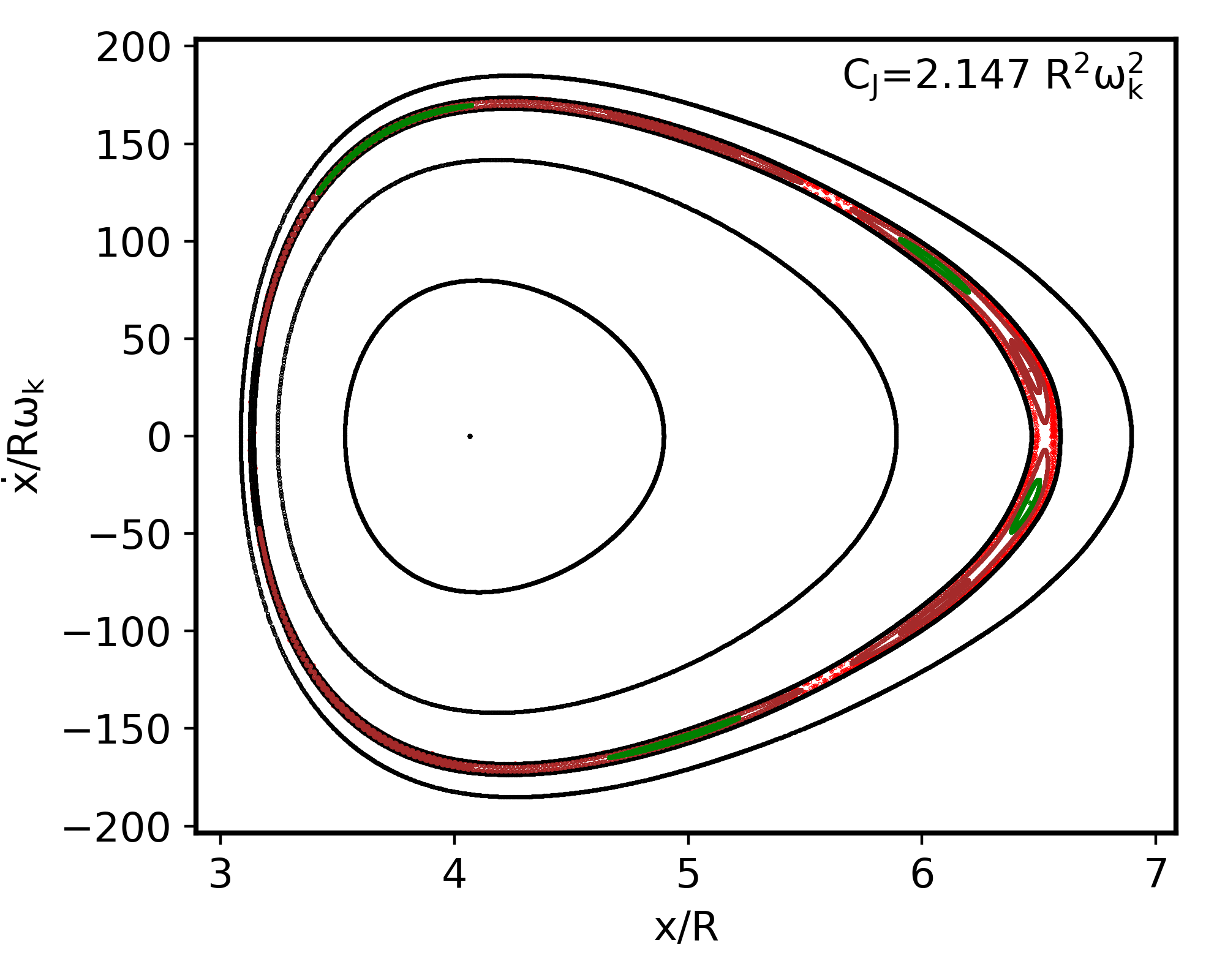}} 
\subfigure[]{\includegraphics[width=0.8\columnwidth]{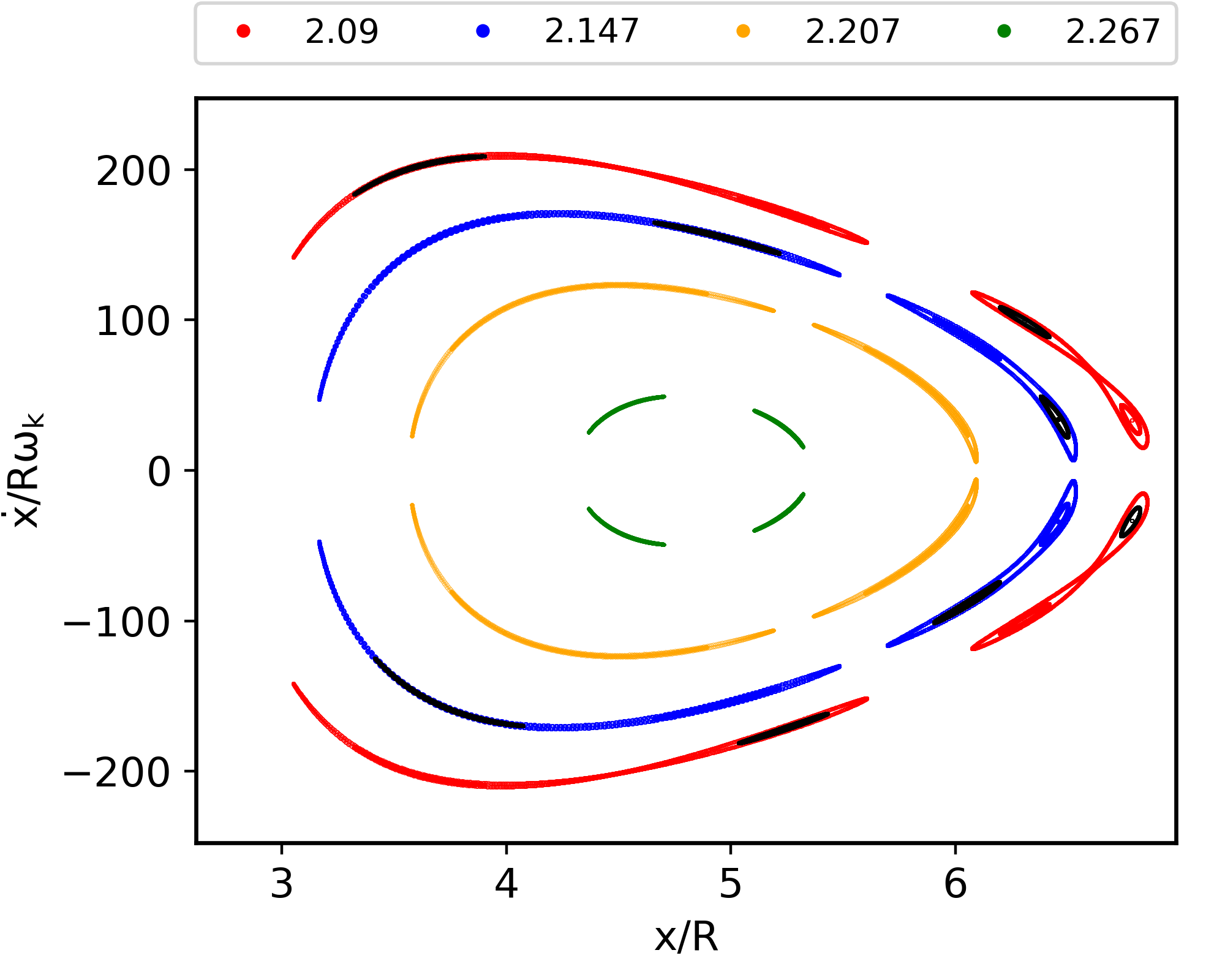}}
\centering
\subfigure[]{\includegraphics[width=0.8\columnwidth]{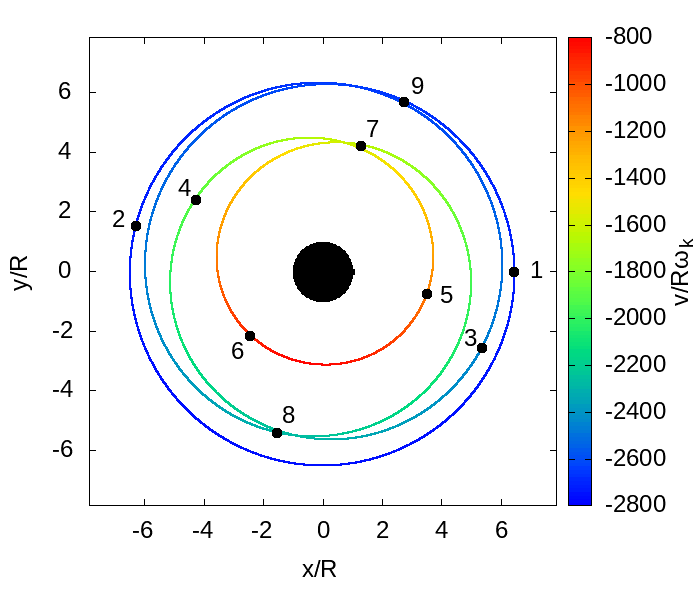}}
\caption{a) Poincar\'e surface of section for $C_J=2.147~{\rm R^2\omega_k^2}$. The periodic/quasi-periodic orbits of first kind are in black, the 1:5 resonance orbits are in brown and green and chaotic ones in red. b) Resonance islands for different values of $C_J$. The label on the panel gives the colour of the largest island for each value of $C_J$. c) Central orbit in the rotating frame of one of the families associated with the 1:5 resonance (in brown in the top panel) for $C_J=2.147~{\rm R^2\omega_k^2}$. The numbers and colours on the panel provide time evolution and the velocity in the rotating frame, respectively. \label{fig:15}}
\end{figure}

\begin{figure}
\centering
\subfigure[]{\includegraphics[width=0.8\columnwidth]{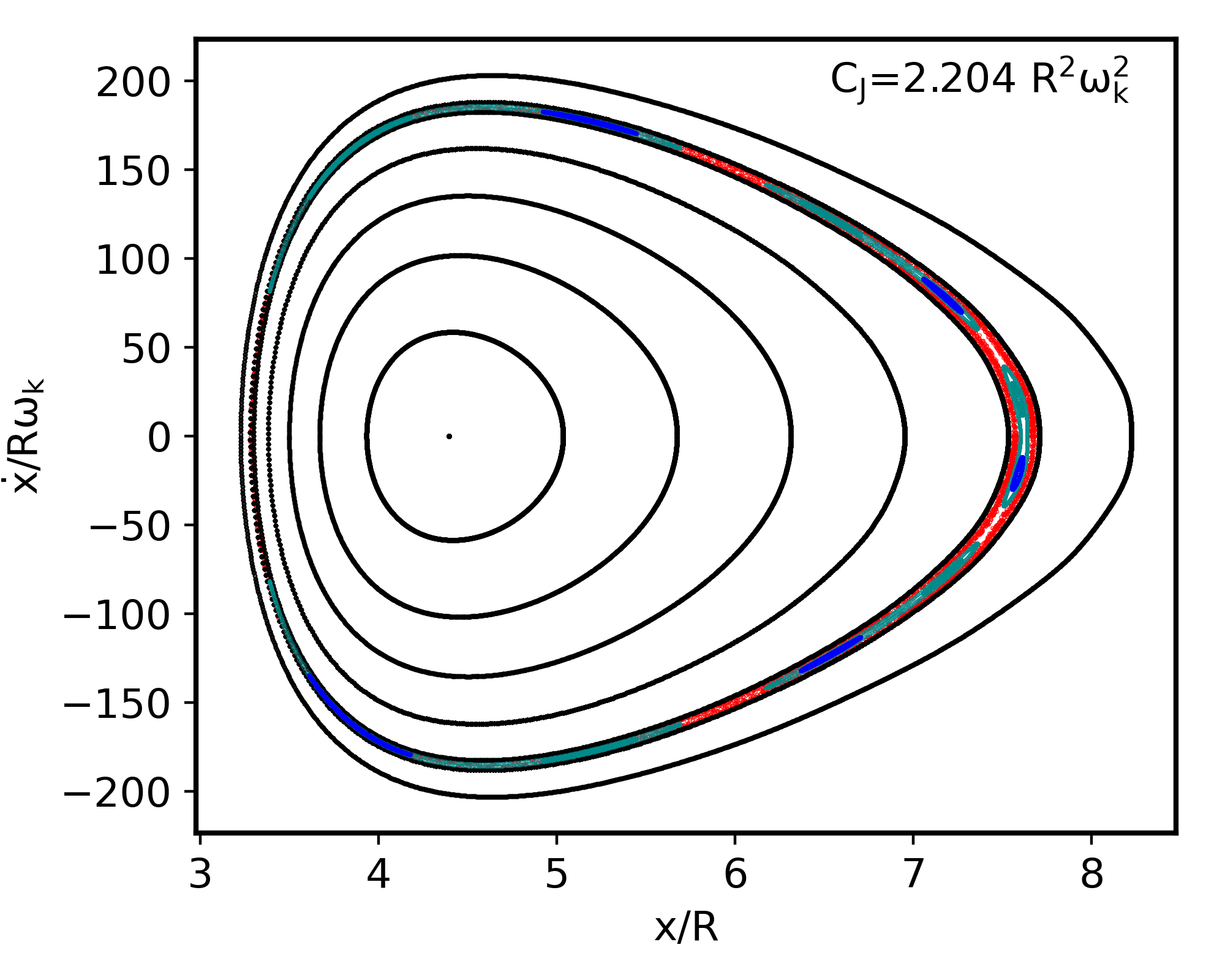}} 
\subfigure[]{\includegraphics[width=0.8\columnwidth]{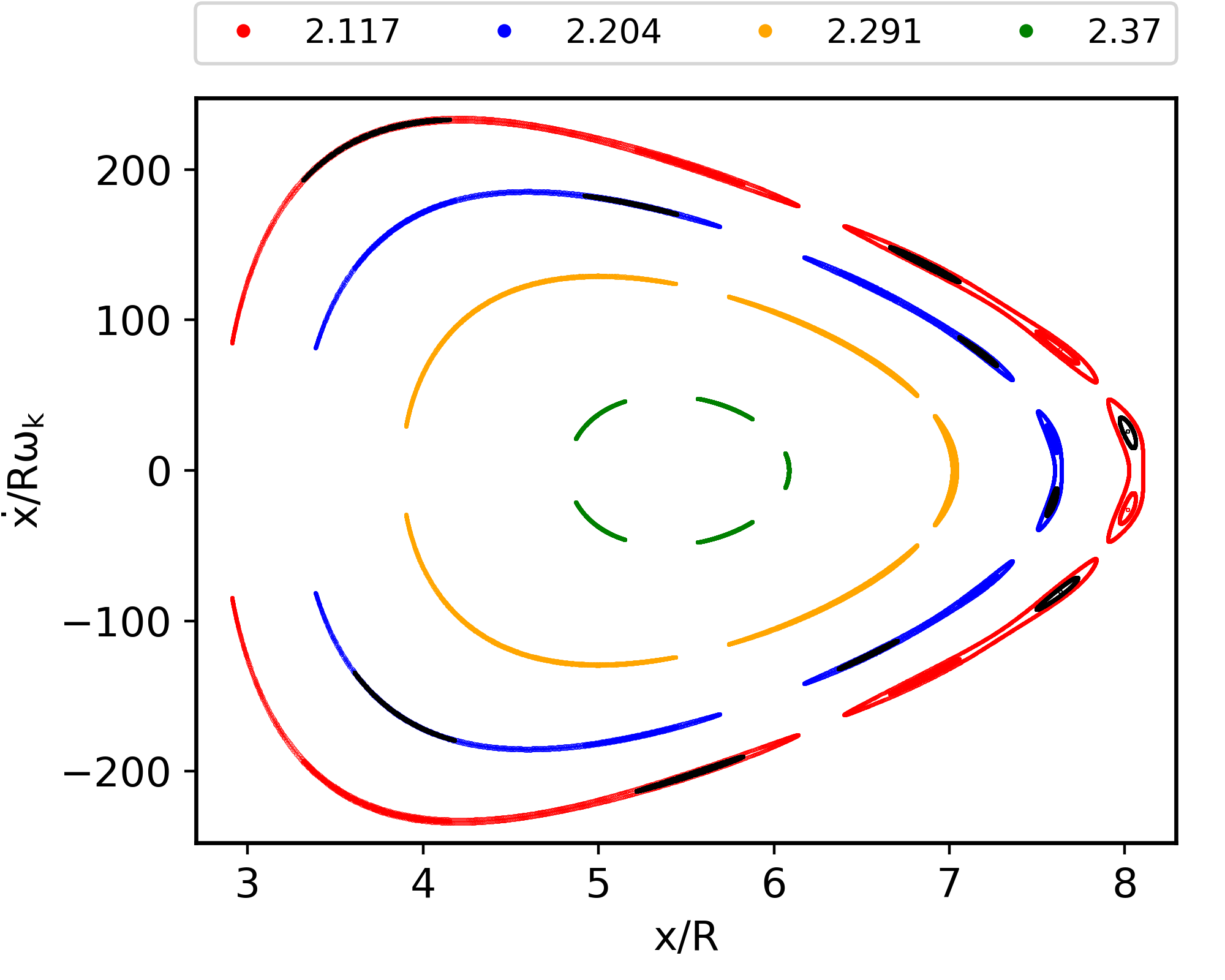}}
\centering
\subfigure[]{\includegraphics[width=0.8\columnwidth]{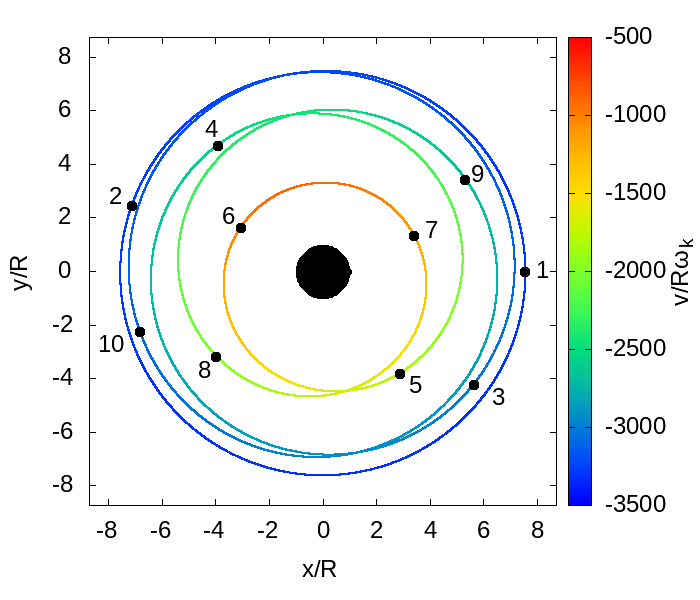}}
\caption{a) Poincar\'e surface of section for $C_J=2.204~{\rm R^2\omega_k^2}$. The periodic/quasi-periodic orbits of first kind are in black, the 1:6 resonance orbits are in cyan and blue and chaotic ones in red. b) Resonance islands for different values of $C_J$. The label on the panel gives the colour of the largest island for each value of $C_J$. c) Central orbit in the rotating frame of one of the families associated with the 1:6 resonance (in cyan in the top panel) for $C_J=2.204~{\rm R^2\omega_k^2}$. The numbers and colours on the panel provide time evolution and the velocity in the rotating frame, respectively. \label{fig:16}}
\end{figure}

In this appendix, we present Poincar\'e surface of sections for all resonances in the stable region of the reference case -- $\lambda=0.471$ and $\mu=10^{-3}$. All figures have three panels, following the pattern:
\begin{enumerate}
    \item[a)] the top panel shows the Poincar\'e surface of section of a broad region, with the value of $C_J$ given in the upper right corner of the figure. Black closed curves are periodic/quasi-periodic orbits of first kind. In red is the chaotic motion. Closed coloured curves correspond to islands of the resonance given on the figure label. In the case of $1:1+p$ resonances, we plot some islands with different colours to show the asymmetric libration of the resonance.
    \item[b)] the middle panel shows the entire evolution of the resonance. The label at the top gives the colour of the largest island for each $C_J$ value. In the case of $1:1+p$ resonances, we plotted some islands in black to highlight the asymmetric libration observed in such resonances.
    \item[c)] the bottom panel shows the trajectory of the central orbit of the resonance in the rotating frame. The value of $C_J$ is the same as the top panel. We show the temporal evolution of the orbit by numbers and dots equally spaced in time, while colour-coding gives the velocity in the rotating frame.
\end{enumerate}
The pattern above was not followed only for the 4:9 resonance, for which we do not show the middle panel. We got only one value of $C_J$ in the stable region for this resonance. Figures are given from the resonance closest to the central body to the farthest. Figures~\ref{fig:49}-\ref{fig:16} correspond to 4:9, 3:7, 2:5, 3:8, 1:3, 2:7, 1:4, 1:5, and 1:6 resonances, respectively.

%%%%%%%%%%%%%%%%%%%%%%%%%%%%%%%%%%%%%%%%%%%%%%%%%%

% Don't change these lines
\bsp	% typesetting comment
\label{lastpage}
\end{document}